\documentclass[12pt,english]{article}
\usepackage[T1]{fontenc}
\usepackage[latin9]{inputenc}
\usepackage{setspace}
\usepackage{makecell}
\usepackage{placeins}

\ExplSyntaxOn
\cs_if_exist:NF \expandableinput
  {
    \cs_new:Npn \expandableinput #1
      { \use:c { @@input } { \file_full_name:n {#1} } }
  }
\AddToHook{env/tabular/begin}
  { \cs_set_eq:NN \input \expandableinput }
\ExplSyntaxOff

\makeatletter
\usepackage{zimstyle}
\makeatother
\usepackage{babel}

\begin{document}

\title{\Large Does Peer-Reviewed Research Help Predict Stock Returns?
\footnotetext{First posted to arxiv.org: December 2022. E-mails: andrew.y.chen@frb.gov, Alejandro.Lopez-Lira@warrington.ufl.edu, tom.zimmermann@uni-koeln.de. Code: \url{https://github.com/chenandrewy/flex-mining}.  Data: \url{https://sites.google.com/site/chenandrewy/}.  We thank Alec Erb for excellent research assistance. Initial drafts of this paper relied on data provided by Sterling Yan and Lingling Zheng, to whom we are grateful. For helpful comments, we thank discussants: Leland Bybee, Yufeng Han, Theis Jensen, Jeff Pontiff, Shri Santosh, and Yinan Su. For helpful comments we also thank Svetlana Bryzgalova, Charlie Clarke, Mike Cooper, Albert Menkveld, Ben Knox, Emilio Osambela, Dino Palazzo, Matt Ringgenberg,  Dacheng Xiu, Lingling Zheng, and seminar participants at Auburn, Baruch, Emory, the Fed Board, Georgetown, Louisiana State, Universitat Pompeu Fabra, University of Kentucky, University of Utah, University of Wisconsin-Milwaukee, Virginia Tech,  MSU FCU, AFA, Arrowstreet Capital, NBER SI, and Stanford.  The views in this paper are not necessarily those of the Federal Reserve Board or the Federal Reserve System.} 
}
\author[1]{\normalsize Andrew Y. Chen} 
\author[2]{\normalsize  Alejandro Lopez-Lira} 
\author[3]{\normalsize Tom Zimmermann}

\affil[1]{Federal Reserve Board} 
\affil[2]{University of Florida}
\affil[3]{University of Cologne and Centre for Financial Research}

	% % anon version
	% \title{Does Peer-Reviewed Research Help Predict the Cross-Section of Stock Returns?}
	% \author[1]{} 

\date{\normalsize December 2025}

\maketitle
\thispagestyle{empty}

\setcounter{page}{0}
\begin{abstract}
\singlespacing \noindent \small Mining 29,000 accounting ratios for t-statistics $> 2.0$ leads to cross-sectional return predictability similar to the peer review process. For both, $\approx50\%$ of predictability remains after the original sample periods. This finding holds for many categories of research, including research with risk or equilibrium foundations. Only research agnostic about the theoretical explanation for predictability shows signs of outperformance. Our results imply that inferences about post-sample performance depend little on whether the predictor is peer-reviewed or data mined. They also have implications for the importance of empirical vs theoretical evidence, investors' learning from academic research, and the effectiveness of data mining.
\end{abstract}



\pagebreak{}

% !TEX root = ../risk_vs.tex
\section{Introduction}\label{sec:intro}

% Set context of peer review, different from McLean Pontiff, emphasize the efforts and thus importance
Academic finance has documented more than 200 cross-sectional stock return predictors (\citealt{ChenZimmermann2021}). The peer review process ensures these findings are supported by high quality evidence. It involves professors from prestigious universities and requires roughly five years to complete on average. 

% question is post-sample prediction (1). make question painfully clear (2).
This paper compares the post-sample performance of peer-reviewed predictors with data-mined benchmarks. Our goal is to estimate
\begin{equation}
	\label{eq:painfully_clear}
	E\left[\text{Post-Sample Performance}  
    \mid
    \text{In-Sample $t$-stat $>$ 2.0, Predictor Origin, Controls}
  \right],
\end{equation}
and measure how Predictor Origin (e.g. peer-reviewed vs data-mined) affects this conditional expectation. Estimates of Equation \eqref{eq:painfully_clear} are important to investors. They are also important to academics who care about post-sample robustness.

% discuss the data mined benchmark, explain why it is innocent (5)
To data mine, we search 29,000 accounting ratios for statistical evidence of predictive power. These ratios include all simple ratios and scaled first differences of variables that satisfy data availability requirements. Despite its naivete, this process generates out-of-sample returns of similar magnitude to those of the academic literature. This result replicates \citet{yan2017fundamental}, though our functional forms are not drawn from the modern predictability literature. 

In fact, data mining uncovers many of the same themes as peer-reviewed research. Mining the 1963-1980 sample, the most statistically-significant accounting ratios are related to investment (\citealt{titman2004capital}), debt issuance (\citealt{spiess1999long}), equity issuance (\citealt{loughran1995new}), accruals (\citealt{sloan1996stock}), inventory growth (\citealt{thomas2002inventory}), and earnings surprise (\citealt{watts1978systematic}). Notably, data mining could have uncovered most of these themes before they were published, sometimes long before.

\begin{figure}[!h]
  \caption{Does Peer-Reviewed Research Help Predict Returns?}
  % Shaded area shows one standard error, clustered by calendar time and predictor.
  \label{fig:intro}
  \vspace{0.00in}
  \centering
  \includegraphics[width=0.7\textwidth]{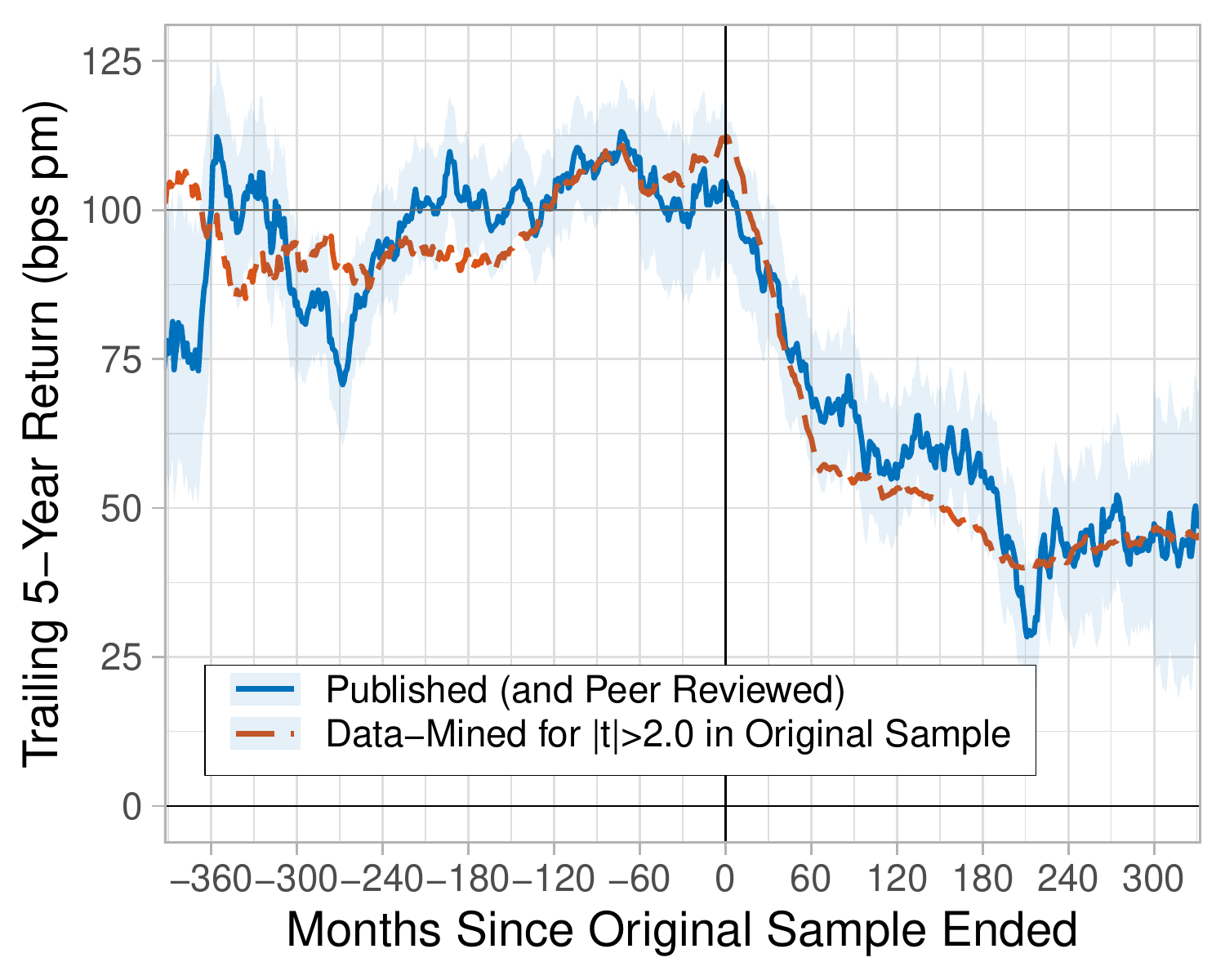} 
\end{figure}

% Finally, explain fig:intro
For a rigorous comparison, we construct data-mined benchmarks that control for sample periods and details of the performance measurement. Figure \ref{fig:intro} illustrates the result. The solid line plots the trailing 5-year long-short returns of published predictors from the \citet{ChenZimmermann2021} (CZ) dataset in event time, where the event is the end of the original sample period. The dashed line plots data-mined benchmarks. All strategies are normalized to have 100 bps mean return in the original samples, so both lines  hover around 100 before the original samples end.

Post-sample, the performance of both types of predictors decays to about 50\% of the original sample means. Data-mined returns decay a bit more than the published returns but the difference is small, both economically and statistically. For most of the plot, the data-mined benchmark is within one standard error of the published predictors (shaded area, clustered by calendar time and predictor). We find similar results controlling for CAPM and Fama-French 3 (FF3) + momentum exposure, among many other controls. Overall, the post-sample performance of peer-reviewed and data-mined predictors is remarkably similar.

Figure \ref{fig:intro} compares the \emph{average} publication to data mining. But peer-reviewed research is heterogeneous in many dimensions, including theoretical foundations, academic discipline (finance or accounting), and journal ranking. Could this heterogeneity lead to heterogeneous outperformance compared to data mining?

To answer this question, we categorize research along the theoretical explanation for predictability (risk, mispricing, or agnostic), equilibrium modeling (no model, stylized, dynamic, or quantitative), academic discipline (finance or accounting), and journal ranking (top 3 finance, top 3 accounting, other). These categorizations are made manually, by reading the papers. Quotes that justify our classifications are in our GitHub repository.\footnote{Research categories and justifications are found at \url{https://github.com/chenandrewy/flex-mining/blob/main/DataInput/SignalsTheoryChecked.csv}.} 

Based on manual reading, peer review attributes 60\% of findings to mispricing and 20\% to risk. For the remaining 20\%, peer review is agnostic about the theoretical explanation. The low prevalence of equilibrium modeling is consistent with mispricing being the most common explanation. 15\% of peer-reviewed predictability findings are supported with an equilibrium model of any sort. 

Of the 11 research categories, only research that is agnostic about the theoretical origin of predictability shows consistent outperformance compared to data mining. But even for this category, the outperformance is sensitive to the performance metric and modest overall. In terms of post-sample FF3 + momentum alphas, agnostic research retains an additional 31 percentage points of its original-sample performance compared to data mining. But this improvement falls to 9 p.p. when using raw long-short returns. While 31 p.p. may appear notable, it is the largest outperformance out of 33 estimates, and should be shrunk toward the average of about zero to account for multiple comparisons (e.g. \citealt{chen2020publication}).

Research that takes a stand on the theoretical origin of predictability shows little, if any, outperformance. In half of our tests, research that takes a stand \emph{under}performs data mining.

Additionally, we investigate whether predictability studied in \citet{fama1992cross} (B/M), \citet{jegadeesh1993returns} (momentum), and \citet{banz1981relationship} (size) outperform data mining. These findings are not only among the most renowned, but are arguably the ones with the strongest supporting evidence, both theoretical  (e.g., \citealt{gomes2003equilibrium}; \citealt{hong1999unified}; \citealt{berk1995critique}), and empirical (e.g., \citealt{fama1993common}; \citealt{asness2013value}). Despite this supporting evidence, the post-sample performance of these findings is on average similar to data mining.

The remainder of this section discusses implications and related literature. Section \ref{sec:dm} characterizes the data mining process. Section \ref{sec:pubvs} compares the average peer-reviewed predictor with data-mining. Section \ref{sec:hetero} examines heterogeneity in research methods, including theoretical foundations. Section \ref{sec:app-most-famous} compares \citet{fama1992cross}, \citet{jegadeesh1993returns}, and \citet{banz1981relationship} with data mining. Section \ref{sec:conclusion} concludes. Robustness is in the Appendix.

\subsection*{Implications and Relation to Literature}

% put all of the interpretations here, after we establish the main empirical results
% focus on lit for likely referees
The statistical implication is straightforward: Whether a trading strategy is found in a journal or is data mined has little effect on mean inferences about post-sample performance (Equation \eqref{eq:painfully_clear}). But a closer look leads to four deeper implications.

% since this implication is very in-your-face, carefully walk through the logic.
\textit{1. Predictive Content of Empirical vs Theoretical Evidence.} 
While the peer-reviewed predictors are supported by modern empirical and theoretical evidence, the data-mined predictors combine modern empirical evidence with the most basic of financial theories: that accounting ratios may predict firm performance. This idea goes back to the 1930s (\citealt{horrigan1968short}),%
\footnote{% highlight the conceptual contrast
Some accounting ratios go back to the 1890s, but \citet{horrigan1968short} credits \citet{wall1919credit} for inspiring a ``virtual explosion of publications on the subject of ratio analysis.'' By the 1930s, there were several studies of ratios' predictive content, including \citepos{smith1935changes} study of 21 ratios.} %
decades before the modern ideas of risk, psychology, and frictions that inform the peer-reviewed predictors. Comparing the post-sample performance of peer-reviewed and data-mined predictors, then, gives a sense of the relative importance of the two types of evidence. We provide a model of this comparison in Appendix \ref{sec:app-model}.

% mention that at least some of the theories are rigorous in the modern sense.
Our estimates imply that empirical evidence is more informative than theoretical evidence for identifying stable cross-sectional predictability. Having the support of modern, peer-reviewed theory does not predict higher post-sample performance compared to using the most basic of financial theories. This result applies to all types of theory in the CZ dataset, including quantitative equilibrium models, which are typically considered the gold standard. In contrast, modern empirical evidence appears to be a requirement for post-sample robustness. Indeed, the only type of research that consistently outperforms data mining is atheoretical research, which likely provides unusually strong empirical evidence, to make up for the lack of theory. 

A priori, the importance of empirics vs theory is unclear. Some argue that theory is important (\citealt{cochrane2009asset}; \citealt{harvey2017presidential}; \citealt{fama2018choosing}), motivated by concerns about data mining bias. This concern is compelling given the volatility of stock returns and the vast number of potential predictors. However, the flexibility of theory and ``model dredging'' may limit the helpfulness of theory (\citealt{sonnenschein1972market}; \citealt{fama1991efficient}). Even if theory does identify true predictors, it may identify unstable disequilibrium phenomena that decay with investor learning (\citealt{cochrane1999portfolio}; \citealt{mclean2016does}). These previous works discuss the question theoretically, but they do not empirically compare predictors with varying amounts of theoretical support.

% start with the McLean-Pontiff framework, which it seems is quite agreeable. This is important for this takeaway, since it's quite aggressive.
\textit{2. Investors' Learning from Academic Research.} Previous papers suggest that investors learn about mispricing from academic research. If this were the case, then investors would respond to academic findings by trading on academic strategies, decreasing predictability post-sample. Consistent with this hypothesis, \citet{mclean2016does} find predictability does decay, and more so than could be explained by statistical artifacts. Post-sample changes in trading activity also support this hypothesis (\citealt{mclean2016does}; \citealt{calluzzo2019anomalies}; \citealt{mclean2020taking}). These findings beg the question of whether investors learn about risk. 

Our results suggest that learning is asymmetric: while investors learn about mispricing from academic research, they do not seem to learn about risk. Following the logic of \citet{mclean2016does}, if investors learn about risk, one would see the opposite post-sample effect: investors would avoid academic risk-based strategies, increasing predictability post-publication. We find no evidence supporting this hypothesis and strong evidence against it.

It is possible that investors do learn about risk from academic research, but that these risks decay post-sample in a manner similar to mispricing-based and data-mined predictors. In our view, the simpler explanation is that the risks identified by academic research are not considered risks by investors. This explanation is consistent with survey evidence (\citealt{doran2007really}; \citealt{mukhlynina2020choice}; \citealt{chinco2022new}; \citealt{bender2022millionaires}).

\textit{3. Effectiveness of Data Mining.} Our findings imply that data mining is surprisingly effective for identifying true cross-sectional predictability. Indeed, data mining is competitive with the peer-review process in terms of effectiveness. 

These results add to the debate that began with \citepos{yan2017fundamental} pioneering study of 18,000 accounting ratios and 4,000 past return signals.\footnote{% begin footnote
Earlier studies by \citet{ou1989financial}, \citet{abarbanell1998abnormal}, and \citet{haugen1996commonality} successfully predict cross-sectional returns using many accounting signals. But they consider far fewer signals (up to 68) and do not focus on data mining bias.
} % end footnote
Yan and Zheng reject the null of no predictability based on both bootstrap and out-of-sample tests. However, followups to Yan and Zheng come to contradictory conclusions, based on technical statistical methods (\citealt{Chordia2020Anomalies}; \citealt{harvey2020false}; \citealt{chen2024most}; \citealt{gotofalse}). Our simple data mining process and straightforward out-of-sample tests help provide clarity to this debate.

More broadly, the literature on data mining goes back to \citet{Jensen1970Random}. Earlier studies focused on statistical theory (\citealt{Lo1990Data}) or market timing (\citeauthor{sullivan1999data} \citeyear{sullivan1999data}, \citeyear{sullivan2001dangers}), and found mixed results regarding effectiveness. Other recent studies examine cross-sectional predictability indirectly, using published predictors. Most of these studies find that data mining biases are small (\citealt{mclean2016does}; \citealt{chen2020publication}; \citealt{Jacobs2020Anomalies}; \citealt{jensen2022there};  \citealt{chen2025t}), though a few argue for sizeable biases (\citealt{harvey2016and}; \citealt{hasler2023looking}). By directly examining data-mined predictors, we obtain cleaner inferences. Following up on our paper, \citet{chen2023high} and \citet{marrow2024real} use empirical Bayes to mine data more rigorously.

\textit{4. Risk vs Mispricing in the Cross-Section.} Our findings are consistent with the view that mispricing is the primary driver of cross-sectional stock return predictability. We find that peer review is three times more likely to attribute predictability to mispricing than to risk. Moreover, predictors attributed to risk decay post-sample.

% what do other papers find?
These findings complement recent papers that also point toward mispricing as the primary driver. These papers use a wide range of methods, including announcement effects (\citealt{Engelberg2018Anomalies}; \citealt{frey2023stock}), stochastic dominance (\citealt{holcblat2022anomaly}), machine learning (\citealt{bali2023expected}), and subjective expectations data (\citealt{jensen2024subjective}).

% ===================

% ===================
% !TEX root = ../risk_vs.tex
\section{Data-Mined Predictability}\label{sec:dm}
We describe our data mining procedure and the predictability it uncovers.

\subsection{Data Mining Procedure}\label{sec:dm-data}

We begin with 241 Compustat annual accounting variables used in \citet{yan2017fundamental}. Yan and Zheng select these variables to (1) ensure non-missing values in at least 20 years and (2) that the average number of firms with non-missing values is at least 1,000 per year. We add CRSP market equity, leading to 242 ``ingredient'' variables. A more sophisticated selection would filter on data availability in real time. But given that data availability changes in large, positive, and permanent jumps (\citealt{easterwood2024do}), the more complicated procedure would likely yield similar results.

We then generate 29,315 accounting ratios (signals) using two functional forms: simple ratios ($X/Y$) and first differences scaled by a lagged denominator ($\Delta X / \text{lag}(Y)$). The numerator can use any of the 242 ingredients. The denominator is restricted to the 65 ingredients that are not zero for at least 25\% of firms in 1963 with matched CRSP data. This procedure leads to $\approx 242 \times 65 \times 2 = 31,460$ ratios, but we drop 2,145 ratios that are redundant in ``unsigned'' portfolio sorts.\footnote{For the $65\times65 = 4,225$ ratios where the numerator is also a valid denominator, there are only 65 choose 2 = 2,080 ratios that are in a sense distinct.}  

We lag each signal by six months relative to the fiscal year end, and then form long-short decile strategies by sorting stocks on the lagged signals in each June.  Delisting returns and other data handling methods follow \citet{ChenZimmermann2021}.  For further details, please see \url{https://github.com/chenandrewy/flex-mining}.

% motivations
This procedure aims to be the simplest possible data mining benchmark for peer-reviewed predictors. Accounting data is the modal data source for peer-reviewed predictors, representing roughly 50\% of the CZ dataset. The second most common data source is past returns, which represents only 20\%.  While it is possible to data mine both accounting data and past returns simultaneously (e.g. \citealt{chen2023high}), it requires several additional design choices and significantly complicates the analysis. 

Alternatively, one can motivate this procedure as a data mining benchmark that could have been constructed \emph{before} the bulk of the modern literature. The idea that accounting ratios may be informative about firm value goes back to the 1930s: 17 of the 26 chapters on stock selection in \citet{graham1934securityanalysis} focus on accounting statements. \citet{smith1935changes} examine 21 accounting ratios for their ability to predict financial distress (see also \citealt{ramser1931}; \citealt{fitzpatrick1932}; and \citealt{merwin1942}). These works were available decades before \citet{fama1973risk}. By the 50th anniversary of \citet{graham1934securityanalysis}, the book had become quite influential, and was celebrated in Warren \citepos{buffett1984superinvestors} popular speech and article. Thus, one well could have decided to search accounting ratios in 1984, when only 9 of the 212 predictors in the CZ dataset had been published. Indeed, \citet{ou1989financial} propose selecting stocks based on ``a large number of financial statement attributes,'' using a process in which ``[n]o conscious attempt is made to assess predictive ability on the basis of what we think should work.''

Our selection of accounting ratios contrasts with Yan and Zheng \citeyearpar{yan2017fundamental}, who use functional forms inspired, in part, by the asset pricing literature. Our procedure avoids the concern that such inspiration induces look-ahead bias. Nevertheless, previous versions of this paper used Yan and Zheng's data and found similar results. 

\subsection{Out-of-Sample Returns from Data Mining}\label{sec:dm-datasum}

Our data mining procedure generates notable out-of-sample returns, as seen in Panel (a) of Table \ref{tab:dm-oos}.  Each June, we sort the 29,000 accounting ratios into five bins based on their mean long-short returns over the past 30 years (in-sample) and compute the mean return over the next year within each bin (out-of-sample).  We then average these statistics across each year.  

\begin{table}[!h]
\caption{Descriptive Statistics of Data-Mined Accounting Strategies}
\label{tab:dm-oos}

\begin{singlespace}
\noindent The table describes data-mined accounting strategies using ``out-of-sample'' portfolio sorts (Panel (a)) and PCA (Panel (b)). Panel (a) sorts all ratios each June into 5 bins based on past 30-year long-short returns (in-sample) and computes the mean return over the next year within each bin (out-of-sample). Statistics are calculated by strategy, then averaged within bins, then averaged across sorting years. Decay is the percentage decrease in mean return out-of-sample relative to in-sample. We omit decay for bin 4 because the mean return in-sample is negligible.  Panel (b) applies PCA to ratios that have t-statistics greater than 2.0 in at least 10\% of the in-sample periods from Panel (a). Data-mined predictors resemble published ones in terms of in-sample performance, out-of-sample performance, and covariance structure.
\end{singlespace}
\begin{centering}
\vspace{-1ex}
\par\end{centering}
\centering{}\setlength{\tabcolsep}{0.9ex} \small
\begin{center}
% ==== begin paste
% Table generated by Excel2LaTeX from sheet 'dm summary'
\begin{tabular}{crrrrrrrrrrrr} \toprule 
\multicolumn{13}{c}{Panel (a): ``Out-of-Sample'' Returns of All Ratios 1994-2020} \\ 
\midrule
In- &   & \multicolumn{5}{c}{Equal-Weighted Long-Short Deciles} &   & \multicolumn{5}{c}{Value-Weighted Long-Short Deciles} 
\\ \cmidrule{3-7}\cmidrule{9-13}Sample &   & \multicolumn{2}{c}{Past 30 Years (IS)} &   & \multicolumn{2}{c}{Next Year (OOS)} &   & \multicolumn{2}{c}{Past 30 Years (IS)} &   & \multicolumn{2}{c}{Next Year (OOS)} \\ 
\cmidrule{3-4}\cmidrule{6-7}\cmidrule{9-10}\cmidrule{12-13}Bin &   & \multicolumn{1}{c}{Return} & \multicolumn{1}{c}{\multirow{2}[2]{*}{t-stat}} &   & \multicolumn{1}{c}{Return} & \multicolumn{1}{c}{Decay} &   & \multicolumn{1}{c}{Return} & \multicolumn{1}{c}{\multirow{2}[2]{*}{t-stat}} &   & \multicolumn{1}{c}{Return} & \multicolumn{1}{c}{Decay} \\ 
&   & \multicolumn{1}{c}{(bps pm)} &   &   & \multicolumn{1}{c}{(bps pm)} & \multicolumn{1}{c}{(\%)} &   & \multicolumn{1}{c}{(bps pm)} &   &   & \multicolumn{1}{c}{(bps pm)} & \multicolumn{1}{c}{(\%)} \\ 
\cmidrule{1-1}\cmidrule{3-4}\cmidrule{6-7}\cmidrule{9-10}\cmidrule{12-13}
% latex table generated in R 4.2.3 by xtable 1.8-4 package
% Mon Mar  3 17:09:07 2025
 1 &  & -59.0 & -4.20 &  & -47.3 & 19.8 &  & -37.7 & -2.05 &  & -16.0 & 57.7 \\ 
  2 &  & -29.1 & -2.45 &  & -18.4 & 36.8 &  & -15.9 & -1.03 &  & -5.8 & 63.5 \\ 
  3 &  & -13.5 & -1.23 &  & -4.6 & 65.9 &  & -5.4 & -0.37 &  & -3.0 & 43.4 \\ 
  4 &  & -0.8 & -0.09 &  & 3.7 &  &  & 4.6 & 0.31 &  & -1.0 &  \\ 
  5 &  & 21.3 & 1.40 &  & 14.6 & 31.5 &  & 24.6 & 1.31 &  & 6.2 & 74.7 \\
\bottomrule \end{tabular}% 
\vspace{2ex}

% pca panel
\setlength{\tabcolsep}{1.5ex}
% latex table generated in R 4.2.3 by xtable 1.8-4 package
% Mon Mar  3 17:13:39 2025
\begin{tabular}{clccccccccccc}
  \toprule
 & \multicolumn{11}{c}{Panel (b): PCA Explained Variance of Predictive Ratios (\%)} \\
  \midrule
Number of PCs & 1 & 5 & 10 & 20 & 30 & 40 & 50 & 60 & 70 & 80 & 90 & 100 \\ 
\midrule
  Equal-Weighted & 23 & 50 & 58 & 67 & 72 & 75 & 78 & 81 & 83 & 84 & 86 & 87 \\ 
  Value-Weighted & 16 & 41 & 50 & 61 & 68 & 73 & 77 & 80 & 82 & 85 & 87 & 88 \\ 
   \bottomrule
\end{tabular}

\end{center} 
\end{table}

Using equal-weighted strategies, the in-sample returns of the first bin are on average -59 bps per month, with an average t-stat of -4.2.  These statistics are similar to those of the typical published predictor (\citet{ChenZimmermann2021}).  Out-of-sample, the first bin returns -47 bps per month, implying a  decay of only  20\%, once again resembling published predictability (\citet{mclean2016does}).  Since investors can flip the long and short legs of these strategies, these statistics imply substantial out-of-sample returns.  Similar predictability is seen in bin 5, which decays by 32\%.  

Out-of-sample predictability is also seen in value-weighted strategies but with smaller magnitudes. Still, the out-of-sample returns monotonically increase in the in-sample return, indicating the presence of true predictability.  Moreover, the roughly  60\% decay is far from 100\%, and is in the ballpark of the post-sample decay for published predictors.  Similarly, out-of-sample predictability is much weaker post-2004, though it still exists (see Appendix Table \ref{tab:dm-sum-2004}). The concentration of predictability in small stocks and the pre-2004 sample is also found in published predictors (\citealt{chen2022zeroing}). 

\subsection{Data-Mined Predictability Themes}\label{sec:dm-themes}

Since there are 29,000 data-mined signals, Panel (a) of Table \ref{tab:dm-oos} implies thousands of strategies with notable out-of-sample predictability. But how many distinct themes are in these strategies? 

Panel (b) of Table \ref{tab:dm-oos} helps address this question. It applies PCA to the predictive ratios: ratios that have t-statistics greater than 2.0 in at least 10\% of the 30-year in-sample periods from Panel (a). 

PCA results in a non-trivial factor structure: the first five PCs explain about 50\% of total variance. However, many dozens of PCs are required to span 80\% of total variance. A similar variance decomposition is seen in published predictors (\citealt{kozak2018interpreting}; \citealt{bessembinder2021time}; \citealt{chen2023missing}). Pairwise correlations lead to a similar conclusion (Internet Appendix Table \ref{tab:dm-cor}).

% named themes methods
One can alternatively examine the themes by examining the numerators that generate the very largest t-statistics. Table \ref{tab:dm-theme} does this by reporting the 20 numerator and stock weight (equal- or value-) combinations that produce the largest mean t-stats, where the mean is taken across the 65 possible denominators. The table uses the 1963-1980 sample, but similar themes are found in other samples (Internet Appendix \ref{sec:intapp-themes}).

\begin{table}
\caption{Themes from Mining Accounting Ratios in 1980}\label{tab:dm-theme}
\small

Table reports the 20 accounting ratio numerator and stock weight (equal- or value-) combinations with the largest mean t-stats using returns in the years 1963-1980 (IS). `ew' is equal-weight, `vw' is value-weight. We manually group numerators into themes from the literature. Strategies are signed to have positive mean returns IS. `Pct Short' is the share of strategies that short stocks with high  ratios.  `t-stat' and  `Mean Return' are averages across the 65 possible denominators. `Mean Return' is in bps per month. `Mean return OOS/IS' is the mean in either 1981-2004 or 2005-2022 (OOS),  divided by the mean IS. Data mining can uncover themes from the literature before they are published.
\vspace{2ex}

\centering

\setlength{\tabcolsep}{2.0ex}

\begin{tabular}{lrrrlrr}
\toprule
 & \multicolumn{3}{c}{1963-1980 (IS)} &  & \multicolumn{1}{c}{1981-2004} & \multicolumn{1}{c}{2005-2023} \\ \cmidrule{2-4} \cmidrule{6-7}
Numerator (Stock Weight) & \multicolumn{1}{c}{Pct} & \multirow{2}{*}{t-stat} & \multicolumn{1}{c}{Mean} &  & \multicolumn{2}{c}{Mean Return} \\  
 & \multicolumn{1}{c}{Short} &  & \multicolumn{1}{c}{Return} &  & \multicolumn{2}{c}{OOS / IS} \\ 
\midrule
\multicolumn{7}{l}{Investment / Investment Growth (Titman-Wei-Xie \citeyear{titman2004capital}; Cooper-Gulen-Schill \citeyear{cooper2008asset})} \\ \hline
$\Delta$Assets (ew) & 100 & 4.0 & 0.86 &  & 1.05 & 0.32\\
$\Delta$PPE net (ew) & 98 & 4.0 & 0.79 &  & 1.08 & 0.20\\
$\Delta$Intangible assets (ew) & 100 & 4.0 & 0.52 &  & 1.04 & 0.26\\
$\Delta$PPE gross (ew) & 98 & 3.8 & 0.76 &  & 1.00 & 0.14\\
$\Delta$Invested capital (ew) & 100 & 3.5 & 0.73 &  & 1.35 & 0.34\\
$\Delta$Capital expenditure (ew) & 100 & 3.2 & 0.43 &  & 1.54 & 0.46 \bigstrut[b] \\ 
\multicolumn{7}{l}{External Financing (\citealt{loughran1995new}; \citealt{spiess1999long})} \\ \hline
$\Delta$Common stock (ew) & 100 & 5.1 & 0.81 &  & 0.66 & 0.34\\
$\Delta$Liabilities (ew) & 100 & 4.7 & 0.80 &  & 0.79 & 0.28\\
$\Delta$Capital surplus (ew) & 100 & 4.2 & 0.61 &  & 1.19 & 0.99\\
$\Delta$Long-term debt (ew) & 100 & 3.6 & 0.47 &  & 1.43 & 0.23\\
$\Delta$Capital surplus (vw) & 98 & 3.0 & 0.54 &  & 0.93 & 0.54 \bigstrut[b] \\ 
\multicolumn{7}{l}{Accruals / Inventory Growth (\citealt{sloan1996stock}; \citealt{thomas2002inventory})} \\ \hline
$\Delta$Inventories (ew) & 100 & 4.2 & 0.66 &  & 1.22 & 0.22\\
$\Delta$Notes payable st (ew) & 100 & 3.8 & 0.44 &  & 0.57 & 0.25\\
$\Delta$Receivables (ew) & 100 & 3.7 & 0.67 &  & 0.59 & 0.33\\
$\Delta$Debt in current liab (ew) & 100 & 3.7 & 0.43 &  & 0.73 & 0.28\\
$\Delta$Current liabilities (ew) & 100 & 3.7 & 0.51 &  & 1.32 & 0.22 \bigstrut[b] \\ 
\multicolumn{7}{l}{Earnings Surprise (\citealt{watts1978systematic}; \citealt{foster1984earnings})} \\ \hline
$\Delta$Cost of goods sold (ew) & 100 & 3.7 & 0.60 &  & 0.87 & 0.23\\
$\Delta$Operating expenses (ew) & 100 & 3.5 & 0.58 &  & 0.99 & 0.35\\
$\Delta$SG\&A (ew) & 100 & 3.3 & 0.62 &  & 1.04 & 0.25\\
$\Delta$Interest expense (ew) & 98 & 3.3 & 0.47 &  & 1.38 & 0.73\\
\bottomrule
\end{tabular}

\end{table}

% describe themes
All of the top 20 numerators fit into themes from the cross-sectional literature.  These themes include investment (\citet{titman2004capital}), debt issuance (\citet{spiess1999long}), share issuance (\citet{loughran1995new}), accruals (\citet{sloan1996stock}),  inventory growth (\citet{thomas2002inventory}), and earnings surprise (\citet{watts1978systematic}). For all of these themes, the sign of predictability obtained from  data mining is the same as the sign from the literature (e.g. short stocks with high investment).  Notably, most of these themes are published after 1980, sometimes long after.

% discuss decay
The predictive power of these themes persists out-of-sample (``OOS/IS'' columns). All data-mined numerators produce positive mean returns after 1980. In fact, the return decay in the 1981-2004 out-of-sample period is on average zero. 

% hint at the main results
Taken together, these results hint at our main finding. Data-mined predictability resembles that of peer-reviewed research. This resemblance is seen in performance both in- and out-of-sample (Table \ref{tab:dm-oos}, Panel (a)), covariance structure (Table \ref{tab:dm-oos}, Panel (b)), and themes (Table \ref{tab:dm-theme}).

% !TEX root = ../risk_vs.tex
\section{Research vs Data Mining}\label{sec:pubvs}

We compare in detail the post-sample returns of peer-reviewed research to data mining. 

\subsection{Peer-Reviewed Predictor Data} \label{sec:pubvs-data-pub}
% see DataCounts.md

Peer-reviewed predictors come from the October 2024 release of the \citet{ChenZimmermann2021}  (CZ) dataset.  This dataset is built from 212 firm-level variables that were shown to predict returns cross-sectionally in academic journals. It covers the vast majority of firm-level predictors that can be created from widely-available data and were published before 2016. The CZ data is a uniquely accurate representation of the literature: unlike other large-scale replications, CZ show that their t-stats are generally a good match for the t-stats in the original papers.

CZ select their predictors to provide comprehensive coverage of predictors examined in previous  meta-studies (\citealt{mclean2016does}; \citealt{harvey2016and}; \citealt{green2017characteristics}; \citealt{Hou2020Replicating}). These meta-studies, in turn, aim for comprehensive coverage of academic cross-sectional stock return predictors. 

We drop five predictors that have fewer than 9 years of post-sample returns.  These predictors use specialized data sources that have been discontinued (e.g., the \citet{gompers2003corporate} governance index). This restriction makes our charts easier to interpret.

We drop an additional 29 predictors to ensure each paper is represented by at most 2 predictors. For papers that present more than 2 predictors, we only include the two predictors with the largest in-sample t-statistics. This filter ensures our results are representative of the literature, and do not over-weight papers that present numerous implementations of the same theme (e.g., \citepos{heston2008seasonality} seasonal momentum).\footnote{Previous versions of our study without this filter show very similar results.}

\subsection{Post-Sample Performance: Research vs Data Mining}\label{sec:pubvs-results}

We can now answer the question posed on page \pageref{sec:intro}. Does peer-reviewed research help predict cross-sectional returns compared to data mining?  

In our first test, we measure performance using the mean long-short return of strategies formed following the ``original paper'' specifications from CZ. These specifications match the original papers' predictability tests in terms of stock weighting and portfolio sorting (e.g., equal-weighted quintiles). We keep only predictors that produce mean long-short return t-stats $>$ 2.0 in the original samples. All strategies are signed to be positive and normalized to have 100 bps mean return in the original samples for ease of interpretation.

For each of these published strategies, we construct a data-mined benchmark by applying the same statistical treatment to the 29,000 accounting ratios. For each ratio, we construct t-stats based on the original papers' stock weighting and sample periods, filter for t-stats $>$ 2.0, sign and normalize to have $+100$ bps mean return in the original samples. Since finding t-stats $>$ 2.0 is rather common (Table \ref{tab:dm-oos}), on average, this process selects roughly 6,000 data-mined strategies for each published predictor. 

Figure \ref{fig:intro} compares the post-sample performance of the published strategies to their data-mined benchmarks. It plots trailing 5-year mean returns in event time, where the event is the end of the original sample periods.  

Post-sample, peer-reviewed (solid line) and data-mined (long-dash) predictors perform similarly. This similarity is seen both in the average post-sample performance, as well as in the event-time decay patterns. This result suggests that peer-reviewed research provides limited additional information about post-sample performance compared to data mining. 

\begin{figure}[!h]
  \caption{Factor Adjustments and Alternative Data Mining Methods}
  \label{fig:pub-vs-dm-2}
  Top panels repeat Figure \ref{fig:intro} using abnormal returns $r^{a}_{i,t} = r_{i,t}- \hat{\beta}_i^{(s)} f_t$, where $\hat{\beta}_i^{(s)}$ is estimated using sample-specific periods (in-sample or post-sample) and $f_t$ is the return of either the market less risk-free rate (CAPM) or the Fama-French three factors plus momentum (FF4). Bottom panels use alternative data mining methods: mining 3,000 ticker-based strategies (dashed line) or filtering for the top 5\% of t-stats (Panel (d)). Each predictor is normalized so that its mean original-sample return is 100 bps per month. Shaded area shows one standard error for the published predictors, clustered by calendar month and predictor. Figure \ref{fig:intro} is robust to factor adjustments. The statistical screen and number of strategies used for data-mining are unimportant but the type of data being mined is critical.

  \vspace{0.15in}
  
  \centering
  \subfloat[CAPM]{
  \includegraphics[width=0.48\textwidth]{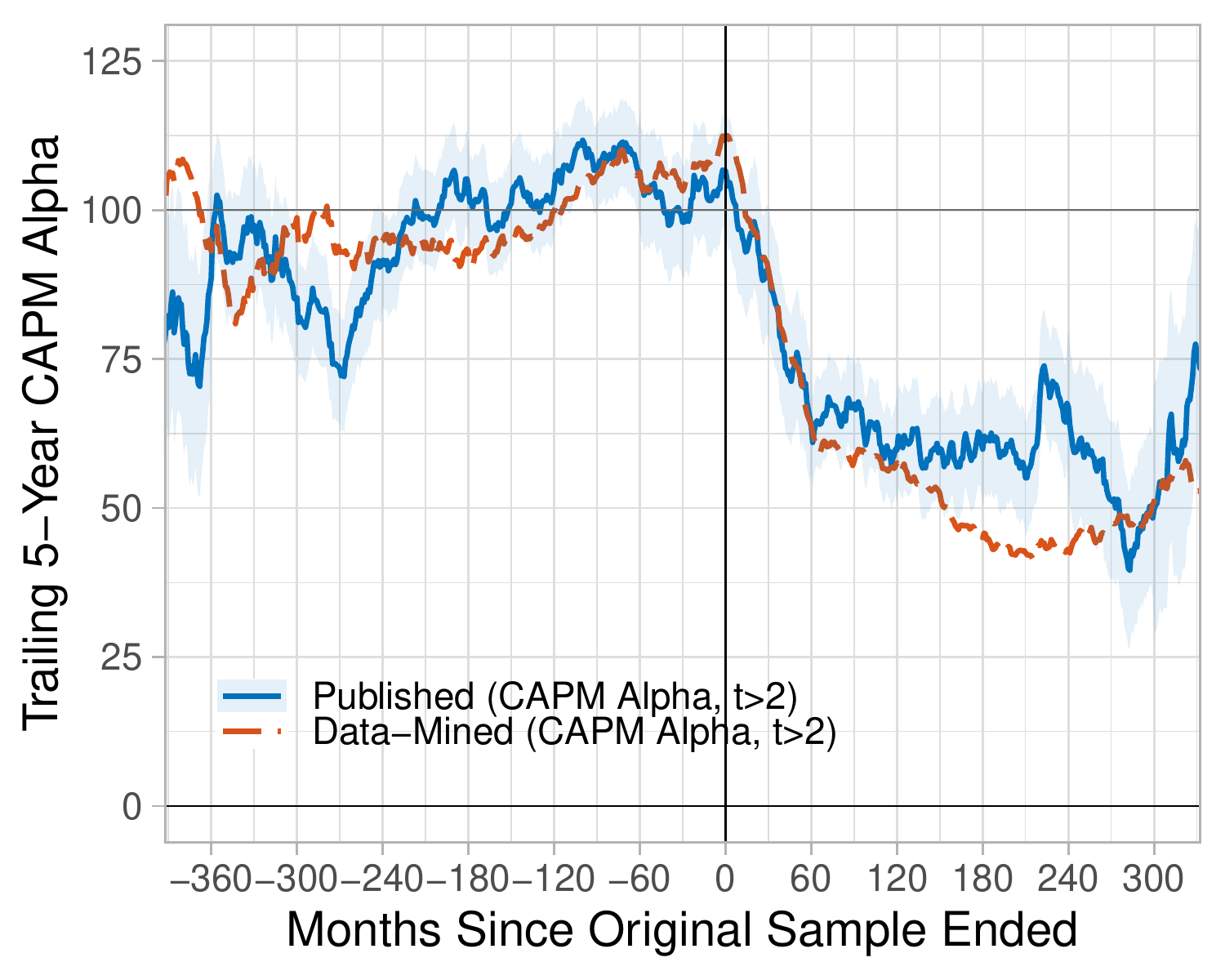} 
  }
  \subfloat[FF3+Momentum]{
  \includegraphics[width=0.48\textwidth]{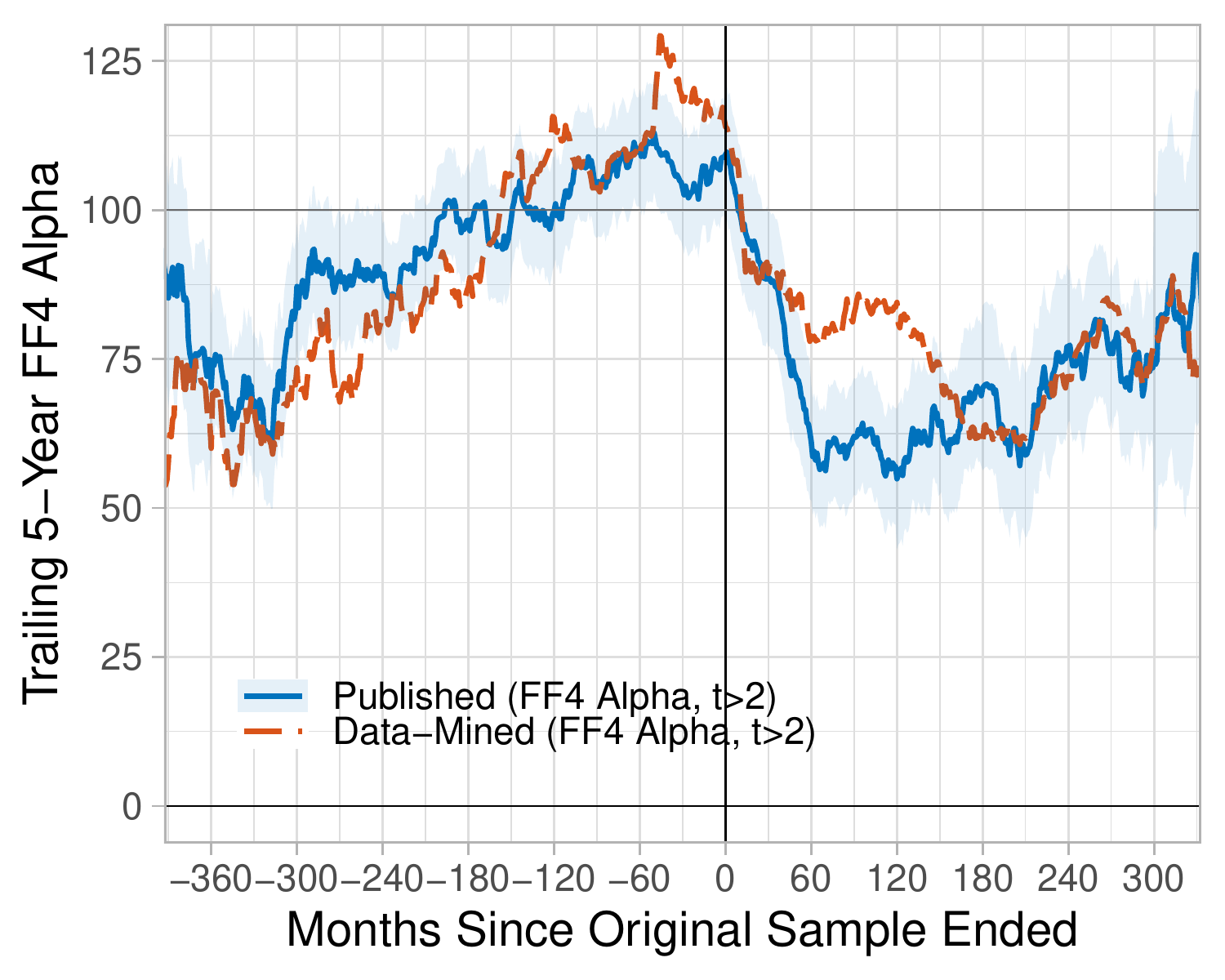}
  }\\
  \subfloat[Mining for t-stats $>$ 2.0 ]{
  \includegraphics[width=0.48\textwidth]{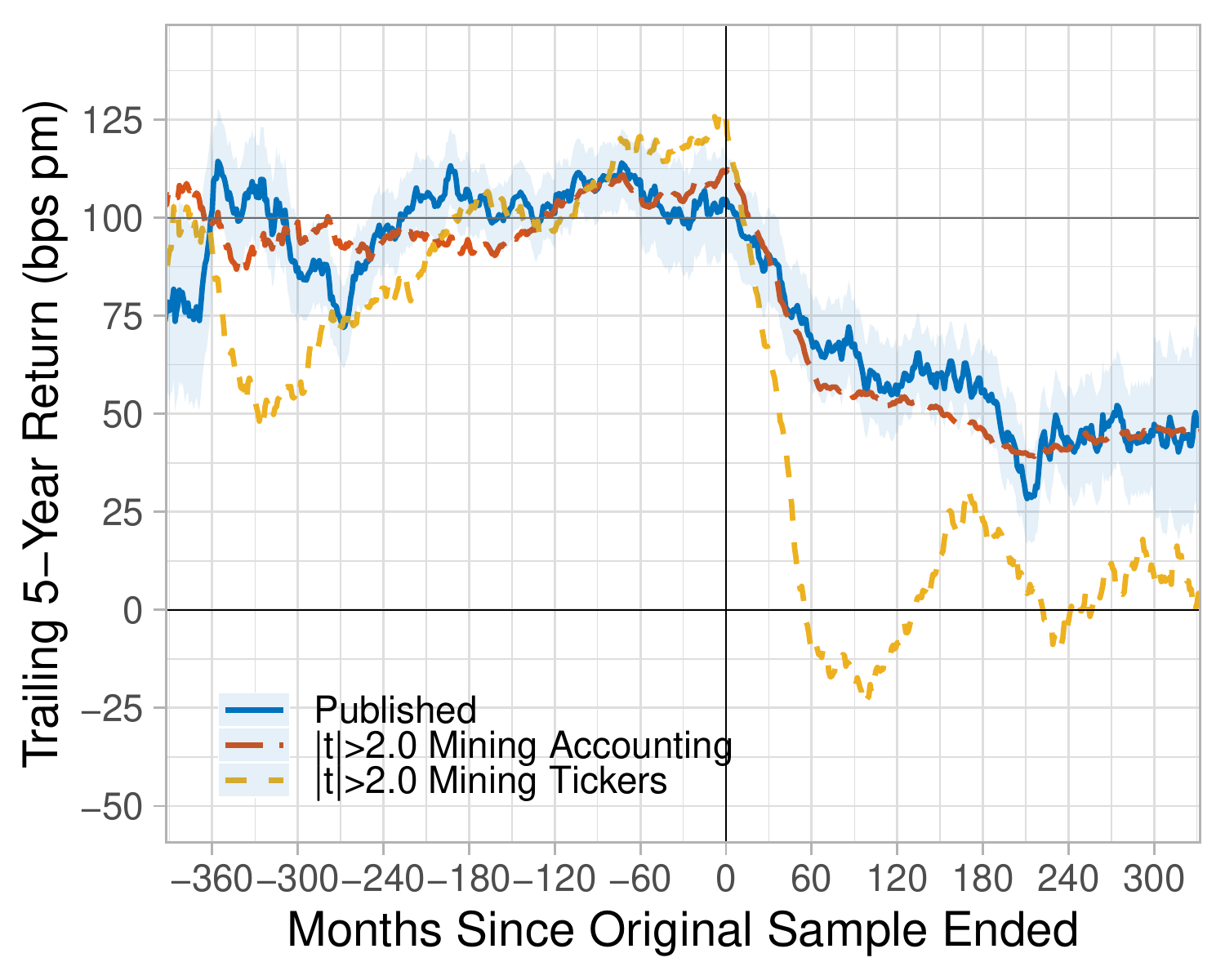} 
  }
  \subfloat[Mining for top 5\% of t-stats]{
  \includegraphics[width=0.48\textwidth]{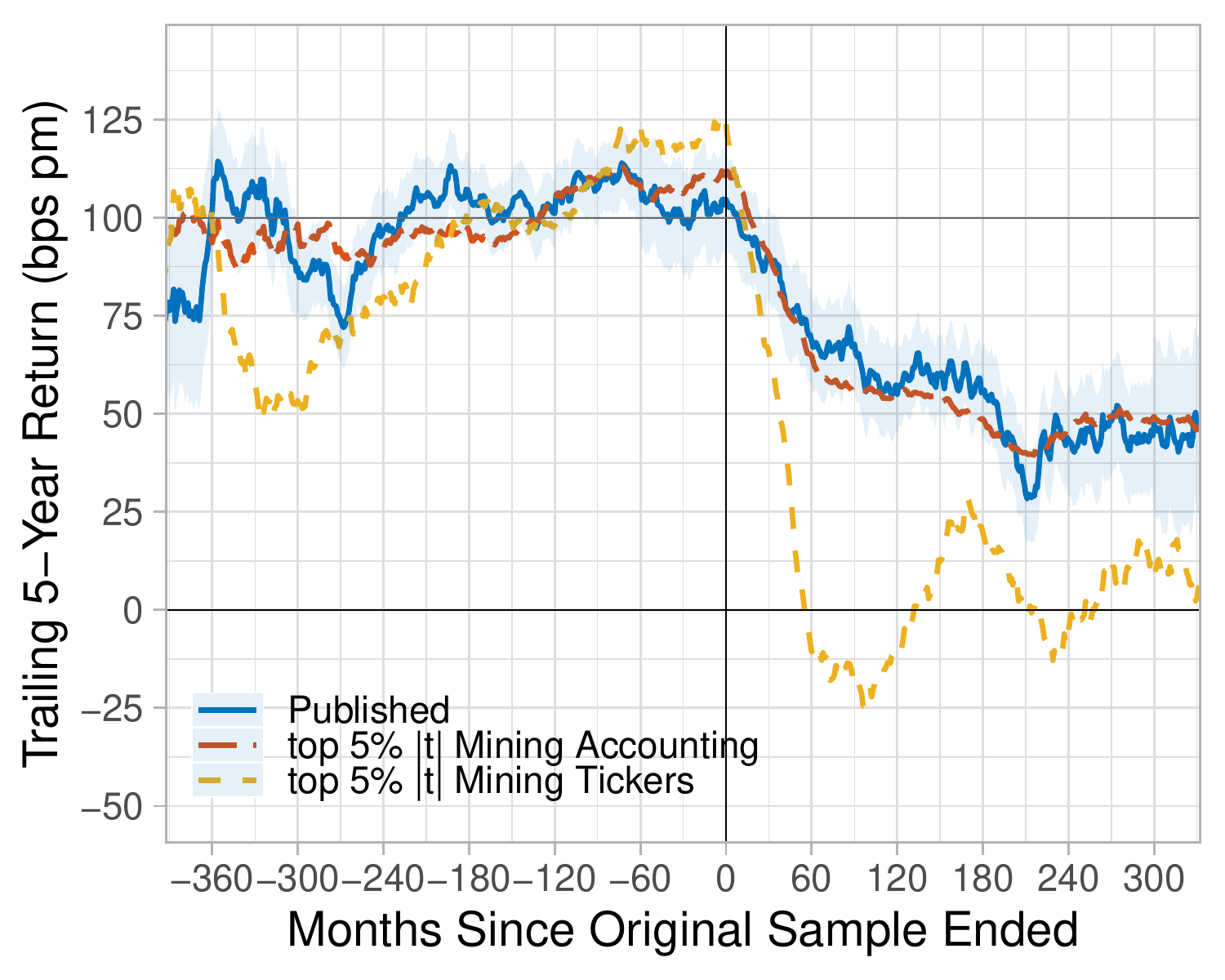}
  }
\end{figure}

Figure \ref{fig:pub-vs-dm-2} shows robustness to factor exposure. We calculate the abnormal return $r^{a}_{i,t}$ of strategy $i$ in month $t$:
\begin{align}
  r^{a}_{i,t} = r_{i,t}- \hat{\beta}_{i}^{(s)} f_t,
  \label{eq:abnormal_return}
\end{align}
where $r_{i,t}$ is the raw long-short return, $f_t$ is a vector of factor returns, and $\hat{\beta}_{i}^{(s)}$ is a row vector of betas estimated using sample $s$. The sample $s$ is either the original sample or the post-sample period. This separation ensures our t-stat filters do not produce look-ahead bias, but using full-sample betas leads to similar results (Internet Appendix \ref{sec:intapp-fullsamplerisk}). We then repeat the test in Figure \ref{fig:intro} using abnormal returns in place of the raw returns. This replacement is made throughout the analysis, implying that the t-stat $>$ 2.0 filters also use abnormal rather than raw returns.

Panel (a) of Figure \ref{fig:pub-vs-dm-2} shows that published predictors exhibit some outperformance relative to data-mined benchmarks in terms of CAPM-adjusted returns. However, the outperformance is modest. Post-sample, published predictors retain 60\% of their original-sample performance, compared to 52\% for data-mined benchmarks. Indeed, Panel (b) shows that if we measure abnormal returns relative to the Fama-French three factors + momentum, data mining slightly \emph{out}performs research. Overall, post-sample alphas are similar for published and data-mined predictors.

Further robustness is seen using many other controls. These include controls for research methods and quality (Sections \ref{sec:hetero} and \ref{sec:app-most-famous}), in-sample mean returns, in-sample t-statistics, and data sources (Appendix \ref{sec:app-robust}). We also find robustness to excluding data-mined predictors that are highly correlated with published predictors (Appendix \ref{sec:app-robust-manycor}).

\subsection{Alternative Data Mining Methods}\label{sec:pubvs-alt}

Our data mining process (Section \ref{sec:dm-data}) searches accounting ratios for t-stats $>$ 2.0. This method avoids look-ahead bias and is arguably the most straightforward way to mine data. However, one can think of alternative data mining methods, and use these to inspect the mechanism behind Figure \ref{fig:intro}. 

For example, one could use the data mining method proposed in \citet{harvey2017presidential}. Harvey asks his research assistant to ``form portfolios based on the first, second, and third letters of the ticker symbol,'' leading to 3,160 long-short portfolios. We interpret his instructions as follows:  Generate 26 portfolios by going long all stocks with a  first ticker letter of ``A,'' ``B,'' ``C,'' ..., and ``Z.''  Generate 26 portfolios by doing the same for the second ticker letter, and add a 27th portfolio for tickers that have no second ticker letter.  Apply the same to the third ticker.  This process results in $26 + 27 + 27 = 80$ long portfolios.  Finally, form $\binom{80}{2} = 3,160$ long-short portfolios by selecting all distinct pairs of the 80 long portfolios.  

The bottom panels of Figure \ref{fig:pub-vs-dm-2} compare published strategies to strategies based on ticker mining. Panel (c) applies the same predictability screen as in Figure \ref{fig:intro}: we screen for t-stats $>$ 2.0 in the original sample periods. Post-sample, the mean returns from mining tickers (short-dash line) are approximately zero. Thus, peer-reviewed research (solid line) is much more helpful for predicting returns than mining tickers. 

Panel (d) applies an alternative predictability screen: we screen for the top 5\% of t-stats in the original sample periods. Screening accounting ratios this way still leads to very similar returns to research. Screening tickers this way still leads to post-sample returns of around zero.

Figure \ref{fig:pub-vs-dm-2} illustrates two lessons about data mining. The first is that the data being mined is important. Accounting data are helpful for predicting returns, while ticker data are not. The second lesson is that mining more data does not necessarily mean worse out-of-sample performance. The accounting dataset is almost 10 times as large as the ticker dataset, yet it produces much stronger post-sample returns.  This result is consistent with false discovery controls, which typically depend on the distribution of test statistics rather than the number of tests (\citealt{benjamini1995controlling}; \citealt{chen2025t}).

% Summarize and interpret a bit
In summary, the average peer-reviewed publication leads to post-sample returns that are similar to those of a simple data mining process.

% ===============================================================

% !TEX root = ../risk_vs.tex
\section{Heterogeneous Research and Outperformance}\label{sec:hetero}

Peer-reviewed publications differ in many dimensions, including theoretical foundations, discipline of origin, and outlet quality. This section examines these differences and whether they matter for predictive outperformance compared to data mining.

% == data summary == #
\subsection{Research Categorization Methods\label{sec:hetero-method}}

We categorize research at the predictor-paper level, along four dimensions:
\begin{enumerate}
   \item Theoretical foundation: is the theoretical explanation for the predictor risk or mispricing? Or is there no clear theoretical explanation (agnostic)?
   \item Equilibrium modeling: is the theoretical explanation supported by a stylized model, dynamic model, or quantitative model?
   \item Academic Discipline: is the paper published in a finance, accounting, or other discipline's journal?
   \item Quality: is the research published in a top-3 finance or top-3 accounting journal?
\end{enumerate}
These dimensions are key characteristics of asset pricing research. Conferences and societies are organized around the first three. Journal quality is important for tenure. 

We apply these classifications based on manual reading of the original papers. All classifications can be found at \url{https://github.com/chenandrewy/flex-mining/blob/main/DataInput/SignalsTheoryChecked.csv}, which also contains quotes justifying classifications for the first two dimensions (theory and equilibrium modeling).

We categorize predictor-papers, rather than papers, because similar predictors appear across multiple papers, and the peer review process sometimes arrives at different theoretical foundations. Similar predictors appear in the CZ dataset, on occasion, due to their goal of comprehensive coverage and the fact that it's difficult to judge if two predictors are indeed duplicates.

For example, predictors related to profitability appear in both \citet{fama2006profitability} and \citet{balakrishnan2010post}. Fama and French use annual earnings and scale by book equity, while Balakrishnan et al. use quarterly earnings and scale by total assets.\footnote{% begin footnote
While one might argue that these predictors are duplicates, Balakrishnan et al say their predictor is ``incremental to, and more pronounced than previously documented earnings-related anomalies.''
} % end footnote 
Fama and French provide an agnostic explanation: ``We take no stance on whether the patterns in average returns observed here are rational or irrational.'' In contrast, Balakrishnan et al. argue for mispricing: ``We document a market failure to fully respond to loss/profit quarterly announcements.'' Thus, the broader concept of profitability appears as both agnostic and mispricing in our analysis, and is evaluated at the predictor-paper level. 

In rare instances, predictors from the same paper may have distinct theoretical foundations. \citet{ang2006cross} provide a risk-based explanation for the predictive power of the VIX beta (``innovations in aggregate volatility carry a statistically significant negative price of risk''), but are agnostic about why idiosyncratic volatility predicts returns (``our results on idiosyncratic volatility represent a substantive puzzle''). Thus, the Ang et al. paper appears as both risk-based and agnostic, and one needs to delve into the predictor-paper level to have a clean description of the theoretical ideas.

As these quotes illustrate, categorizing predictor-papers into risk, mispricing, and agnostic is typically straightforward. However, a handful of predictor-papers focus on liquidity mechanisms.  We categorize these predictor-papers as mispricing if the argument focuses on stock-specific measures of liquidity (\citealt{amihud2002illiquidity})  and risk if the argument focuses on a market-wide component (\citealt{pastor2003liquidity}). This method gives the risk category the best chance at finding post-sample returns, since idiosyncratic liquidity has improved over time (\citealt{chen2022zeroing}). Nevertheless, this issue affects only seven predictor-papers, and has little impact on our main results.

\subsection{The Distribution of Research Categories}\label{sec:hetero-distribution}

Table \ref{tab:hetero-count} illustrates the distribution of predictor-papers across the four categories. Among the 173 predictor-papers we examine, 20\% (34/173) are attributed to risk by the peer review process, 61\% (105/173) are attributed to mispricing, and 20\% (34/173) have uncertain or agnostic origins. Top finance journals show a similar pattern, with 22\% (23/105) of predictors attributed to risk. Interestingly, accounting journals rarely attribute predictability to risk. A detailed breakdown by journal is in the Internet Appendix (Table \ref{tab:journal_theory}).

% table: research approach summary 
\begin{table}[!h]\caption{Research Categories in Cross-Sectional Predictability}
\label{tab:hetero-count}

\begin{singlespace}
\noindent  We categorize predictor-papers by reading the original papers. Justification for each ``Theoretical Explanation'' is at \url{https://github.com/chenandrewy/flex-mining/blob/main/DataInput/SignalsTheoryChecked.csv}. ``JF, JFE, RFS'' includes only predictors published in the Journal of Finance, Journal of Financial Economics, or Review of Financial Studies. ``AR, JAE, JAR'' includes only predictors published in the Accounting Review, the Journal of Accounting and Economics or the Journal of Accounting Research. Peer review attributes little cross-sectional predictability to risk. Few predictor-papers are explained with an equilibrium model.
\end{singlespace}
\begin{centering}
\vspace{-1ex}
\par\end{centering}
\centering{}\setlength{\tabcolsep}{1ex} \footnotesize
\begin{center}
   \begin{tabular}{lrrrrr}
      \toprule
      \multicolumn{1}{l}{\textbf{Category}} 
        & \multicolumn{5}{c}{\textbf{Journal Category}} \\
      \cmidrule(lr){2-6}
      & Any & JF, JFE, RFS & AR, JAE, JAR & Finance & Accounting \\
      \midrule
      \multicolumn{6}{l}{\textbf{Theoretical Explanation}} \\
      % latex table generated in R 4.4.1 by xtable 1.8-4 package
% Mon Dec  8 10:27:44 2025
 Risk &  34 &  23 &   0 &  28 &   1 \\ 
  Mispricing & 105 &  59 &  24 &  68 &  32 \\ 
  Agnostic &  34 &  23 &   5 &  23 &   5 \\ 
  
      \multicolumn{6}{l}{\textbf{Equilibrium Modeling}} \\
      % latex table generated in R 4.4.1 by xtable 1.8-4 package
% Mon Dec  8 10:30:02 2025
 No Model & 148 &  87 &  29 &  98 &  38 \\ 
  Stylized &  14 &  10 &   0 &  12 &   0 \\ 
  Dynamic &   5 &   4 &   0 &   5 &   0 \\ 
  Quantitative &   6 &   4 &   0 &   4 &   0 \\ 
  
      \midrule
      % latex table generated in R 4.4.1 by xtable 1.8-4 package
% Mon Dec  8 10:31:09 2025
 \textbf{Total} & 173 & 105 &  29 & 119 &  38 \\ 
  
      \bottomrule
   \end{tabular}
\end{center} 
\end{table}

This distribution suggests a consensus about the origins of cross-sectional predictability: peer review attributes little of this predictability to risk. We emphasize that this attribution is chosen not only by the authors, but also by the editors and referees. This consensus is consistent with factor model measures of risk (Appendix \ref{sec:intapp-risk-factor}). It contrasts with recent reviews of empirical asset pricing, which are typically agnostic about the origins of predictability (e.g. \citealt{bali2016empirical,zaffaroni2022asset}). 

The distribution also shows that relatively few predictor-papers are justified with an equilibrium model. Only 14.5\% of the predictors-papers have a mathematical theory of any sort. And among these models, most are stylized. This result may be related to the high prevalence of mispricing explanations, which can be thought of as disequilibrium phenomena. Indeed, of the 25 papers with a mathematical theory, only five are theories of mispricing, and most of these center around trading frictions. Alternatively, this result may be due to the technical challenges around solving equilibrium models with a cross-section of stocks under uncertainty. 

The papers that lack a mathematical theory vary in the assertiveness of their theoretical explanations. Some papers begin with a verbal theory of investor behavior (sometimes with citations to mathematical theories), and then present empirical evidence consistent with the theory (e.g. \citealt{sloan1996stock}; \citealt{asquith2005short}; \citealt{bali2011maxing}; \citealt{eberhart2004examination}). Others take a more agnostic stance, discussing several possible theories, and then use empirical evidence to argue for one of the theories (\citealt{spiess1999long}; \citealt{titman2004capital}; \citealt{chan1996momentum}). 80\% of predictor-papers end up with a stance on the theoretical origin of predictability, as seen in Table \ref{tab:hetero-bytheory}. 

\subsection{Post-Sample Outperformance by Theoretical Support}\label{sec:hetero-mispricing}

% Describe the table's methods
Table \ref{tab:hetero-bytheory} repeats the post-sample analysis from Figures \ref{fig:intro} and \ref{fig:pub-vs-dm-2}, applied to subsets of predictor-papers based on their theoretical foundation and equilibrium modeling.

% table: outperformance by research approach
\begin{table}[!h]
   \caption{Post-Sample Outperformance by Theoretical Support}
   \label{tab:hetero-bytheory}
   
   \begin{singlespace}
      \noindent Strategies are normalized to have 100 bps per month performance in-sample. `Post-Sample' is the performance of predictor-papers in bps per month. `Versus Data Mining' is `Post-Sample' minus the performance of data-mined benchmarks. `Theoretical Foundation' and `Equilibrium Modeling' are determined by reading the papers (see Table \ref{tab:hetero-count}). CAPM- and FF3+momentum-alphas use regressions specific to the sample periods (in-sample or post-sample). Standard errors (parentheses) are clustered by calendar month and predictor. Research that is agnostic on the origin of predictability shows signs of outperformance relative to data mining. Research that takes a stand on the theory does not.
   \end{singlespace}
   \begin{centering}
   \vspace{0ex}
   \par\end{centering}
   \centering{}\setlength{\tabcolsep}{1.2ex} \small
   \begin{center}
   \begin{tabular}{lcccccc}
      \toprule
      & \multicolumn{2}{c}{Long-Short Return} & \multicolumn{2}{c}{CAPM Alpha} & \multicolumn{2}{c}{FF3 + Mom Alpha} \\
      \cmidrule(lr){2-3}\cmidrule(lr){4-5}\cmidrule(lr){6-7}
      & Post- & \multicolumn{1}{c}{Versus} & Post- & \multicolumn{1}{c}{Versus} & Post- & \multicolumn{1}{c}{Versus} \\
      & Sample & \multicolumn{1}{c}{Data Mining} & Sample & \multicolumn{1}{c}{Data Mining} & Sample & \multicolumn{1}{c}{Data Mining} \\
      \midrule
      % FF4 figures copied from exhibits/ff4/Table_RiskAdjusted_ff4_t2.tex
      \multicolumn{7}{l}{\textbf{Theoretical Foundation}} \\
      Agnostic          & 65 & 9 & 79 & 23 & 110 & 31 \\
                        & (8) & (8) & (8) & (8) & (8) & (9) \\
      Mispricing        & 55 & 4 & 60 & 6 & 59 & -17 \\
                        & (4) & (4) & (4) & (4) & (4) & (5) \\
      Risk              & 43 & 5 & 38 & -4 & 49 & -21 \\
                        & (8) & (8) & (8) & (8) & (6) & (9) \\                        
      \addlinespace
      % FF4 figures copied from exhibits/ff4/Table_RiskAdjusted_ff4_t2.tex
      \multicolumn{7}{l}{\textbf{Equilibrium Modeling}} \\
      No Model          & 56 & 5 & 62 & 9 & 71 & -4 \\
                        & (3) & (3) & (3) & (3) & (3) & (4) \\
      Stylized          & 63 & 15 & 49 & -5 & 51 & -42 \\
                        & (12) & (13) & (13) & (14) & (7) & (11) \\
      Dynamic or        & 34 & -2 & 50 & 4 & 39 & -54 \\
      Quantitative      & (14) & (14) & (14) & (19) & (14) & (28) \\
      \addlinespace
      \textbf{Overall}  & 56 & 5 & 60 & 8 & 68 & -8 \\
                        & (3) & (3) & (3) & (3) & (3) & (4) \\
      \bottomrule
   \end{tabular}
   \end{center} 
\end{table}

% walk through results. begin with the simple pattern that agnostic performs best
The top panel shows that research with agnostic theoretical foundations has the strongest post-sample performance. As seen in the `Post-Sample' columns, agnostic research produces 65 to 110 bps per month post-sample, depending on the performance metric. In comparison, mispricing foundations produce 55 to 60 bps per month post-sample, and risk foundations produce 38 to 49 bps.

% results: main question - does anything outperform data-mining?
The ``Versus Data Mining'' columns report the difference between the post-sample performance of the predictor-papers and their data-mined benchmarks, and thereby address our main question: does peer-reviewed research with a particular theoretical foundation outperform data mining?

Research with theoretical foundations in mispricing or risk show little to no outperformance. Predictor-papers with mispricing foundations outperform their data-mined benchmarks by only 4 bps per month in terms of long-short returns and 6 bps in terms of CAPM alphas. In terms of FF3 + momentum alphas, mispricing foundations \emph{under}perform data mining, by 17 bps per month. The performance of predictor-papers with risk foundations is even worse. By most metrics, risk-based predictor-papers underperform data mining.

% main result: agnostic methods outperform
Agnostic theoretical foundations show some signs of outperformance, but the magnitudes are modest. Agnostic predictor-papers outperform data mining by 9 to 31 bps per month, depending on the performance metric. Since performance is normalized to 100 bps in-sample, this means agnostic predictors retain up to an additional 31 percentage points of their original-sample performance, compared to data mining. While 31 percentage points may appear notable, it is the largest estimate out of many, and should be shrunk toward the average of about zero to account for multiple comparisons (e.g. \citealt{chen2020publication}). This finding, that agnostic research outperforms somewhat while risk-based and mispricing-based research do not, is robust to alternative construction methods of the data-mined benchmarks, including directly controlling for t-stats and mean returns of published strategies (Appendix \ref{sec:app-robust}). 

% results: equilibrium modeling
Grouping research by equilibrium modeling leads to a similar pattern, as seen in the bottom panel of Table \ref{tab:hetero-bytheory}. By most metrics, predictor-papers with no equilibrium model have stronger post-sample performance than predictor-papers with equilibrium modeling. The performance of model-free research is not as strong as research with agnostic theoretical foundations, however. Predictor-papers with no model produce 56 to 71 bps per month post-sample, compared to 65 to 110 bps per month for agnostic predictor-papers. Indeed, when compared to data mining, model-free research shows minimal outperformance. 

Notably, Table \ref{tab:hetero-bytheory} implies that the support of a stylized or dynamic equilibrium model leads to little if any outperformance relative to data mining. Additional robustness checks are in the Internet Appendix (\ref{sec:intapp-riskalt} and \ref{sec:intapp-pub-additional}).

\subsection{Post-Sample Outperformance by Discipline and Journal Ranking}\label{sec:hetero-discipline}

Table \ref{tab:hetero-byjournal} examines how discipline and journal ranking affect outperformance. The top panel shows that predictor-papers in finance journals have stronger post-sample performance compared to accounting journals. Finance predictor-papers produce 59 to 76 bps per month post-sample, compared to 43 to 56 bps per month for accounting journals. 

\begin{table}[!h]
   \caption{Post-Sample Outperformance by Discipline and Journal Ranking}
   \label{tab:hetero-byjournal}
   
   \begin{singlespace}
   \noindent Strategies are normalized to have 100 bps per month performance in-sample. `Post-Sample' is the performance of predictor-papers in bps per month. `Versus Data Mining' is `Post-Sample' minus the performance of data-mined benchmarks. `Discipline' categorizes the journal of publication. `JF,' `JFE,' and `RFS' includes papers in the Journal of Finance, Journal of Financial Economics, and Review of Financial Studies. `AR,' `JAR,' and `JAE' includes papers in the Accounting Review, Journal of Accounting Research, and Journal of Accounting and Economics. CAPM and FF3 + momentum alphas use regressions specific to the sample periods (in-sample or post-sample). Standard errors (parentheses) are clustered by calendar month and predictor. Finance, and particularly top-ranked finance journals show some signs of outperformance relative to data mining, but the improvement is modest. 
   \end{singlespace}
   \begin{centering}
   \vspace{0ex}
   \par\end{centering}
   \centering{}\setlength{\tabcolsep}{1.2ex} \small
   \begin{center}
   \begin{tabular}{lcccccc}
      \toprule
      & \multicolumn{2}{c}{Long-Short Return} & \multicolumn{2}{c}{CAPM Alpha} & \multicolumn{2}{c}{FF3 + Mom Alpha} \\
      \cmidrule(lr){2-3}\cmidrule(lr){4-5}\cmidrule(lr){6-7}
      & Post- & \multicolumn{1}{c}{Versus} & Post- & \multicolumn{1}{c}{Versus} & Post- & \multicolumn{1}{c}{Versus} \\
      & Sample & \multicolumn{1}{c}{Data Mining} & Sample & \multicolumn{1}{c}{Data Mining} & Sample & \multicolumn{1}{c}{Data Mining} \\
      \midrule
      % FF4 figures copied from exhibits/ff4/Table_RiskAdjusted_DisciplineJournal_ff4_t2.tex
      \multicolumn{7}{l}{\textbf{Discipline}} \\
      Finance      & 59 & 8 & 65 & 12 & 76 & -2 \\
                   & (4) & (4) & (4) & (4) & (4) & (5) \\
      Accounting   & 43 & -6 & 45 & -8 & 43 & -27 \\
                   & (6) & (7) & (6) & (6) & (5) & (7) \\
      \addlinespace
      \multicolumn{7}{l}{\textbf{Journal Rank}} \\
      JF, JFE, RFS & 60 & 8 & 68 & 16 & 81 & 3 \\
                   & (4) & (4) & (4) & (5) & (4) & (5) \\
      AR, JAR, JAE & 43 & -6 & 45 & -8 & 43 & -27 \\
                   & (6) & (7) & (6) & (6) & (5) & (7) \\
      Other        & 53 & 8 & 55 & 2 & 61 & -20 \\
                   & (6) & (6) & (6) & (7) & (7) & (9) \\
      \bottomrule
   \end{tabular}
   \end{center} 
\end{table}

This performance is less impressive when compared to data mining, however. Predictor-papers in finance journals outperform data mining by between -2 and +12 bps per month.  Predictor-papers in accounting journals underperform data mining, regardless of the performance metric.

The top 3 finance journals (Journal of Finance, Journal of Financial Economics, Review of Financial Studies) perform a bit better than finance journals overall. Nevertheless, the outperformance relative to data mining is small. Predictor-papers in these top journals outperform data mining by 3 to 16 bps per month, depending on the performance metric. Given the normalization, this means top finance journals retain an additional 3 to 16 percentage points of their original-sample performance, relative to simple data mining. 

% !TEX root = ../risk_vs.tex

% ==============================
\section{The Most Renowned Research vs Data Mining}\label{sec:app-most-famous}

Section \ref{sec:hetero} showed that journal ranking has relatively little effect. But perhaps one needs to go beyond journal ranking, and focus on the very best research, to find notable outperformance relative to data mining.

To examine this possibility, we take a closer look at the predictability studied in \citet{fama1992cross}, \citet{jegadeesh1993returns}, and \citet{banz1981relationship}. These papers are renowned for studying the predictive power of B/M, momentum, and size, respectively. These findings are not only among the most renowned, but are arguably the ones with the strongest supporting evidence, both theoretical and empirical.  

Theoretical foundations for B/M include \citet{berk1999optimal}, \citet{gomes2003equilibrium}, \citet{campbell2004bad}, \citet{zhang2005value}, and \citet{lettau2007long}. Many of these papers also provide a theoretical foundation for size (see also \citealt{berk1995critique}). Theoretical foundations for momentum include \citet{hong1999unified}, \citet{brav2002competing}, \citet{holden2002news}, and \citet{da2014frog}.\footnote{% 
Other theoretical foundations for B/M and size include \citet{gabaix2008variable}, \citet{papanikolaou2011investment}, and \citet{chen2018general}. For other theoretical foundations for momentum, see \citealt{subrahmanyam2018equity}.
} % end footnote
Many of these theories are themselves award-winning and renowned papers.

Almost all of these theories provide equilibrium foundations---that is, stable relationships between firm characteristics and expected returns. One might argue that the behavioral equilibria are unstable, but others will argue that psychological biases are fundamental, as are limits to arbitrage. And while the multiplicity of theories could be viewed as ``model dredging,'' it could also be viewed as robustness. Theoretical robustness is a feature of physics and statistics, in which core phenomena can be derived from multiple perspectives.\footnote{% begin footnote
Thermodynamic phenomena (e.g. ideal gas law) can be derived from the laws of thermodynamics,  classical mechanics, or quantum mechanics. Core statistical formulas (e.g. regression coefficients), can be derived from method of moments, maximum likelihood, Bayesian assumptions, or data fitting. 
} % end footnote

These renowned papers provide robust empirical evidence. Indeed, \citet{fama1992cross} is in essence a robustness check on \citet{stattman1980book} and \citet{banz1981relationship}. But other empirical papers provide even more robustness (e.g., \citealt{fama1993common}; \citealt{lakonishok1994contrarian}; \citealt{chan1996momentum}; \citealt{asness2013value}). 

Tables \ref{tab:closer_bmdec}-\ref{tab:closer_size} compare these renowned findings to data-mined alternatives. The alternatives come from searching the 29,000 accounting ratios for t-stats and mean returns that are within 10\% and 30\% of the original findings. This filtering ensures the alternatives are similar to the original findings in terms of statistical support. As before,  data-mined t-stats and mean returns use the original papers' stock weighting and sample periods.

\begin{table}[p]
\caption{Data-Mined Predictors that Performed Similarly to Fama-French's B/M (1992)}
\label{tab:closer_bmdec}

\begin{singlespace}
\noindent Table lists 20 of the 163 data-mined predictors that performed similarly to Fama and French's \citeyearpar{fama1992cross} B/M in the original sample period. It includes predictors with t-stats within 10\% and mean returns within 30\% of the original findings. Signals are ranked by the absolute difference in mean return. Sign = -1 indicates that a high signal implies a lower mean return in-sample. Data mining performs similarly to trading on \citepos{fama1992cross} B/M.
\end{singlespace}
\begin{centering}
\vspace{0ex}
\par\end{centering}
\centering{}\setlength{\tabcolsep}{0.7ex} \small
\begin{center}
% ==== begin paste
% Table generated by Excel2LaTeX from sheet 'inspect BMdec' 
\begin{tabular}{rlrrr} \toprule 
\multicolumn{1}{c}{Similarity} 
    & \multirow{2}[2]{*}{Signal} 
    & \multicolumn{1}{l}{\multirow{2}[2]{*}{Sign}} & \multicolumn{2}{c}{Mean Return (\% p.m.)} \\   
\multicolumn{1}{c}{Rank} 
    &   &   
    & \multicolumn{1}{c}{1963-1990} & \multicolumn{1}{c}{1991-2023} \\ \midrule \multicolumn{5}{l}{\textit{Peer-Reviewed}} \\ 
\midrule   
% latex table generated in R 4.2.3 by xtable 1.8-4 package
% Mon Mar  3 17:42:07 2025
  & Book / Market (Fama-French 1992) & 1 & 0.96 & 0.61 \\ 
  \midrule 
\multicolumn{5}{l}{\textit{Data-Mined}} \\ 
\midrule 
 1 & $\Delta$[PPE net]/lag[Sales] & -1 & 0.96 & 0.73 \\ 
  2 & $\Delta$[Assets]/lag[Cost of goods sold] & -1 & 0.95 & 0.80 \\ 
  3 & $\Delta$[Assets]/lag[Operating expenses] & -1 & 0.95 & 0.84 \\ 
  4 & [Depreciation (CF acct)]/[Capex PPE sch V] & 1 & 0.97 & 0.68 \\ 
  5 & [Stock issuance]/[Debt in current liab] & -1 & 0.94 & 0.73 \\ 
  6 & $\Delta$[Assets]/lag[SG\&A] & -1 & 0.94 & 0.78 \\ 
  7 & $\Delta$[PPE net]/lag[Gross profit] & -1 & 0.98 & 0.45 \\ 
  8 & $\Delta$[PPE net]/lag[Current liabilities] & -1 & 0.94 & 0.85 \\ 
  9 & [Stock issuance]/[Capex PPE sch V] & -1 & 0.94 & 1.00 \\ 
  10 & $\Delta$[PPE (gross)]/lag[Gross profit] & -1 & 0.93 & 0.33 \\ 
   & . . . &  &  &  \\ 
  101 & $\Delta$[Assets]/lag[Assets other sundry] & -1 & 0.75 & 0.95 \\ 
  102 & $\Delta$[Liabilities]/lag[Invest tax credit inc ac] & -1 & 0.74 & 0.14 \\ 
  103 & $\Delta$[PPE net]/lag[Capital expenditure] & -1 & 0.74 & 0.79 \\ 
  104 & $\Delta$[PPE net]/lag[Interest expense] & -1 & 0.75 & 0.63 \\ 
  105 & $\Delta$[Receivables]/lag[Assets] & -1 & 0.74 & 0.59 \\ 
   & . . . &  &  &  \\ 
  159 & $\Delta$[Assets]/lag[IB adjusted for common s] & -1 & 0.67 & -0.02 \\ 
  160 & $\Delta$[Assets]/lag[Income bf extraordinary] & -1 & 0.67 & -0.03 \\ 
  161 & $\Delta$[Assets]/lag[Net income] & -1 & 0.67 & -0.01 \\ 
  162 & $\Delta$[Cost of goods sold]/lag[Current liabilities] & -1 & 0.67 & 0.65 \\ 
  163 & $\Delta$[Inventories]/lag[Curr assets other sundry] & -1 & 0.67 & 0.63 \\ 
  \midrule 
  & Mean Data-Mined &  & 0.83 & 0.65 \\ 
  
\bottomrule \end{tabular}% 
% ==== end paste
\end{center} 
\end{table}

\begin{table}[p]
\caption{Data-Mined Predictors That Performed Similarly to Jegadeesh and Titman's
12-Month Momentum (1993) }
\label{tab:closer_mom12m}

\begin{singlespace}
\noindent Table lists 20 of the 44 data-mined predictors that performed similarly to Jegadeesh and Titman's \citeyearpar{jegadeesh1993returns} 12-month momentum in the original sample period. It includes predictors with t-stats within 10\% and mean returns within 30\% of the original findings. Signals are ranked by the absolute difference in mean return. Sign = -1 indicates that a high signal implies a lower mean return in-sample. Data mining somewhat underperforms \citepos{jegadeesh1993returns} momentum.
\end{singlespace}
\begin{centering}
\vspace{0ex}
\par\end{centering}
\centering{}\setlength{\tabcolsep}{0.5ex} \small
\begin{center}
% ==== begin paste
% Table generated by Excel2LaTeX from sheet 'inspect Mom12m' 
\begin{tabular}{rlrrr} \toprule 
\multicolumn{1}{c}{Similarity} 
    & \multirow{2}[2]{*}{Signal} 
    & \multicolumn{1}{l}{\multirow{2}[2]{*}{Sign}} & \multicolumn{2}{c}{Mean Return (\% p.m.)} \\   
\multicolumn{1}{c}{Rank}  &  &   & \multicolumn{1}{c}{1964-1989} & \multicolumn{1}{c}{1990-2023} \\ 
\midrule 
\multicolumn{5}{l}{\textit{Peer-Reviewed}} \\ 
\midrule   
% === insert 
% latex table generated in R 4.2.3 by xtable 1.8-4 package
% Mon Mar  3 17:42:07 2025
  & 12-Month Momentum (Jegadeesh-Titman 1993) & 1 & 1.36 & 0.72 \\ 
  \midrule 
\multicolumn{5}{l}{\textit{Data-Mined}} \\ 
\midrule 
 1 & [Retained earnings unadj]/[Liabilities other] & 1 & 1.37 & 0.21 \\ 
  2 & [Retained earnings unadj]/[Market equity FYE] & 1 & 1.38 & -0.02 \\ 
  3 & [Retained earnings unadj]/[Assets other sundry] & 1 & 1.40 & 0.20 \\ 
  4 & [PPE and machinery]/[Current liabilities] & 1 & 1.42 & 0.46 \\ 
  5 & [Retained earnings unadj]/[Cash \& ST investments] & 1 & 1.42 & 0.31 \\ 
  6 & [PPE and machinery]/[Capital expenditure] & 1 & 1.50 & 0.69 \\ 
  7 & [Retained earnings unadj]/[Invest \& advances other] & 1 & 1.51 & 0.08 \\ 
  8 & [Income taxes paid]/[PPE net] & 1 & 1.22 & 0.22 \\ 
  9 & [Current assets]/[Market equity FYE] & 1 & 1.19 & 0.84 \\ 
  10 & [Investing activities oth]/[Nonop income] & 1 & 1.53 & 0.08 \\ 
   & . . . &  &  &  \\ 
  21 & $\Delta$[PPE (gross)]/lag[Operating expenses] & -1 & 1.09 & 0.62 \\ 
  22 & [Operating expenses]/[Market equity FYE] & 1 & 1.08 & 0.83 \\ 
  23 & $\Delta$[PPE (gross)]/lag[Num employees] & -1 & 1.07 & 0.66 \\ 
  24 & [Sales]/[Market equity FYE] & 1 & 1.08 & 0.88 \\ 
  25 & [SG\&A]/[Market equity FYE] & 1 & 1.07 & 0.84 \\ 
   & . . . &  &  &  \\ 
  40 & [Income taxes paid]/[Debt in current liab] & 1 & 1.75 & 0.29 \\ 
  41 & $\Delta$[Invested capital]/lag[Current assets] & -1 & 0.97 & 1.19 \\ 
  42 & $\Delta$[PPE net]/lag[Num employees] & -1 & 0.96 & 0.83 \\ 
  43 & $\Delta$[PPE net]/lag[Operating expenses] & -1 & 0.96 & 0.74 \\ 
  44 & $\Delta$[Assets]/lag[Operating expenses] & -1 & 0.96 & 0.84 \\ 
  \midrule 
  & Mean Data-Mined &  & 1.26 & 0.52 \\ 
  
\bottomrule 
\end{tabular}% 
% ==== end paste
\end{center} 
\end{table}

\begin{table}[p]
  \caption{Data-Mined Predictors That Performed Similarly to Banz's Size (1981)}
    \label{tab:closer_size}
    
    \begin{singlespace}
    \noindent Table lists 20 of the 220 data-mined predictors that performed similarly to Banz's \citeyearpar{banz1981relationship} size in the original sample period. It includes predictors with t-stats within 10\% and mean returns within 30\% of the original findings. Signals are ranked by the absolute difference in mean return. Sign = -1 indicates that a high signal implies a lower mean return in-sample. Data mining outperforms \citepos{banz1981relationship} size.
    \end{singlespace}
    \begin{centering}
    \vspace{-1ex}
    \par\end{centering}
    \centering{}\setlength{\tabcolsep}{1.0ex} \footnotesize
    \begin{center}
    % ==== begin paste
    % Table generated by Excel2LaTeX from sheet 'inspect Size' 
    \begin{tabular}{rlrrr} \toprule 
    \multicolumn{1}{c}{Similarity} 
        & \multirow{2}[2]{*}{Signal} 
        & \multicolumn{1}{l}{\multirow{2}[2]{*}{Sign}} & \multicolumn{2}{c}{Mean Return (\% Monthly)} \\   
    \multicolumn{1}{c}{Rank} 
        &  &  & \multicolumn{1}{c}{1926-1975} & \multicolumn{1}{c}{1976-2023} \\ \midrule \multicolumn{5}{l}{\textit{Peer-Reviewed}} \\ \midrule   
    % latex table generated in R 4.2.3 by xtable 1.8-4 package
% Mon Mar  3 17:42:08 2025
  & Size (Banz 1981) & -1 & 0.50 & 0.15 \\ 
  \midrule 
\multicolumn{5}{l}{\textit{Data-Mined}} \\ 
\midrule 
 1 & $\Delta$[Equity liq value]/lag[Sales] & -1 & 0.50 & 0.72 \\ 
  2 & [Invested capital]/[Market equity FYE] & 1 & 0.50 & 0.83 \\ 
  3 & $\Delta$[Assets]/lag[Pref stock liq value] & -1 & 0.49 & 0.18 \\ 
  4 & $\Delta$[Equity liq value]/lag[Current liabilities] & -1 & 0.48 & 0.79 \\ 
  5 & $\Delta$[Receivables]/lag[Pref stock redemp val] & -1 & 0.48 & 0.10 \\ 
  6 & $\Delta$[Current assets]/lag[Invest tax credit inc ac] & -1 & 0.52 & 0.35 \\ 
  7 & $\Delta$[Assets]/lag[Pref stock redemp val] & -1 & 0.47 & 0.23 \\ 
  8 & $\Delta$[Equity liq value]/lag[Curr assets other sundry] & -1 & 0.48 & 0.69 \\ 
  9 & $\Delta$[Common equity tangible]/lag[SG\&A] & -1 & 0.47 & 0.40 \\ 
  10 & $\Delta$[Invested capital]/lag[PPE (gross)] & -1 & 0.47 & 0.90 \\ 
   & . . . &  &  &  \\ 
  101 & $\Delta$[Depreciation \& amort]/lag[Common equity tangible] & -1 & 0.39 & 0.40 \\ 
  102 & $\Delta$[Depreciation \& amort]/lag[Invest \& advances other] & -1 & 0.38 & 0.52 \\ 
  103 & $\Delta$[Depreciation depl amort]/lag[Interest expense] & -1 & 0.39 & 0.07 \\ 
  104 & $\Delta$[Num employees]/lag[Long-term debt] & -1 & 0.39 & 0.55 \\ 
  105 & $\Delta$[Num employees]/lag[Invest \& advances other] & -1 & 0.39 & 0.45 \\ 
   & . . . &  &  &  \\ 
  216 & $\Delta$[Pref stock nonredeemable]/lag[PPE (gross)] & -1 & 0.35 & 0.69 \\ 
  217 & $\Delta$[Receivables]/lag[Curr assets other sundry] & -1 & 0.35 & 0.60 \\ 
  218 & $\Delta$[Operating expenses]/lag[Invested capital] & -1 & 0.35 & 0.62 \\ 
  219 & [Acquisitions]/[Nonop income] & -1 & 0.65 & 0.15 \\ 
  220 & [Acquisitions]/[Operating expenses] & -1 & 0.64 & 0.34 \\ 
  \midrule 
  & Mean Data-Mined &  & 0.44 & 0.42 \\ 
  
    \bottomrule \end{tabular}% 
    % ==== end paste
    \end{center} 
\end{table}

% walk through one table
Table \ref{tab:closer_bmdec} applies this exercise to \citepos{fama1992cross} B/M. It lists 20 of the 163 data-mined predictors that performed similarly to B/M in \citepos{fama1992cross} 1963-1990 sample period. The predictors are sorted by the absolute difference in the original sample mean return. By this metric, the most similar predictor to B/M is $\Delta \text{[PPE net]} / \text{lag [Sales]}$, which can be thought of as a measure of investment. This predictor earned 96 bps per month in the original sample period, identical to B/M. The `1991-2023' column shows that $\Delta \text{[PPE net]} / \text{lag [Sales]}$ slightly outperforms B/M post-sample, earning 73 bps per month compared to 61 bps for B/M. 

Other data-mined alternatives to B/M include those related to equity issuance ([Stock issuance]/ [Debt in current liab]) and accruals ($\Delta$ [Receivables]/ lag [Assets]). The post-sample performance of these data-mined alternatives varies. But on average, they are quite similar to \citet{fama1992cross}. Trading on Fama and French's finding would have earned 61 bps per month post-sample, compared to 65 bps for the typical data mined alternative.

Table \ref{tab:closer_mom12m} applies the same exercise to \citepos{jegadeesh1993returns} 12-month momentum. Since momentum has a much higher mean return, only 44 data-mined alternatives are found. Here, data mining underperforms on average, earning 52 bps compared to 72 bps for \citepos{jegadeesh1993returns} finding. However, Table \ref{tab:closer_size} shows data mining outperforming, earning 42 bps compared to 15 bps for \citepos{banz1981relationship} size.

Averaging across the three tables, data mining performs similarly to these renowned findings. Though the samples are small, they suggest that focusing on the best research does not significantly affect our results.  

% ===================
% !TEX root = ../risk_vs.tex
\section{Conclusion and Limitations}\label{sec:conclusion}

We show that the post-sample performance of published cross-sectional return predictors is remarkably similar to that of data-mined benchmarks. This result holds for most types of research we examine, including research that is risk-based or includes the support of a mathematical equilibrium model. Research that is agnostic about the theoretical foundation for predictability shows some signs of outperformance, but the magnitude is modest. The statistical implication is that whether a predictor is found in a journal or is data mined has little effect on mean inferences about post-sample performance (Equation \eqref{eq:painfully_clear}).

Beyond statistical inference, our findings suggest four deeper implications about cross-sectional stock return predictability: (1) empirical evidence is more informative than theoretical evidence for post-sample prediction, (2) investors do not learn about risk from academic research, (3) data mining is effective, and (4) mispricing is the primary driver. These implications come from analyzing the theoretical foundations of published predictors and their relationship with post-sample performance.

A limitation of our study is that we cannot identify the economic mechanism behind the lack of outperformance. The noisiness of predictor returns makes it difficult to determine the exact timing of decay, and thus makes it hard to separate publication effects (\citealt{mclean2016does}) from technological changes (\citealt{Chordia2014Have}). We illustrate this difficulty in the Internet Appendix \ref{sec:intapp-struct-break}.

A second limitation is that we study single-predictor strategies. For strategies that use many predictors, the factor structure and spanning are central questions. Table \ref{tab:dm-theme} suggests that data-mined accounting predictors are to a significant extent spanned by the ideas in the CZ dataset, but a more systematic investigation is needed. 

Last, our study is limited to the peer review process as characterized by the CZ dataset. This dataset is composed of papers that study cross-sectional return predictability, published between the years 1973 and 2016. Each literature has its own norms and practices, which evolve over time. The extent to which our findings generalize is an important question for future research.

\clearpage

\clearpage{}
\newpage{}

\begin{appendices}

% Appendix ==============================
\setcounter{table}{0}
\setcounter{figure}{0}
\renewcommand*\thetable{\Alph{section}.\arabic{table}}
\renewcommand*\thefigure{\Alph{section}.\arabic{figure}}

% \hypersetup{bookmarksdepth=1} % use to reset bookmark depth
% !TEX root = ../risk_vs.tex

\section{A Model of Post-Sample Performance}\label{sec:app-model}
We present a simple model for interpreting our results. $\mathcal{D}$ is a set of data-mined return signals (e.g. 29,000 accounting ratios). For signal $i\in \mathcal{D}$, the in-sample and post-sample returns follow
\begin{align}
\bar{r}_{i}^{IS} & = \mu_i + \bar{\varepsilon}_i^{IS} \\
  \bar{r}_{i}^{PS} & = \mu_i + \Delta \mu_i + \bar{\varepsilon}_i^{PS}, \label{eq:r-ps}
\end{align}
where $\mu_i$ is the stable component of expected returns, $\Delta \mu_i$ is an unstable component of expected returns, and $\bar{\varepsilon}_i^{IS}$ and $\bar{\varepsilon}_i^{PS}$ are unpredictable. 

Peer review replaces $\mathcal{D}$ with a different set $\mathcal{P}$ (e.g. signals consistent with neoclassical Q-theory). Controlling for $\bar{r}_{i}^{IS}$, the expected post-sample returns differ by:
\begin{align}
  E\left(\bar{r}_{i}^{PS}  \mid i\in\mathcal{P}, \bar{r}_{i}^{IS}\right) 
  &- E\left(\bar{r}_{i}^{PS}  \mid i\in\mathcal{D}, \bar{r}_{i}^{IS}\right) 
  \label{eq:ps-diff}
  \\
  &= E\left(\mu_i  \mid i\in\mathcal{P}, \bar{r}_{i}^{IS}\right) 
    - E\left(\mu_i  \mid i\in\mathcal{D}, \bar{r}_{i}^{IS}\right) 
    \label{eq:ps-diff-1}
    \\
  &\hphantom{=}+ E\left(\Delta \mu_i  \mid i\in\mathcal{P}, \bar{r}_{i}^{IS}\right) 
    - E\left(\Delta \mu_i  \mid i\in\mathcal{D}, \bar{r}_{i}^{IS}\right),
    \label{eq:ps-diff-2}
\end{align}
where the $\bar{\varepsilon}_i^{PS}$ terms vanish because they are unpredictable. 

Thus, there are two reasons why $\mathcal{P}$ may have stronger post-sample performance than $\mathcal{D}$: (1) $\mathcal{P}$ finds a larger stable expected return component $\mu_i$ or (2) $\mathcal{P}$ finds a more positive unstable expected return component $\Delta \mu_i$. 

Ideally, the modern theoretical evidence in the peer review process helps with both. When researchers observe $\bar{r}_{i}^{IS}$, theory should help determine whether it reflects stable expected returns $\mu_i$ or unpredictable noise $\bar{\varepsilon}_i^{IS}$. The core of modern theory is finding equilibrium, which means solving for $\mu_i$ in a model market. Theory should also predict how expected returns change. For example, if returns compensate for risk, we may expect $\Delta \mu_i \ge 0$ as investors will either not adjust their portfolios or adjust to reduce risk exposure. This potential of theory is described in Chapter 7 of \citet{cochrane2009asset}, which states ``In my opinion, the best hope for finding pricing factors that are robust out of sample and across different markets, is to try to understand the fundamental macroeconomic sources of risk.''

However, there are reasons why peer-reviewed theoretical evidence may not help. If the theory is so flexible that it can accommodate \emph{any} potential predictor, then it cannot separate $\mu_i$ from $\bar{\varepsilon}_i^{IS}$ (\citealt{fama1991efficient}). This problem can be formalized as $\mathcal{P} = \mathcal{D}$, in which case expected post-sample returns are identical. But even if $\mathcal{P}$ is in some sense smaller than $\mathcal{D}$, there is the chance that theory is unable to separate $\mu_i$ from $\bar{\varepsilon}_i^{IS}$. If the theoretical investors are concerned about risks that are irrelevant to real-world investors, then the theory may mistake $\mu_i$ for $\bar{\varepsilon}_i^{IS}$, or incorrectly predict $\Delta \mu_i \ge 0$. 

Perhaps most importantly, the peer-reviewed process may lead to a more negative $\Delta \mu_i$ by publicizing mispricing. In contrast, data-mined predictors may avoid this effect, if information about the mispricing must diffuse from the data directly to investors, without the assistance of academics. 

Overall, the model suggests that the relative importance of theoretical evidence is unclear, a priori. Thus, it is important to conduct an empirical analysis.

\section{Robustness}\label{sec:app-robust}

\subsection{Controlling for Data Source}\label{sec:app-robust-data} 

\begin{figure}[!h]
  \caption{Published annual accounting predictors against data-mined benchmarks}
  \label{fig:accounting-vs-dm}
  We repeat Figures \ref{fig:intro} and \ref{fig:pub-vs-dm-2} but now limit published predictors to those that use annual accounting data. 

  \vspace{0.15in}
  
  \centering
  \subfloat[Raw Returns, $|t|>2$]{
    \includegraphics[width=0.48\textwidth]{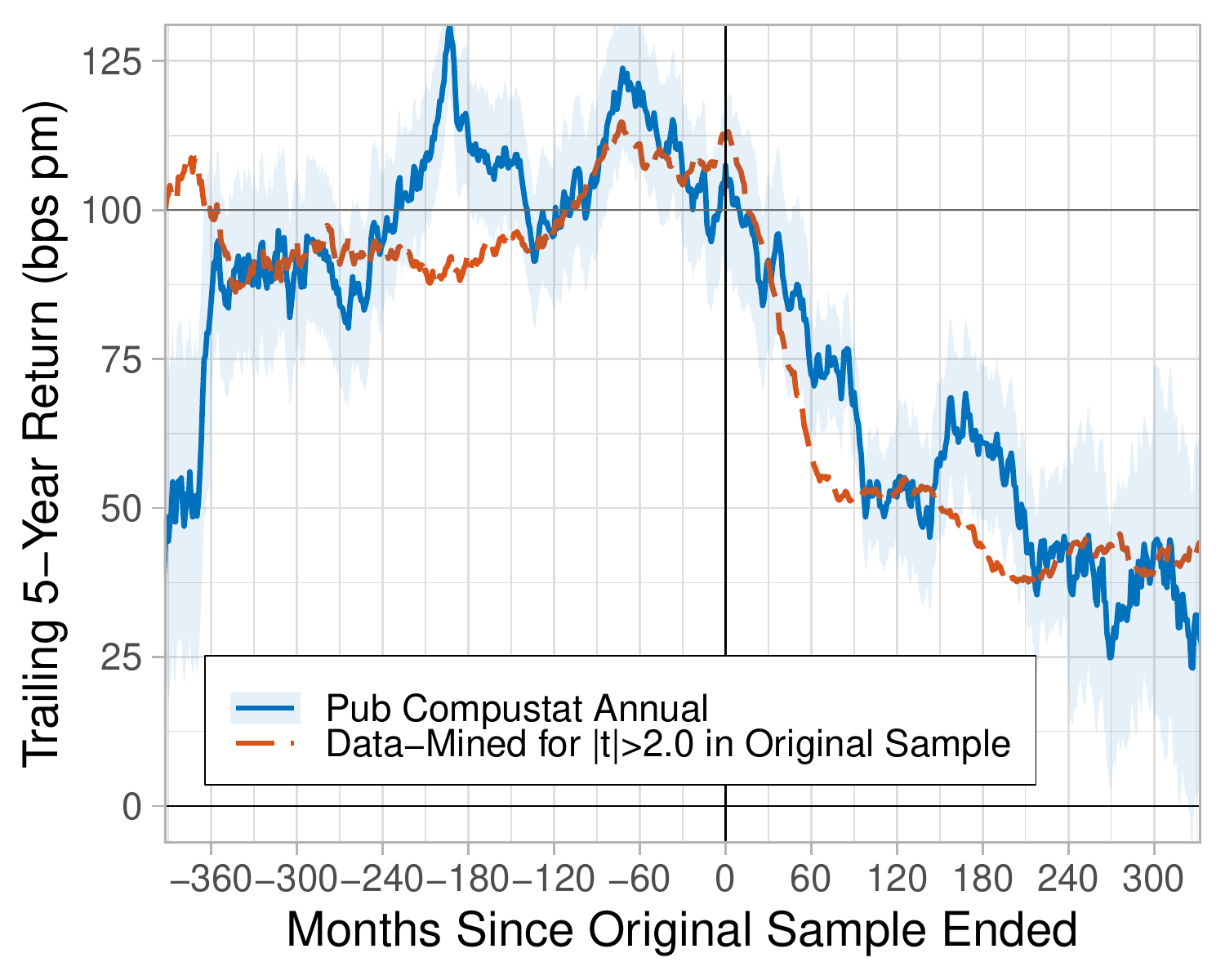}
  }
  \subfloat[Raw Returns, Top 5\% $|t|$]{
    \includegraphics[width=0.48\textwidth]{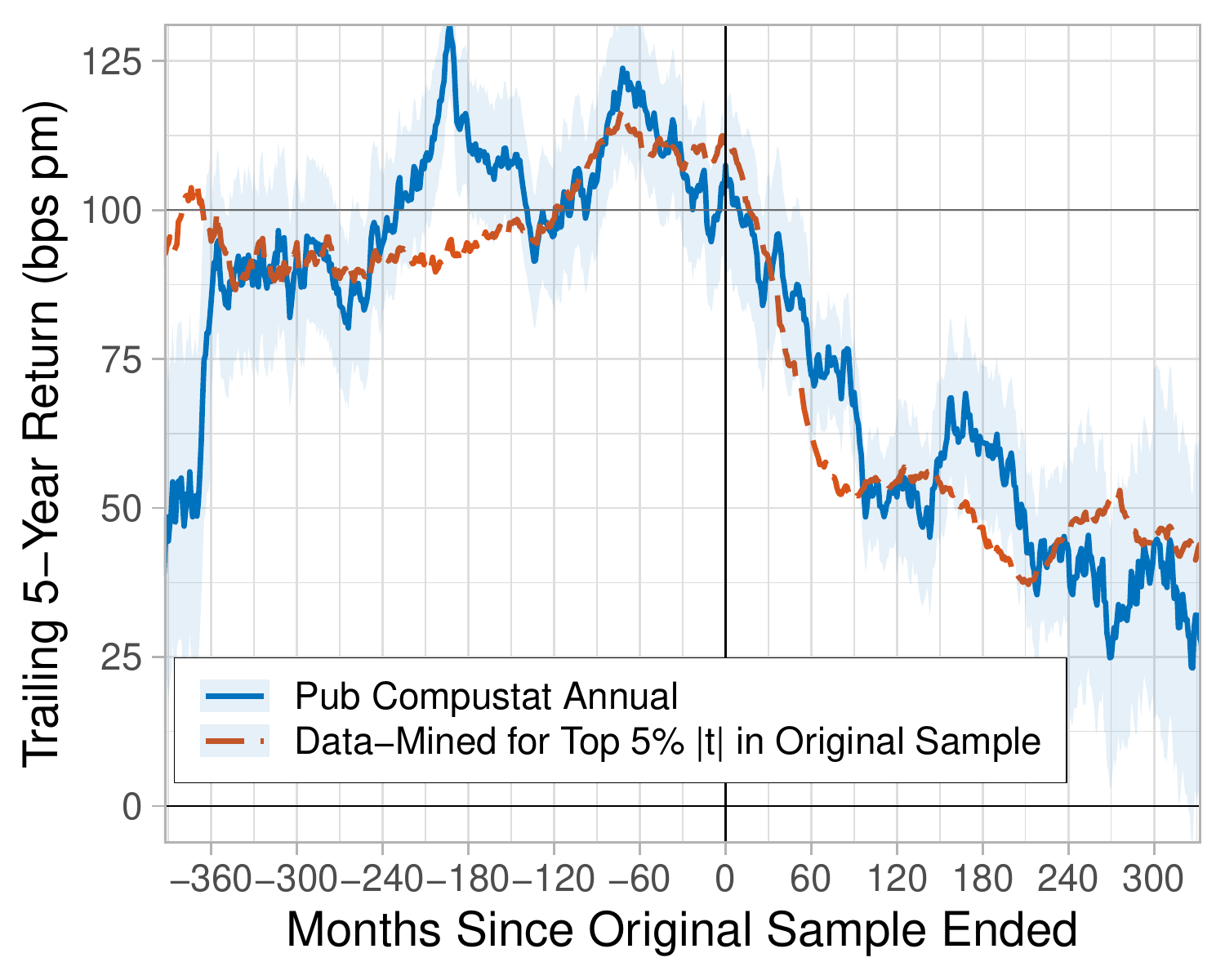}
  }
  
  \subfloat[CAPM Alpha, $|t|>2$]{
    \includegraphics[width=0.48\textwidth]{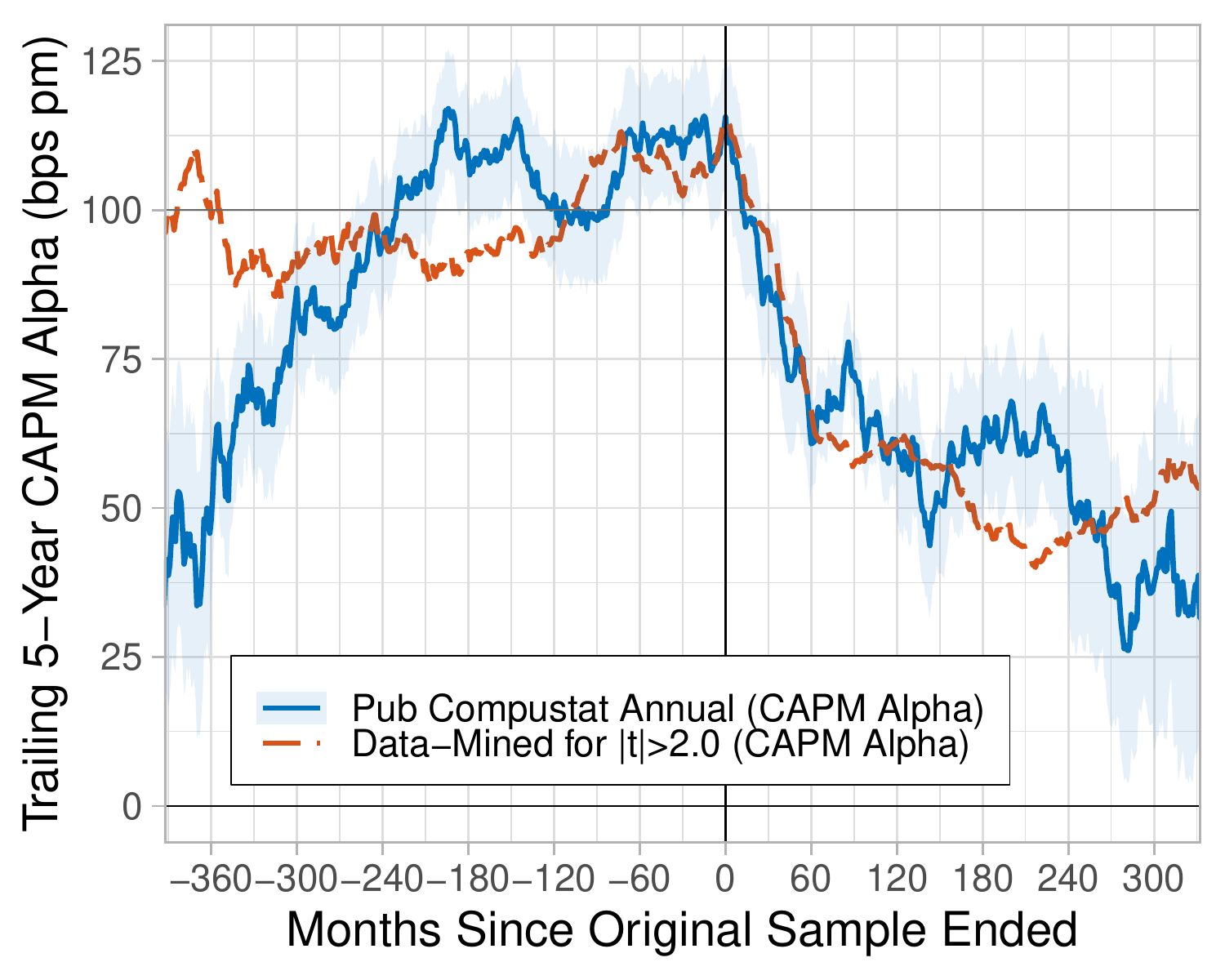}
  }
  \subfloat[FF4 Alpha, $|t|>2$]{
    \includegraphics[width=0.48\textwidth]{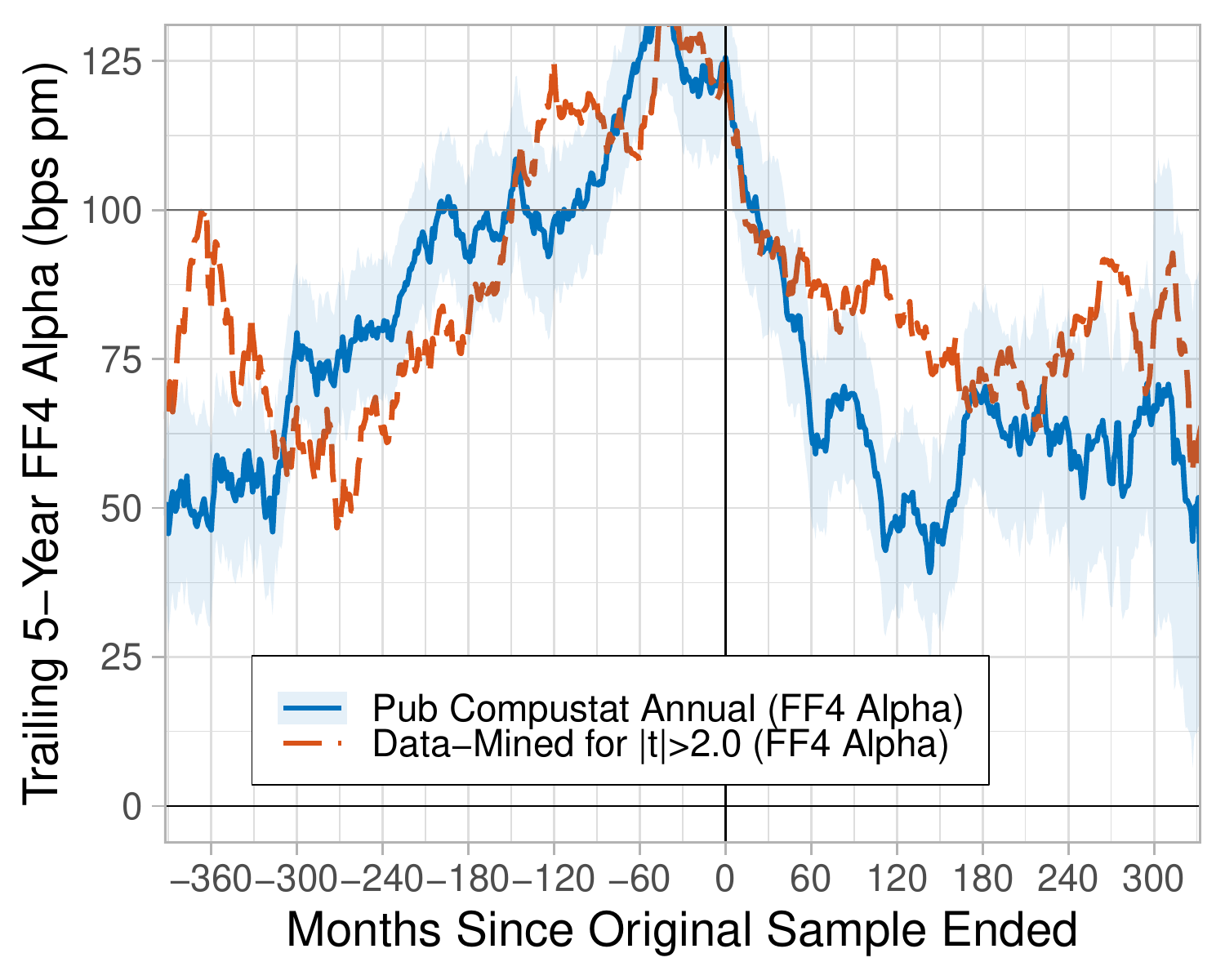}
  }

\end{figure}

The data-mined predictors use annual accounting data. A natural question is whether Equation \eqref{eq:painfully_clear} is affected by controlling for this data source. Figure \ref{fig:accounting-vs-dm} limits the published predictors to those that use annual accounting data. This filter drops roughly 50\% of the published predictors. The post-sample patterns are similar to our baseline Figures \ref{fig:intro} and \ref{fig:pub-vs-dm-2}.

\subsection{Controlling for Sample Mean Returns, and $t$-stats}\label{sec:app-robust-stats} 

Predictors with stronger statistical evidence may have stronger post-sample performance. Additionally, our normalization to 100 bps per month in-sample may fail to control for leverage effects. 

\begin{figure}[!h]\caption{Controlling for Sample Mean Returns, t-stats, Correlations}
  \label{fig:pub-vs-dm-controls}
  We repeat Figure \ref{fig:intro} but now we drop data-mined predictors if they have t-stats that differ by more than 10\% or mean returns that differ by more than 30\% (long-dash). We additionally drop data-mined strategies that are more than 10\% correlated with published strategies in the original sample (short-dash). 

  \vspace{2 ex}
  \centering
  \subfloat[All]{
  \includegraphics[width=0.45\textwidth]{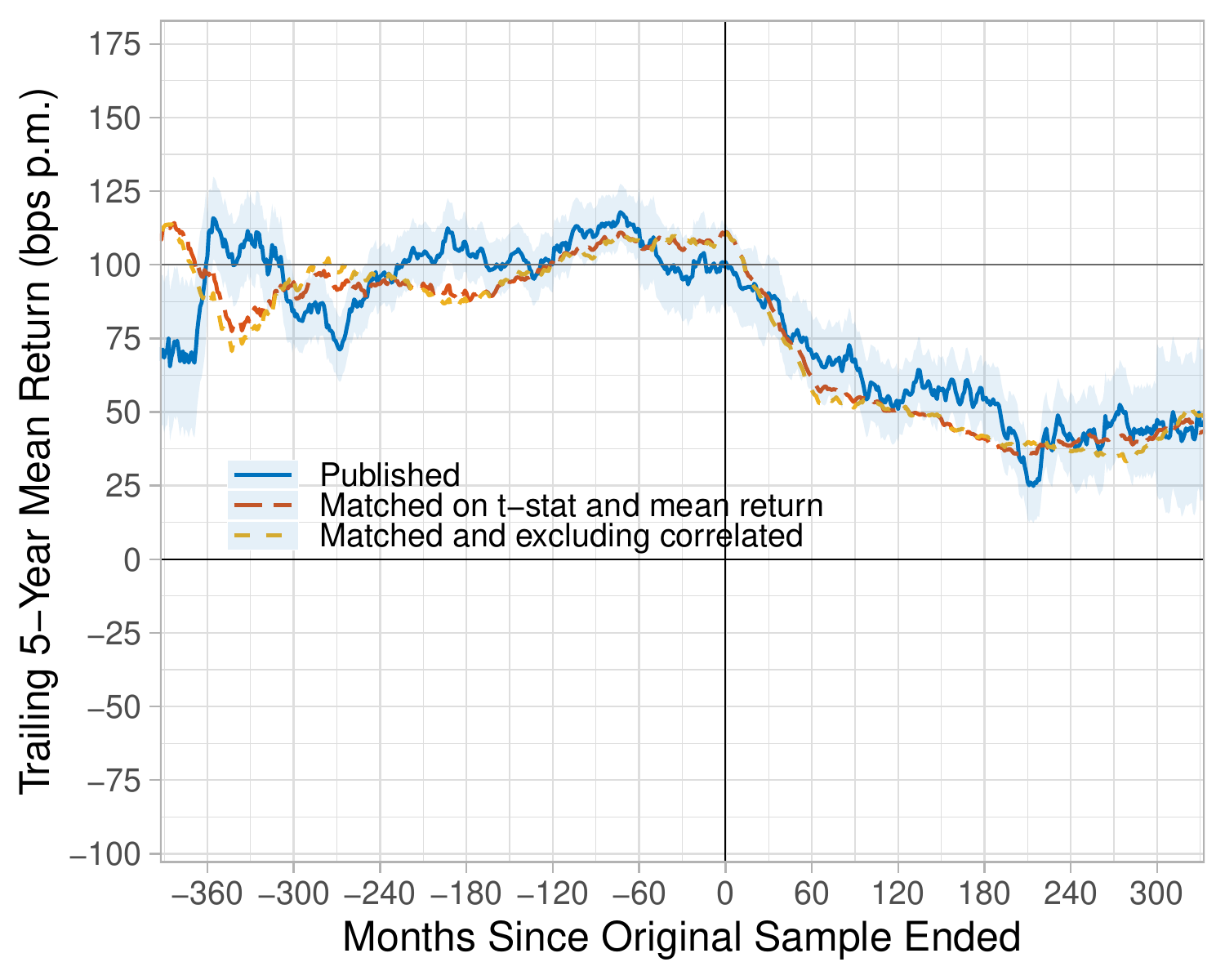} 
  }
  \subfloat[Risk-based]{
  \includegraphics[width=0.45\textwidth]{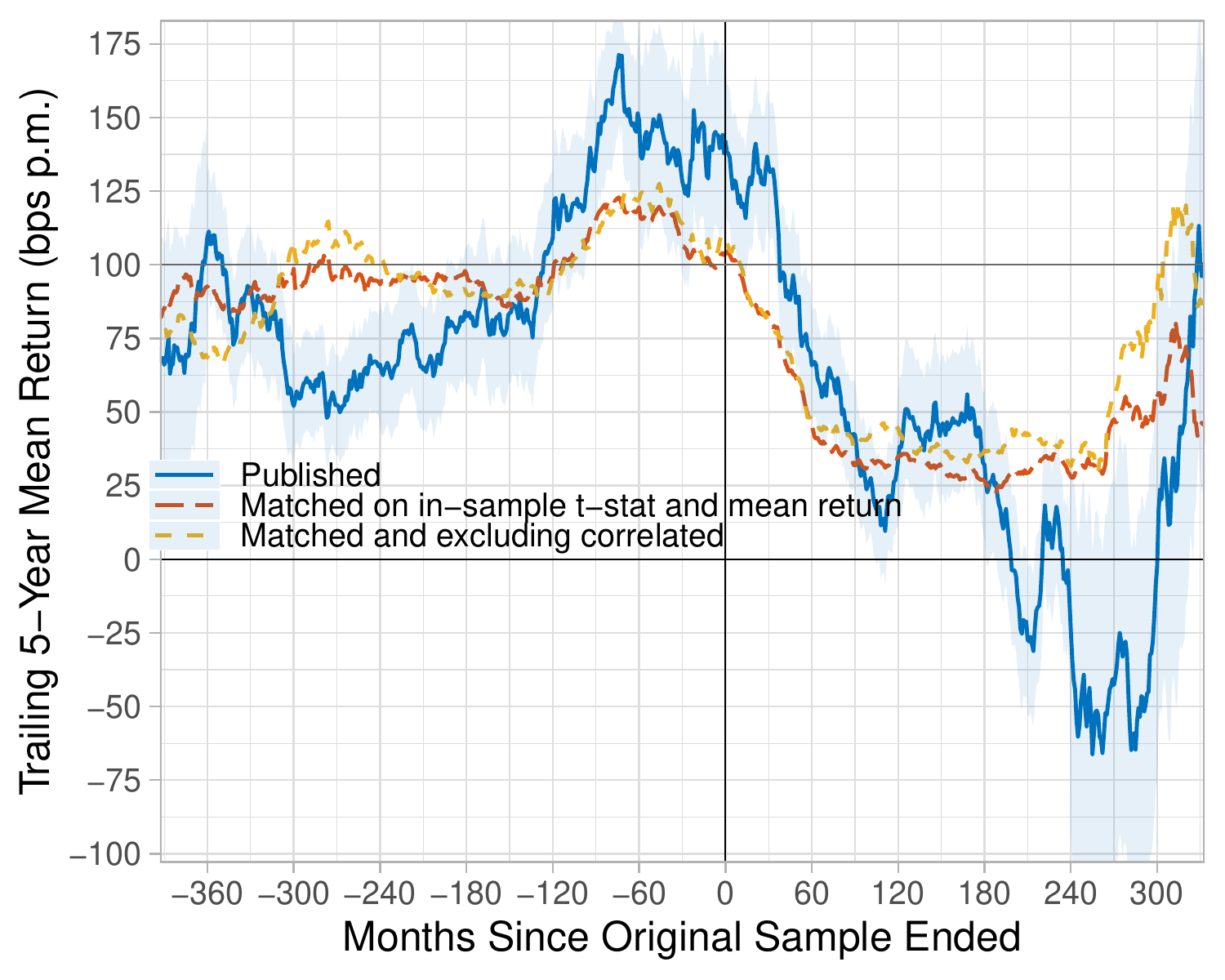}
  }\\
  \subfloat[Mispricing-based]{
  \includegraphics[width=0.45\textwidth]{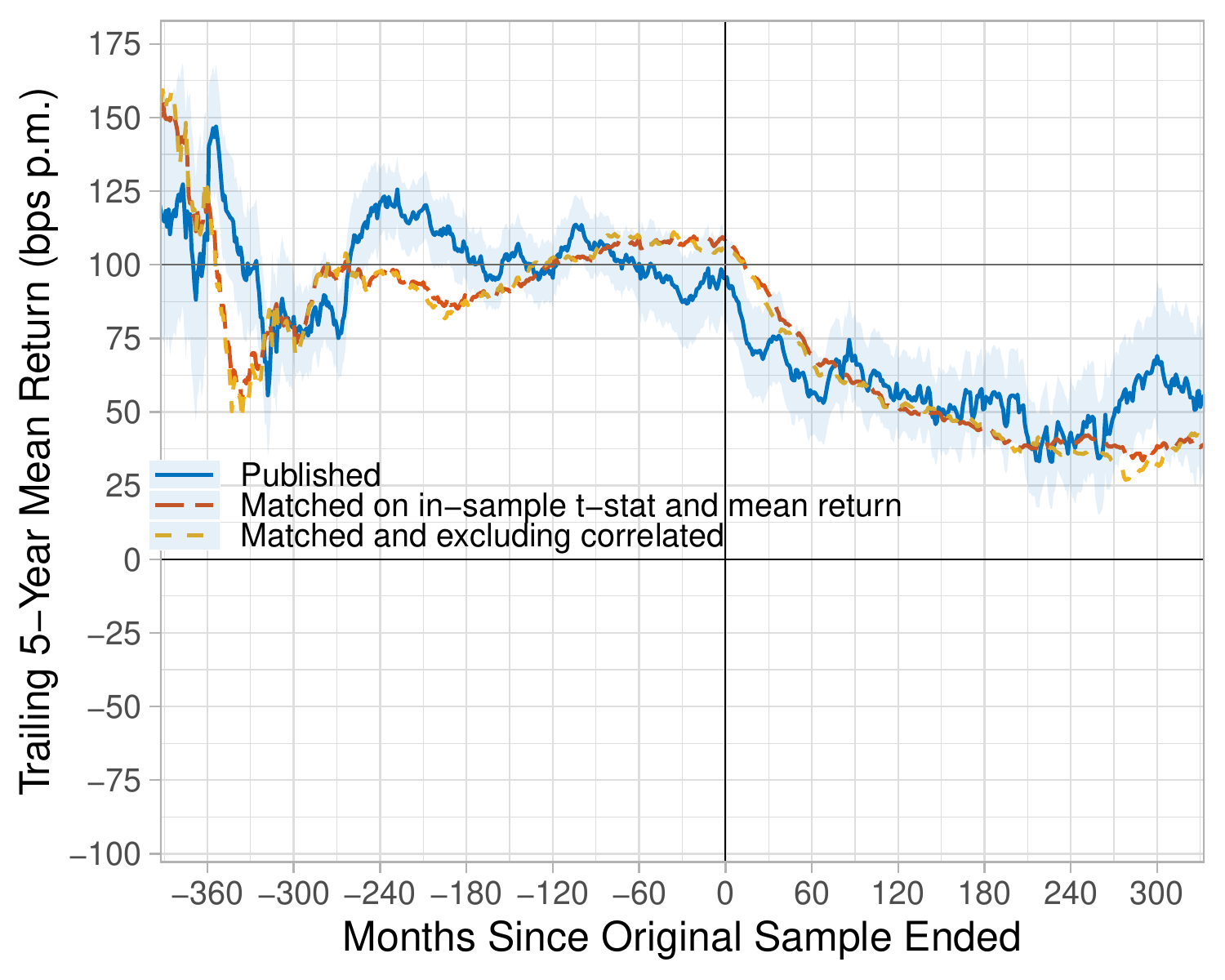} 
  }
  \subfloat[Agnostic]{
  \includegraphics[width=0.45\textwidth]{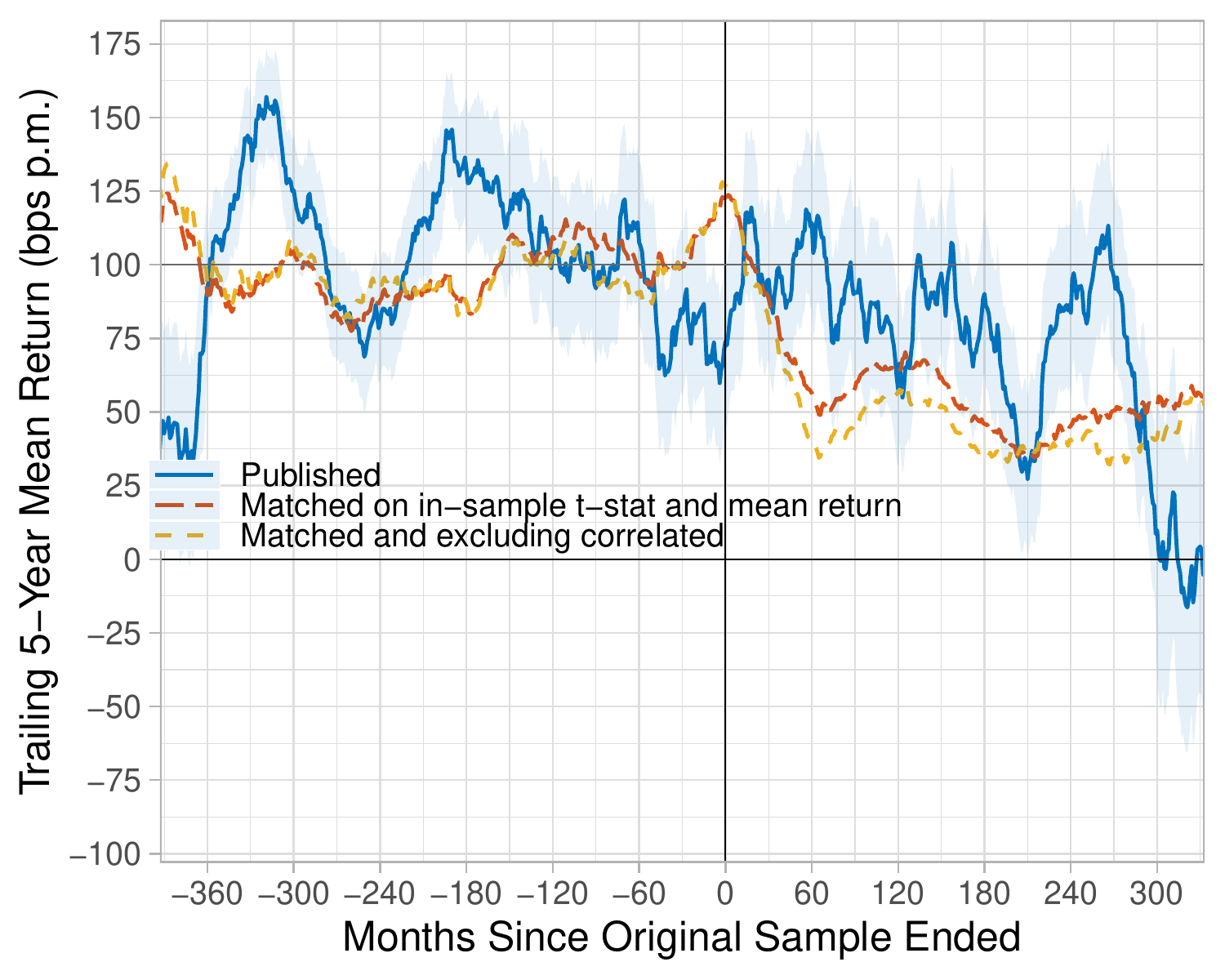}
  }
\end{figure}

\begin{figure}[!h]\caption{Controlling for Sample Mean Returns, t-stats, Correlations}
  \label{fig:pub-vs-dm-controls_robust}
  We repeat Figure \ref{fig:intro} but now we drop data-mined predictors if they have t-stats that differ by more than 10\% or mean returns that differ by more than 10\% (long-dash). We additionally drop data-mined strategies that are more than 10\% correlated with published strategies in the original sample (short-dash).

  \vspace{2 ex}
  \centering
  \subfloat[All]{
  \includegraphics[width=0.45\textwidth]{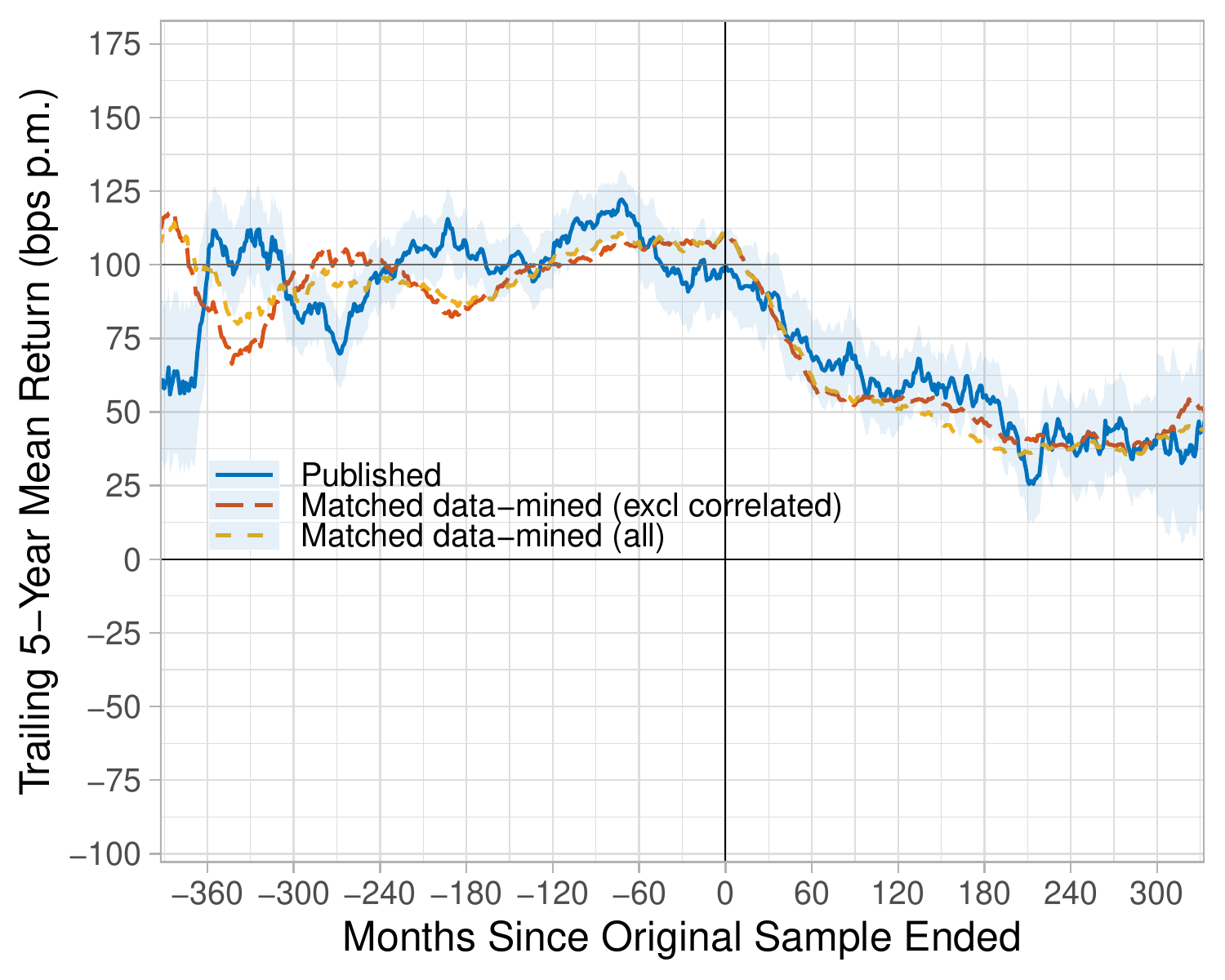} 
  }
  \subfloat[Risk-based]{
  \includegraphics[width=0.45\textwidth]{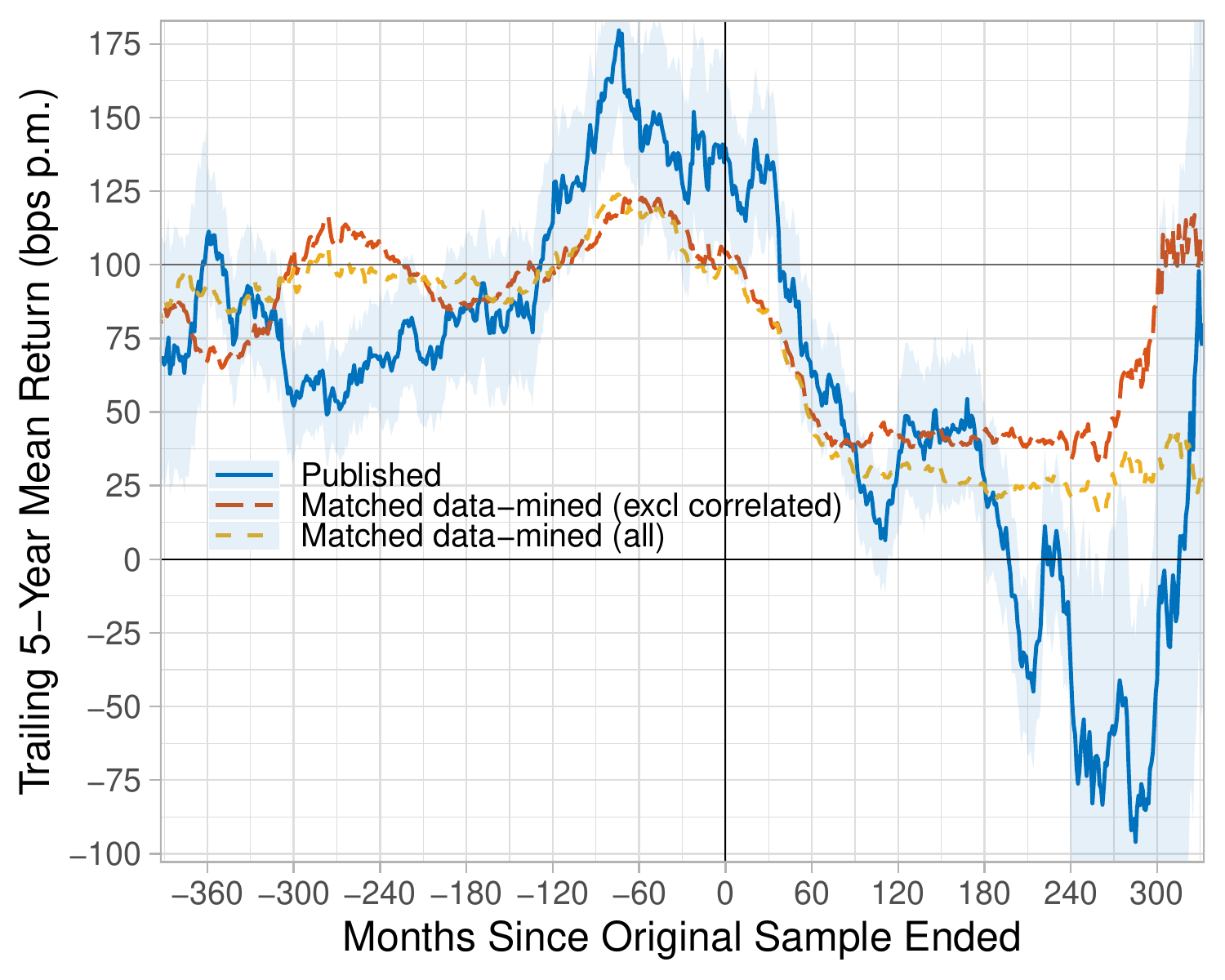}
  }\\
  \subfloat[Mispricing-based]{
  \includegraphics[width=0.45\textwidth]{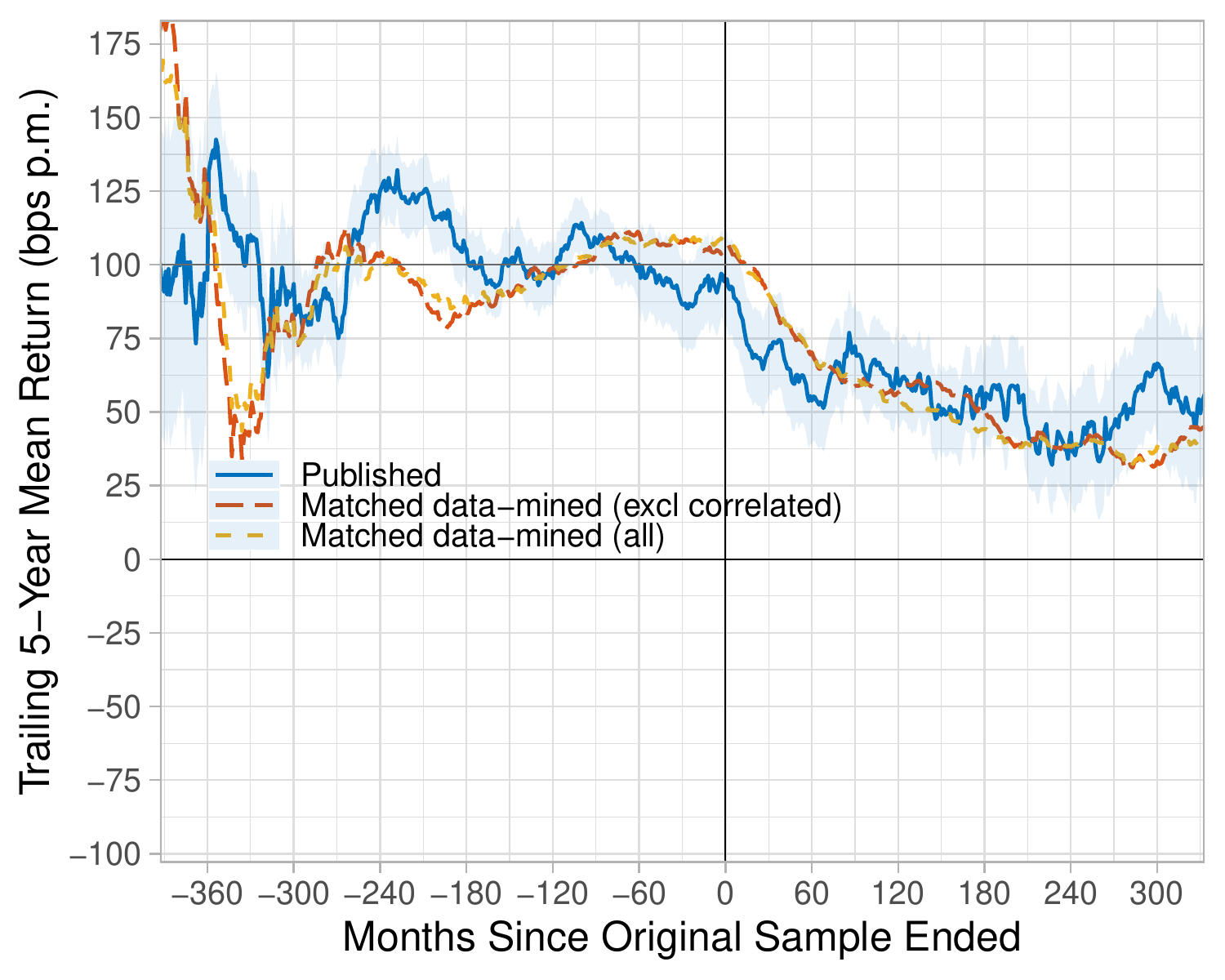} 
  }
  \subfloat[Agnostic]{
  \includegraphics[width=0.45\textwidth]{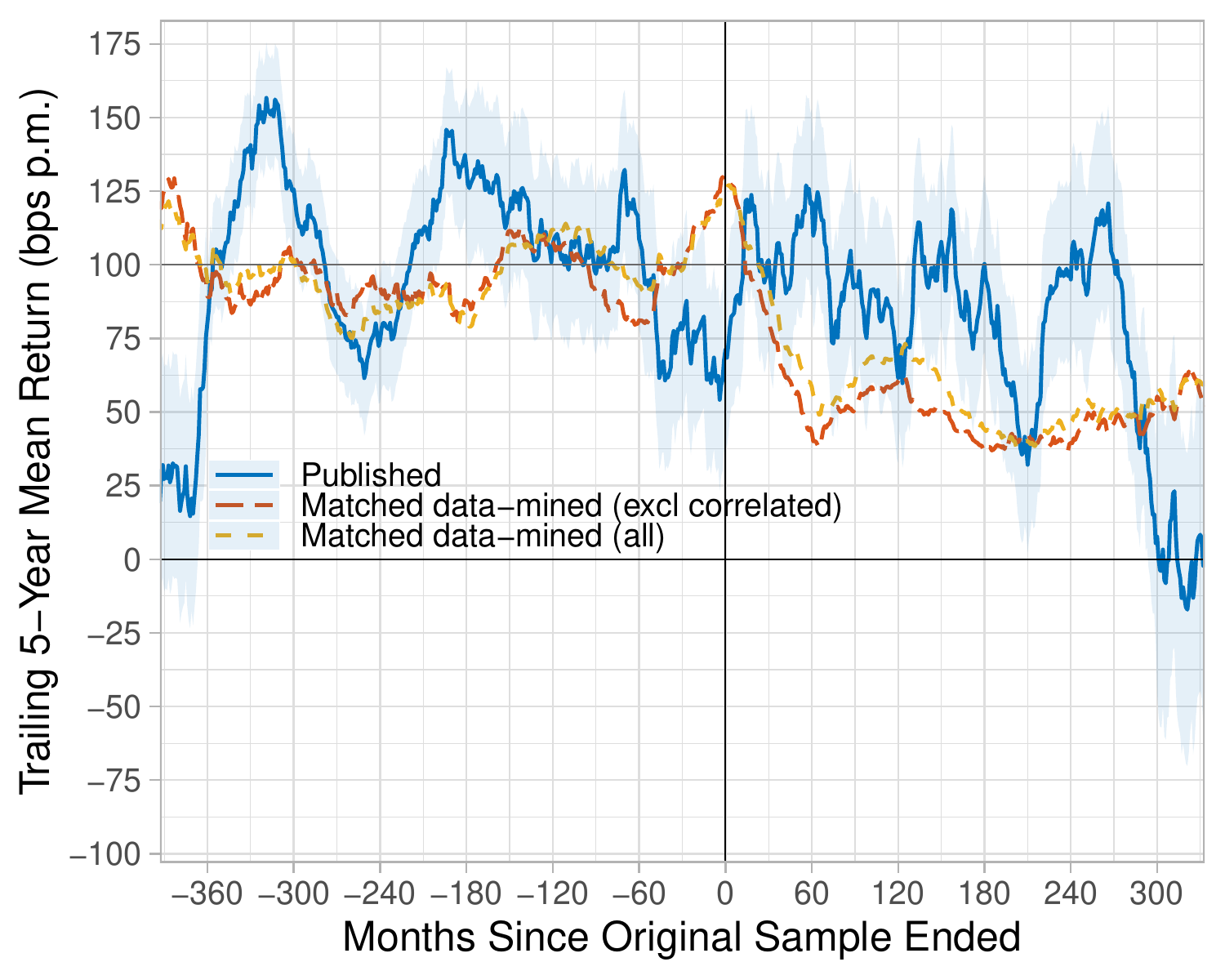}
  }
\end{figure}

The long-dash lines in Figure \ref{fig:pub-vs-dm-controls} control for these issues. These lines include only data-mined predictors with t-statistics within 10\% and mean returns within 30\% of the published predictors.  Figure \ref{fig:pub-vs-dm-controls_robust} uses a tighter mean return filter of 10\%. This analysis requires dropping published predictors, as some published predictors lack data-mined counterparts that meet this filter. Figure \ref{fig:pub-vs-dm-controls} drops 12 published predictors, while Figure \ref{fig:pub-vs-dm-controls_robust} drops 33. Overall, the patterns are similar to those in Sections \ref{sec:pubvs} and \ref{sec:hetero}. 

\subsection{Removing Correlated Data-Mined Predictors}\label{sec:app-robust-manycor}

One may be interested in whether our results are robust to excluding data-mined predictors that are highly correlated with published predictors. This question is perhaps natural given the central role of correlations in classical asset pricing theory. 

However, excluding highly correlated data-mined predictors is not important for our main question. Equation \eqref{eq:painfully_clear} implies \emph{matching} covariance patterns across data-mined and published predictors---that is, we should exclude data-mined predictors that have \emph{low} correlation with published predictors. 

The short-dash lines of Figures \ref{fig:pub-vs-dm-controls} and \ref{fig:pub-vs-dm-controls_robust} exclude data-mined predictors that have pairwise correlations of more than 0.10 with the published predictor in question (short-dash). The results are very similar to the main results.

\begin{figure}\caption{Excluding Data-Mined Predictors Correlated with Any Existing Research}
  \label{fig:decay-unspanned}
  We compare published predictors (solid) to data-mined predictors made with t-stats $>$ 2.0, but separate predictors by correlation and in-sample t-stats. Panel (a) uses the maximum pairwise correlation with any existing published predictor. Panel (b) uses the $R^2$ from regressing the data-mined return on 5 principal components of existing predictors computed using probabilistic PCA. 
  
  \vspace{2 ex}
  \centering
  \subfloat[Any Correlation\label{fig:abnormal_published_1}]{
      \includegraphics[width=0.48\textwidth]{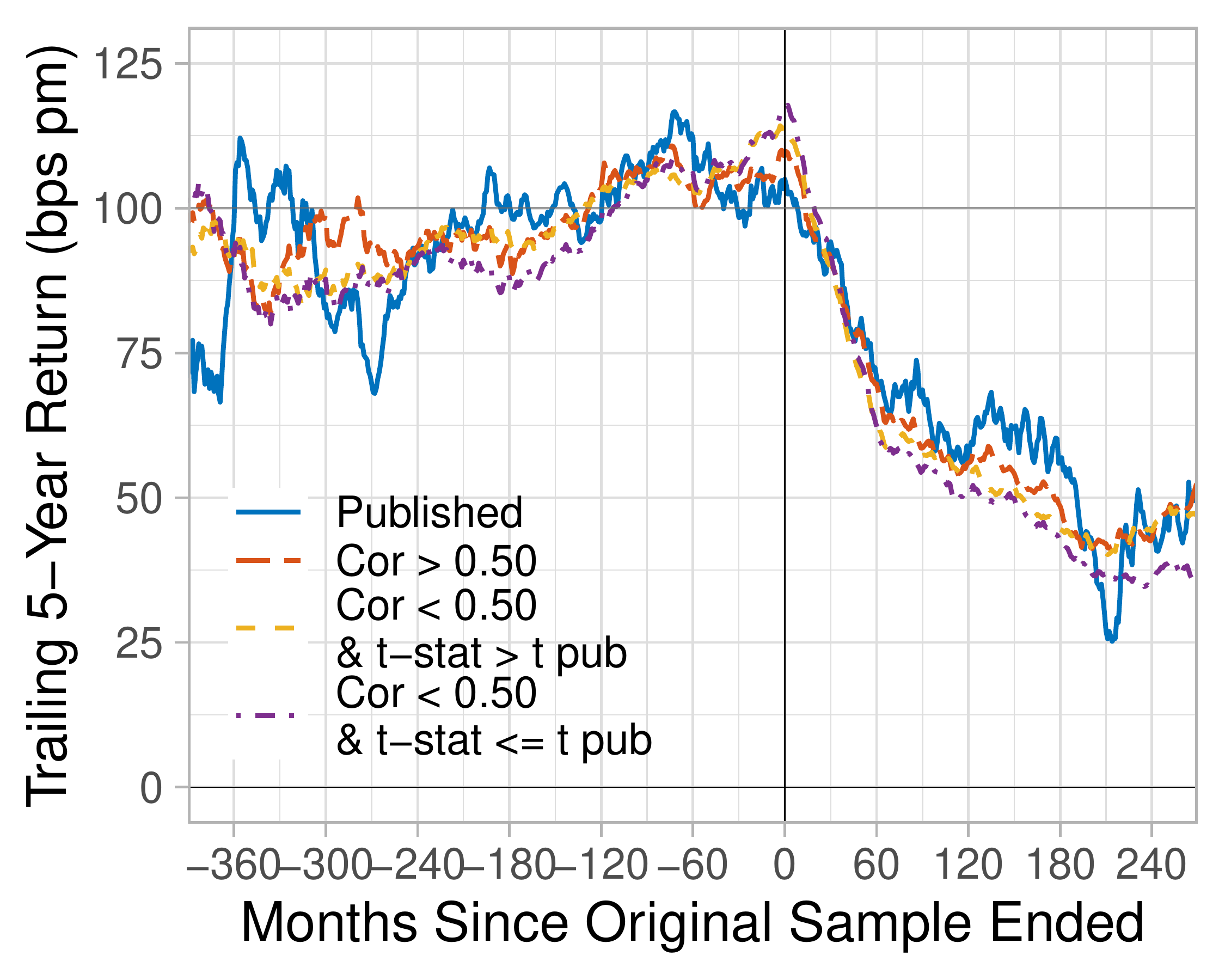}
  }    
  \subfloat[PPCA\label{fig:abnormal_published_2}]{
      \includegraphics[width=0.48\textwidth]{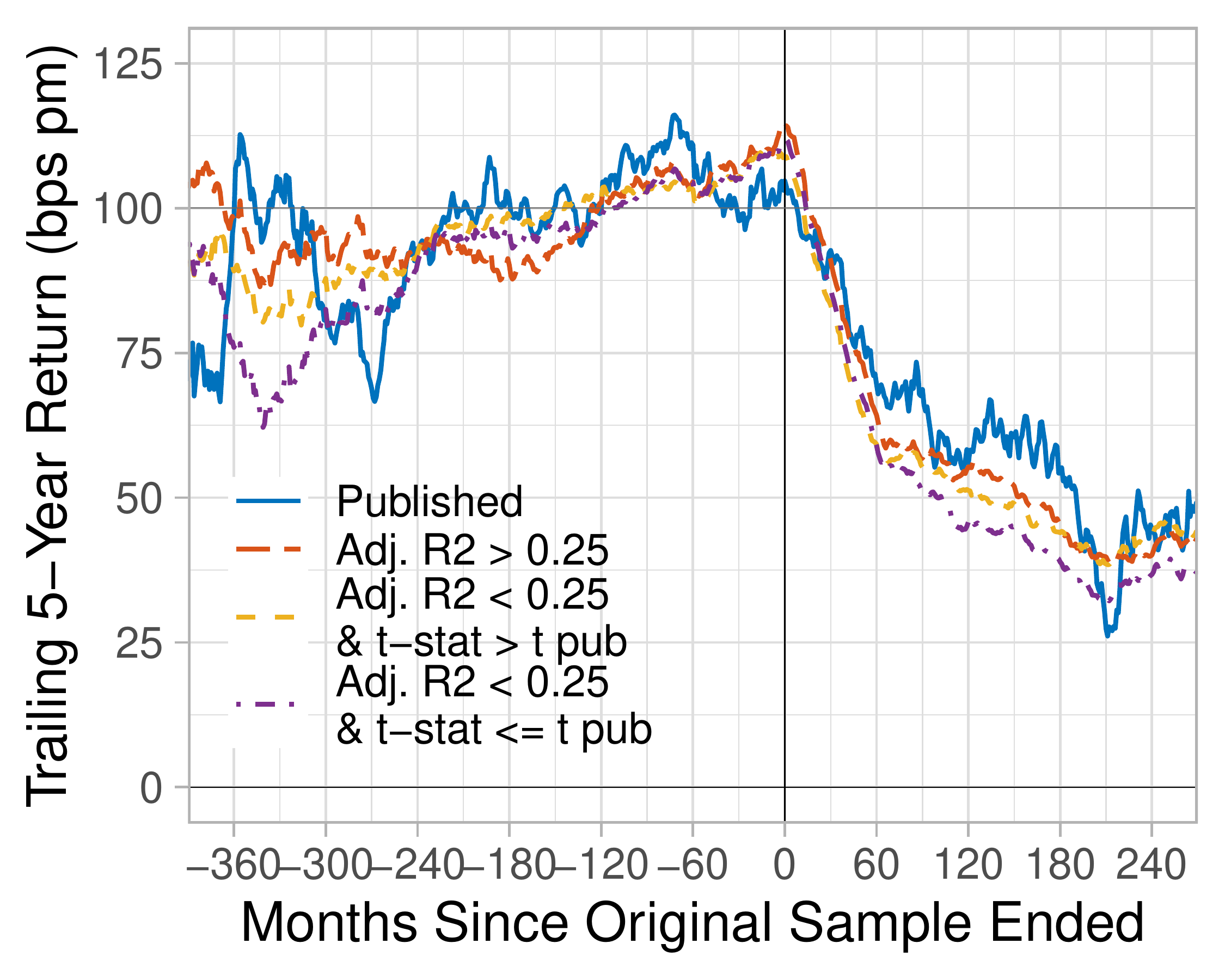}
  }
\end{figure}

Figure \ref{fig:decay-unspanned} aims to remove data-mined predictors that are highly correlated with all previously-published predictors, defined by sample end dates. Panel (a) measures this idea using the data-mined return's maximum pairwise correlation with any previously-published predictor. Panel (b) uses the $R^2$ from regressing the data-mined return on the first five factors extracted from previously-published returns via probabilistic principal component analysis (PPCA, \citet{roweis1997algorithms}; \citealt{chen2023missing}). For signals with low correlation, we further separate data-mined returns based on their in-sample t-stats. 

Excluding data-mined predictors that are highly correlated with all previously-published predictors leads to slightly worse post-sample performance. A worsening is natural, as over time the published literature will capture more and more of the strong predictive components of the data, of which there is a finite number (Table \ref{tab:dm-oos}, Panel (b)). The magnitude of the effect is modest, however, and is only noticeable if we focus on data-mined predictors with low in-sample t-stats. Overall, the similarity in post-sample performance of published and data-mined predictors is robust, and holds even if we purposefully make the data-mined predictors less similar to published ones.

An interesting feature of Figure \ref{fig:decay-unspanned} is that the trailing 5-year returns are correlated, even across groups of returns that have low monthly return correlations.  Given the near-zero autocorrelation in monthly returns, this is likely due to slow movements in long-run expected returns like the post-2004 decay in Appendix Table \ref{tab:dm-sum-2004} (see also \citealt{stambaugh2012short}; \citealt{yan2017fundamental}; \citealt{chen2022zeroing}).

% References ==============================
\clearpage
\pdfbookmark[0]{References}{References}
\printbibliography

@article{lakonishok1994contrarian,
  title={Contrarian investment, extrapolation, and risk},
  author={Lakonishok, Josef and Shleifer, Andrei and Vishny, Robert W},
  journal={The journal of finance},
  volume={49},
  number={5},
  pages={1541--1578},
  year={1994},
  publisher={Wiley Online Library}
}

@article{easterwood2024do,
  author       = {Sara Easterwood},
  title        = {Do Investors Have Data Blind Spots? The Role of Data Vendors in Capital Markets},
  journal      = {SSRN Electronic Journal},
  year         = {2024},
  note         = {Available at SSRN: \url{https://ssrn.com/abstract=4940766}}
}

@article{ou1989financial,
  title={Financial statement analysis and the prediction of stock returns},
  author={Ou, Jane A and Penman, Stephen H},
  journal={Journal of accounting and economics},
  volume={11},
  number={4},
  pages={295--329},
  year={1989},
  publisher={Elsevier}
}

@article{watts1978systematic,
  title={Systematic `abnormal' returns after quarterly earnings announcements},
  author={Watts, Ross L},
  journal={Journal of financial Economics},
  volume={6},
  number={2-3},
  pages={127--150},
  year={1978},
  publisher={Elsevier}
}

@article{loughran1995new,
  title={The new issues puzzle},
  author={Loughran, Tim and Ritter, Jay R},
  journal={The Journal of finance},
  volume={50},
  number={1},
  pages={23--51},
  year={1995},
  publisher={Wiley Online Library}
}

@article{chan1996momentum,
  title={Momentum strategies},
  author={Chan, Louis KC and Jegadeesh, Narasimhan and Lakonishok, Josef},
  journal={The Journal of Finance},
  volume={51},
  number={5},
  pages={1681--1713},
  year={1996},
  publisher={Wiley Online Library}
}

@article{bali2011maxing,
  title={Maxing out: Stocks as lotteries and the cross-section of expected returns},
  author={Bali, Turan G and Cakici, Nusret and Whitelaw, Robert F},
  journal={Journal of Financial Economics},
  volume={99},
  number={2},
  pages={427--446},
  year={2011},
  publisher={Elsevier}
}

@article{eberhart2004examination,
  title={An examination of long-term abnormal stock returns and operating performance following R\&D increases},
  author={Eberhart, Allan C and Maxwell, William F and Siddique, Akhtar R},
  journal={The Journal of Finance},
  volume={59},
  number={2},
  pages={623--650},
  year={2004},
  publisher={Wiley Online Library}
}

@article{asquith2005short,
  title={Short interest, institutional ownership, and stock returns},
  author={Asquith, Paul and Pathak, Parag A and Ritter, Jay R},
  journal={Journal of Financial Economics},
  volume={78},
  number={2},
  pages={243--276},
  year={2005},
  publisher={Elsevier}
}

@article{berk1995critique,
  title={A critique of size-related anomalies},
  author={Berk, Jonathan B},
  journal={The review of financial studies},
  volume={8},
  number={2},
  pages={275--286},
  year={1995},
  publisher={Oxford University Press}
}

@article{fama1973risk,
  title={Risk, return, and equilibrium: Empirical tests},
  author={Fama, Eugene F and MacBeth, James D},
  journal={Journal of political economy},
  volume={81},
  number={3},
  pages={607--636},
  year={1973},
  publisher={The University of Chicago Press}
}

@article{jensen2024subjective,
  title={Subjective Risk and Return},
  author={Jensen, Theis Ingerslev},
  journal={Available at SSRN 4276760},
  year={2024}
}

@techreport{marrow2024real,
  title={Real-Time Discovery and Tracking of Return-Based Anomalies},
  author={Marrow, Benjamin and Nagel, Stefan},
  year={2024},
  institution={Working Paper}
}

@article{fama1991efficient,
  title={Efficient capital markets: II},
  author={Fama, Eugene F},
  journal={The journal of finance},
  volume={46},
  number={5},
  pages={1575--1617},
  year={1991},
  publisher={Wiley Online Library}
}

@article{ang2006cross,
  title={The cross-section of volatility and expected returns},
  author={Ang, Andrew and Hodrick, Robert J and Xing, Yuhang and Zhang, Xiaoyan},
  journal={The journal of finance},
  volume={61},
  number={1},
  pages={259--299},
  year={2006},
  publisher={Wiley Online Library}
}

@article{balakrishnan2010post,
  title={Post loss/profit announcement drift},
  author={Balakrishnan, Karthik and Bartov, Eli and Faurel, Lucile},
  journal={Journal of Accounting and Economics},
  volume={50},
  number={1},
  pages={20--41},
  year={2010},
  publisher={Elsevier}
}

@article{chen2025t,
  title={Do t-statistic hurdles need to be raised?},
  author={Chen, Andrew Y},
  journal={Management Science},
  volume={71},
  number={7},
  pages={5830--5848},
  year={2025},
  publisher={INFORMS}
}

@article{benjamini1995controlling,
  title={Controlling the false discovery rate: a practical and powerful approach to multiple testing},
  author={Benjamini, Yoav and Hochberg, Yosef},
  journal={Journal of the Royal statistical society: series B (Methodological)},
  volume={57},
  number={1},
  pages={289--300},
  year={1995},
  publisher={Wiley Online Library}
}

@article{buffett1984superinvestors,
  title={The superinvestors of Graham-and-Doddsville},
  author={Buffett, Warren E},
  journal={Hermes},
  volume={17},
  year={1984}
}

@article{horrigan1968short,
  title={A short history of financial ratio analysis},
  author={Horrigan, James O},
  journal={The accounting review},
  volume={43},
  number={2},
  pages={284--294},
  year={1968},
  publisher={JSTOR}
}

@article{wall1919credit,
  title={Study of Credit Barometrics},
  author={Wall, Alexander},
  journal={Federal Reserve Bulletin},
  month={March},
  year={1919},
  pages={229--243}
}

@book{graham1934securityanalysis,
  author       = {Graham, Benjamin and Dodd, David L.},
  title        = {Security Analysis},
  publisher    = {Whittlesey House, McGraw-Hill Book Co.},
  address      = {United States},
  year         = {1934},
  pages        = {725},
  isbn         = {0-07-144820-9},
  oclc         = {2140220},
  dewey        = {332.63/2042/0973 22},
  lcclass      = {HG4521 .G67 1934},
  language     = {English}
}

@techreport{smith1935changes,
  title={Changes in the Financial Structure of Unsuccessful Industrial Corporations},
  author={Smith, Raymond F. and Winakor, Arthur H.},
  number={Bulletin No. 51},
  institution={University of Illinois, Bureau of Business Research},
  address={Urbana, Illinois},
  year={1935}
}

@book{fitzpatrick1932,
  author    = {Fitzpatrick, Paul J.},
  title     = {A Comparison of the Ratios of Successful Industrial Enterprises with Those of Failed Companies},
  publisher = {The Accountants Publishing Company},
  year      = {1932},
  address   = {Washington},
  note      = {Reprint of articles appearing in \textit{The Certified Public Accountant}, October, November and December 1932}
}

@techreport{ramser1931,
  author      = {Ramser, J. R. and Foster, Louis O.},
  title       = {A Demonstration of Ratio Analysis},
  institution = {University of Illinois, Bureau of Business Research},
  year        = {1931},
  type        = {Bulletin},
  number      = {40},
  address     = {Urbana, IL}
}

@book{merwin1942,
  author    = {Merwin, Charles L.},
  title     = {Financing Small Corporations: In Five Manufacturing Industries, 1926--36},
  publisher = {National Bureau of Economic Research},
  year      = {1942},
  address   = {New York},
  isbn      = {0-87014-130-9}
}

@article{gompers2003corporate,
  title={Corporate governance and equity prices},
  author={Gompers, Paul and Ishii, Joy and Metrick, Andrew},
  journal={The quarterly journal of economics},
  volume={118},
  number={1},
  pages={107--156},
  year={2003},
  publisher={MIT Press}
}

@article{bai1998estimating,
  title={Estimating and testing linear models with multiple structural changes},
  author={Bai, Jushan and Perron, Pierre},
  journal={Econometrica},
  pages={47--78},
  year={1998},
  publisher={JSTOR}
}

@article{gabaix2008variable,
  title={Variable rare disasters: A tractable theory of ten puzzles in macro-finance},
  author={Gabaix, Xavier},
  journal={American Economic Review},
  volume={98},
  number={2},
  pages={64--67},
  year={2008},
  publisher={American Economic Association}
}

@article{lettau2007long,
  title={Why is long-horizon equity less risky? A duration-based explanation of the value premium},
  author={Lettau, Martin and Wachter, Jessica A},
  journal={The journal of finance},
  volume={62},
  number={1},
  pages={55--92},
  year={2007},
  publisher={Wiley Online Library}
}

@article{hasler2023looking,
  title={Looking under the hood of data-mining},
  author={Hasler, Mathias},
  journal={Available at SSRN 4279944},
  year={2023}
}

@article{subrahmanyam2018equity,
  title={Equity market momentum: A synthesis of the literature and suggestions for future work},
  author={Subrahmanyam, Avanidhar},
  journal={Pacific-Basin Finance Journal},
  volume={51},
  pages={291--296},
  year={2018},
  publisher={Elsevier}
}

@article{holden2002news,
  title={News events, information acquisition, and serial correlation},
  author={Holden, Craig W and Subrahmanyam, Avanidhar},
  journal={The Journal of Business},
  volume={75},
  number={1},
  pages={1--32},
  year={2002},
  publisher={JSTOR}
}

@article{da2014frog,
  title={Frog in the pan: Continuous information and momentum},
  author={Da, Zhi and Gurun, Umit G and Warachka, Mitch},
  journal={The review of financial studies},
  volume={27},
  number={7},
  pages={2171--2218},
  year={2014},
  publisher={Oxford University Press}
}

@article{brav2002competing,
  title={Competing theories of financial anomalies},
  author={Brav, Alon and Heaton, John B},
  journal={The Review of Financial Studies},
  volume={15},
  number={2},
  pages={575--606},
  year={2002},
  publisher={Oxford University Press}
}

@article{papanikolaou2011investment,
  title={Investment shocks and asset prices},
  author={Papanikolaou, Dimitris},
  journal={Journal of Political Economy},
  volume={119},
  number={4},
  pages={639--685},
  year={2011},
  publisher={University of Chicago Press Chicago, IL}
}

@article{campbell2004bad,
  title={Bad beta, good beta},
  author={Campbell, John Y and Vuolteenaho, Tuomo},
  journal={American Economic Review},
  volume={94},
  number={5},
  pages={1249--1275},
  year={2004},
  publisher={American Economic Association}
}

@article{asness2013value,
  title={Value and momentum everywhere},
  author={Asness, Clifford S and Moskowitz, Tobias J and Pedersen, Lasse Heje},
  journal={The journal of finance},
  volume={68},
  number={3},
  pages={929--985},
  year={2013},
  publisher={Wiley Online Library}
}

@article{stambaugh2012short,
  title={The short of it: Investor sentiment and anomalies},
  author={Stambaugh, Robert F and Yu, Jianfeng and Yuan, Yu},
  journal={Journal of financial economics},
  volume={104},
  number={2},
  pages={288--302},
  year={2012},
  publisher={Elsevier}
}

@article{roweis1997algorithms,
  title={EM algorithms for PCA and SPCA},
  author={Roweis, Sam},
  journal={Advances in neural information processing systems},
  volume={10},
  year={1997}
}

@article{titman2004capital,
  title={Capital investments and stock returns},
  author={Titman, Sheridan and Wei, KC John and Xie, Feixue},
  journal={Journal of financial and Quantitative Analysis},
  volume={39},
  number={4},
  pages={677--700},
  year={2004},
  publisher={Cambridge University Press}
}

@article{gomes2003equilibrium,
  title={Equilibrium cross section of returns},
  author={Gomes, Joao and Kogan, Leonid and Zhang, Lu},
  journal={Journal of Political Economy},
  volume={111},
  number={4},
  pages={693--732},
  year={2003},
  publisher={The University of Chicago Press}
}

@article{hong1999unified,
  title={A unified theory of underreaction, momentum trading, and overreaction in asset markets},
  author={Hong, Harrison and Stein, Jeremy C},
  journal={The Journal of finance},
  volume={54},
  number={6},
  pages={2143--2184},
  year={1999},
  publisher={Wiley Online Library}
}

@article{thomas2002inventory,
  title={Inventory changes and future returns},
  author={Thomas, Jacob K and Zhang, Huai},
  journal={Review of Accounting Studies},
  volume={7},
  number={2},
  pages={163--187},
  year={2002},
  publisher={Springer}
}

@article{spiess1999long,
  title={The long-run performance of stock returns following debt offerings},
  author={Spiess, D Katherine and Affleck-Graves, John},
  journal={Journal of Financial Economics},
  volume={54},
  number={1},
  pages={45--73},
  year={1999},
  publisher={Elsevier}
}

@article{heston2008seasonality,
  title={Seasonality in the cross-section of stock returns},
  author={Heston, Steven L and Sadka, Ronnie},
  journal={Journal of Financial Economics},
  volume={87},
  number={2},
  pages={418--445},
  year={2008},
  publisher={Elsevier}
}

@article{fama2018choosing,
  title={Choosing factors},
  author={Fama, Eugene F and French, Kenneth R},
  journal={Journal of financial economics},
  volume={128},
  number={2},
  pages={234--252},
  year={2018},
  publisher={Elsevier}
}

@article{bali2023expected,
  title={Expected Mispricing},
  author={Bali, Turan G and Beckmeyer, Heiner and Wiedemann, Timo},
  journal={Available at SSRN},
  year={2023}
}

@article{holcblat2022anomaly,
  title={Anomaly or possible risk factor? Simple-to-use tests},
  author={Holcblat, Benjamin and Lioui, Abraham and Weber, Michael},
  journal={Simple-To-Use Tests (April 3, 2022)},
  year={2022}
}

@article{frey2023stock,
  title={Which stock return predictors reflect mispricing?},
  author={Frey, Jonas},
  journal={Available at SSRN},
  year={2023}
}

@article{chen2023high,
  title={High-Throughput Asset Pricing},
  author={Chen, Andrew Y and Dim, Chukwuma},
  journal={arXiv preprint arXiv:2311.10685},
  year={2025}
}

@article{chen2018general,
  title={A general equilibrium model of the value premium with time-varying risk premia},
  author={Chen, Andrew Y},
  journal={The Review of Asset Pricing Studies},
  volume={8},
  number={2},
  pages={337--374},
  year={2018},
  publisher={Oxford University Press}
}

@article{fama2006profitability,
  title={Profitability, investment and average returns},
  author={Fama, Eugene F and French, Kenneth R},
  journal={Journal of financial economics},
  volume={82},
  number={3},
  pages={491--518},
  year={2006},
  publisher={Elsevier}
}

@article{sloan1996stock,
  title={Do stock prices fully reflect information in accruals and cash flows about future earnings?},
  author={Sloan, Richard G},
  journal={Accounting review},
  pages={289--315},
  year={1996},
  publisher={JSTOR}
}

@article{cooper2008asset,
  title={Asset growth and the cross-section of stock returns},
  author={Cooper, Michael J and Gulen, Huseyin and Schill, Michael J},
  journal={the Journal of Finance},
  volume={63},
  number={4},
  pages={1609--1651},
  year={2008},
  publisher={Wiley Online Library}
}

@article{stattman1980book,
  title={Book values and stock returns},
  author={Stattman, Dennis},
  journal={The Chicago MBA: A journal of selected papers},
  volume={4},
  number={1},
  pages={25--45},
  year={1980}
}

@article{pastor2003liquidity,
  title={Liquidity risk and expected stock returns},
  author={P{\'a}stor, L'ubo{\v{s}} and Stambaugh, Robert F},
  journal={Journal of Political economy},
  volume={111},
  number={3},
  pages={642--685},
  year={2003},
  publisher={The University of Chicago Press}
}

@article{amihud2002illiquidity,
  title={Illiquidity and stock returns: cross-section and time-series effects},
  author={Amihud, Yakov},
  journal={Journal of financial markets},
  volume={5},
  number={1},
  pages={31--56},
  year={2002},
  publisher={Elsevier}
}

@article{mclean2016does,
  title={Does academic research destroy stock return predictability?},
  author={McLean, R David and Pontiff, Jeffrey},
  journal={The Journal of Finance},
  volume={71},
  number={1},
  pages={5--32},
  year={2016},
  publisher={Wiley Online Library}
}

@article{doran2007really,
  title={What Really Matters When Buying and Selling Stocks?},
  author={Doran, James and Wright, Colbrin},
  journal={Financial Education},
  volume={8},
  number={1},
  pages={35--61},
  year={2007}
}

@article{mukhlynina2020choice,
  title={The Choice of Valuation Techniques in Practice: Education Versus Profession},
  author={Mukhlynina, Liliya and Nyborg, Kjell G},
  journal={Critical Finance Review},
  volume={9},
  number={1-2},
  pages={201--265},
  year={2020},
  publisher={Now Publishers Inc.}
}

@article{bender2022millionaires,
  title={Millionaires speak: What drives their personal investment decisions?},
  author={Bender, Svetlana and Choi, James J and Dyson, Danielle and Robertson, Adriana Z},
  journal={Journal of Financial Economics},
  volume={146},
  number={1},
  pages={305--330},
  year={2022},
  publisher={Elsevier}
}

@article{chinco2022new,
  title={A new test of risk factor relevance},
  author={Chinco, Alex and Hartzmark, Samuel M and Sussman, Abigail B},
  journal={The Journal of Finance},
  volume={77},
  number={4},
  pages={2183--2238},
  year={2022},
  publisher={Wiley Online Library}
}

@article{gotofalse,
  title={False Alpha and Missed Alpha: An Out-of-Sample Mining Expedition},
  author={Goto, Shingo and Yamada, Toru},
  year={2025},
  journal={Working Paper}
}

@article{kozak2018interpreting,
  title={Interpreting factor models},
  author={Kozak, Serhiy and Nagel, Stefan and Santosh, Shrihari},
  journal={The Journal of Finance},
  volume={73},
  number={3},
  pages={1183--1223},
  year={2018},
  publisher={Wiley Online Library}
}

@article{berk1999optimal,
  title={Optimal investment, growth options, and security returns},
  author={Berk, Jonathan B and Green, Richard C and Naik, Vasant},
  journal={The Journal of finance},
  volume={54},
  number={5},
  pages={1553--1607},
  year={1999},
  publisher={Wiley Online Library}
}

@article{haugen1996commonality,
  title={Commonality in the determinants of expected stock returns},
  author={Haugen, Robert A and Baker, Nardin L},
  journal={Journal of financial economics},
  volume={41},
  number={3},
  pages={401--439},
  year={1996},
  publisher={Elsevier}
}

@article{abarbanell1998abnormal,
  title={Abnormal returns to a fundamental analysis strategy},
  author={Abarbanell, Jeffery S and Bushee, Brian J},
  journal={Accounting Review},
  pages={19--45},
  year={1998},
  publisher={JSTOR}
}

@article{mclean2020taking,
  title={Taking sides on return predictability},
  author={McLean, R David and Pontiff, Jeffrey and Reilly, Christopher},
  journal={Georgetown McDonough School of Business Research Paper},
  number={3637649},
  year={2020}
}

@article{calluzzo2019anomalies,
  title={When anomalies are publicized broadly, do institutions trade accordingly?},
  author={Calluzzo, Paul and Moneta, Fabio and Topaloglu, Selim},
  journal={Management Science},
  volume={65},
  number={10},
  pages={4555--4574},
  year={2019},
  publisher={INFORMS}
}

@article{sonnenschein1972market,
  title={Market excess demand functions},
  author={Sonnenschein, Hugo},
  journal={Econometrica: Journal of the Econometric Society},
  pages={549--563},
  year={1972},
  publisher={JSTOR}
}

@article{zaffaroni2022asset,
  title={Asset Pricing: Cross-section Predictability},
  author={Zaffaroni, Paolo and Zhou, Guofu},
  journal={Available at SSRN 4111428},
  year={2022}
}

@book{bali2016empirical,
  title={Empirical asset pricing: The cross section of stock returns},
  author={Bali, Turan G and Engle, Robert F and Murray, Scott},
  year={2016},
  publisher={John Wiley \& Sons}
}

@article{zhang2005value,
  title={The value premium},
  author={Zhang, Lu},
  journal={The Journal of Finance},
  volume={60},
  number={1},
  pages={67--103},
  year={2005},
  publisher={Wiley Online Library}
}

@book{cochrane2009asset,
  title={Asset pricing: Revised edition},
  author={Cochrane, John H},
  year={2009},
  publisher={Princeton university press}
}

@article{cochrane1999portfolio,
  title={Portfolio advice for a multifactor world},
  author={Cochrane, John H},
  journal={Economic Perspectives: Federal Reserve Bank of Chicago},
  volume={23},
  pages={59--78},
  year={1999}
}

@article{foster1984earnings,
  title={Earnings releases, anomalies, and the behavior of security returns},
  author={Foster, George and Olsen, Chris and Shevlin, Terry},
  journal={Accounting Review},
  pages={574--603},
  year={1984},
  publisher={JSTOR}
}

@article{banz1981relationship,
  title={The relationship between return and market value of common stocks},
  author={Banz, Rolf W},
  journal={Journal of financial economics},
  volume={9},
  number={1},
  pages={3--18},
  year={1981},
  publisher={Elsevier}
}

@article{chen2022zeroing,
  title={Zeroing in on the Expected Returns of Anomalies},
  author={Chen, Andrew Y and Velikov, Mihail},
  journal={Journal of Financial and Quantitative Analysis},
  year={2022}
}

@article{fama1993common,
  title={Common risk factors in the returns on stocks and bonds},
  author={Fama, Eugene F and French, Kenneth R},
  journal={Journal of financial economics},
  volume={33},
  number={1},
  pages={3--56},
  year={1993},
  publisher={Elsevier}
}

@article{fama1992cross,
  title={The cross-section of expected stock returns},
  author={Fama, Eugene F and French, Kenneth R},
  journal={the Journal of Finance},
  volume={47},
  number={2},
  pages={427--465},
  year={1992},
  publisher={Wiley Online Library}
}

@article{fama2015five,
  title={A five-factor asset pricing model},
  author={Fama, Eugene F and French, Kenneth R},
  journal={Journal of financial economics},
  volume={116},
  number={1},
  pages={1--22},
  year={2015},
  publisher={Elsevier}
}

@article{harvey2020false,
  title={False (and missed) discoveries in financial economics},
  author={Harvey, Campbell R and Liu, Yan},
  journal={The Journal of Finance},
  volume={75},
  number={5},
  pages={2503--2553},
  year={2020},
  publisher={Wiley Online Library}
}

@article{jegadeesh1993returns,
  title={Returns to buying winners and selling losers: Implications for stock market efficiency},
  author={Jegadeesh, Narasimhan and Titman, Sheridan},
  journal={The Journal of finance},
  volume={48},
  number={1},
  pages={65--91},
  year={1993},
  publisher={Wiley Online Library}
}

@article{green2017characteristics,
  title={The characteristics that provide independent information about average US monthly stock returns},
  author={Green, Jeremiah and Hand, John RM and Zhang, X Frank},
  journal={The Review of Financial Studies},
  volume={30},
  number={12},
  pages={4389--4436},
  year={2017},
  publisher={Oxford University Press}
}

@article{bessembinder2021time,
  title={Time Series Variation in the Factor Zoo},
  author={Bessembinder, Hendrik and Burt, Aaron and Hrdlicka, Christopher M},
  journal={Aaron Paul and Hrdlicka, Christopher M., Time Series Variation in the Factor Zoo},
  year={2023}
}

@misc{chen2023missing,
      title={Missing Values Handling for Machine Learning Portfolios}, 
      author={Andrew Y. Chen and Jack McCoy},
      year={2023},
      eprint={2207.13071},
      archivePrefix={arXiv},
      primaryClass={stat.ME}
}

@article{chen2024most,
  title={Most claimed statistical findings in cross-sectional return predictability are likely true},
  author={Chen, Andrew Y},
  journal={Journal of Finance: Insights and Perspectives},
  year = {\textit{Forthcoming}},
}

@article{harvey2016and,
  title={... and the cross-section of expected returns},
  author={Harvey, Campbell R and Liu, Yan and Zhu, Heqing},
  journal={The Review of Financial Studies},
  volume={29},
  number={1},
  pages={5--68},
  year={2016},
  publisher={Oxford University Press}
}

@article{chen2020publication,
  title={Publication bias and the cross-section of stock returns},
  author={Chen, Andrew Y and Zimmermann, Tom},
  journal={The Review of Asset Pricing Studies},
  volume={10},
  number={2},
  pages={249--289},
  year={2020},
  publisher={Oxford University Press}
}

@article{sullivan2001dangers,
  title={Dangers of data mining: The case of calendar effects in stock returns},
  author={Sullivan, Ryan and Timmermann, Allan and White, Halbert},
  journal={Journal of Econometrics},
  volume={105},
  number={1},
  pages={249--286},
  year={2001},
  publisher={Elsevier}
}

@article{sullivan1999data,
  title={Data-snooping, technical trading rule performance, and the bootstrap},
  author={Sullivan, Ryan and Timmermann, Allan and White, Halbert},
  journal={The journal of Finance},
  volume={54},
  number={5},
  pages={1647--1691},
  year={1999},
  publisher={Wiley Online Library}
}

@article{yan2017fundamental,
  title={Fundamental analysis and the cross-section of stock returns: A data-mining approach},
  author={Yan, Xuemin Sterling and Zheng, Lingling},
  journal={The Review of Financial Studies},
  volume={30},
  number={4},
  pages={1382--1423},
  year={2017},
  publisher={Oxford University Press}
}

@article{ChenZimmermann2021,
  title={Open Source Cross Sectional Asset Pricing},
  author={Chen, Andrew Y. and Tom Zimmermann},
  journal={Critical Finance Review},
  year={2022}
}

@article{harvey2017presidential,
  title={Presidential address: The scientific outlook in financial economics},
  author={Harvey, Campbell R},
  journal={The Journal of Finance},
  volume={72},
  number={4},
  pages={1399--1440},
  year={2017},
  publisher={Wiley Online Library}
}

@article{Chordia2014Have,
  title={Have capital market anomalies attenuated in the recent era of high liquidity and trading activity?},
  author={Chordia, Tarun and Subrahmanyam, Avanidhar and Tong, Qing},
  journal={Journal of Accounting and Economics},
  volume={58},
  number={1},
  pages={41--58},
  year={2014},
  publisher={Elsevier}
}

@article{Chordia2020Anomalies,
  title={Anomalies and false rejections},
  author={Chordia, Tarun and Goyal, Amit and Saretto, Alessio},
  journal={The Review of Financial Studies},
  volume={33},
  number={5},
  pages={2134--2179},
  year={2020},
  publisher={Oxford University Press}
}

@article{Engelberg2018Anomalies,
  title={Anomalies and news},
  author={Engelberg, Joseph and McLean, R David and Pontiff, Jeffrey},
  journal={The Journal of Finance},
  volume={73},
  number={5},
  pages={1971--2001},
  year={2018},
  publisher={Wiley Online Library}
}

@article{Hou2020Replicating,
  title={Replicating anomalies},
  author={Hou, Kewei and Xue, Chen and Zhang, Lu},
  journal={The Review of Financial Studies},
  volume={33},
  number={5},
  pages={2019--2133},
  year={2020},
  publisher={Oxford University Press}
}

@article{Jacobs2020Anomalies,
  title={Anomalies across the globe: Once public, no longer existent?},
  author={Jacobs, Heiko and M{\"u}ller, Sebastian},
  journal={Journal of Financial Economics},
  volume={135},
  number={1},
  pages={213--230},
  year={2020},
  publisher={Elsevier}
}

@article{Jensen1970Random,
 author = {Michael C. Jensen and George A. Benington},
 journal = {The Journal of Finance},
 number = {2},
 pages = {469--482},
 title = {Random Walks and Technical Theories: Some Additional Evidence},
 volume = {25},
 year = {1970}
}

@article{jensen2022there,
  title={Is there a replication crisis in finance?},
  author={Jensen, Theis Ingerslev and Kelly, Bryan and Pedersen, Lasse Heje},
  journal={The Journal of Finance},
  year={2022},
  publisher={Wiley Online Library}
}

@article{Lo1990Data,
  title={Data-snooping biases in tests of financial asset pricing models},
  author={Lo, Andrew W and MacKinlay, A Craig},
  journal={The Review of Financial Studies},
  volume={3},
  number={3},
  pages={431--467},
  year={1990},
  publisher={Oxford University Press}
}

% Internet Appendix ==============================
\newpage
\setcounter{section}{0}
\setcounter{table}{0}
\setcounter{figure}{0}
\renewcommand{\thesection}{IA.\arabic{section}}
\renewcommand*\thetable{IA.\arabic{table}}
\renewcommand*\thefigure{IA.\arabic{figure}}

% Remove appendix formatting - just show section number
\makeatletter
\renewcommand{\@seccntformat}[1]{\csname the#1\endcsname\quad}
\makeatother

% Fix pdf bookmarks
\renewcommand{\theHsection}{IA.\arabic{section}}
\renewcommand{\theHsubsection}{IA.\arabic{section}.\arabic{subsection}}

% Add parent bookmark and shift section depth
% \bookmark[dest=\@currentHref, level=0]{Internet Appendix}
\pdfbookmark[0]{Internet Appendix}{Internet Appendix}

% !TEX root = ../risk_vs.tex

\begin{center}
\Large{\textbf{Internet Appendix for ``Does Peer-Reviewed Research Help Predict Stock Returns''}}
\end{center}

% results that are not needed to justify main results go here
% this includes additional color 
\section{Additional Results on Data-Mined Predictability}\label{sec:intapp-dm-additional}
\begin{table}[!h]
  \caption{Out-of-Sample Returns from Mining Accounting Data: 2004-2020}
  \label{tab:dm-sum-2004}
  
  \begin{singlespace}
  \noindent 
  We sort 29,000 accounting ratios each June into 5 bins based on past 30-year long-short returns (in-sample) and compute the mean return over the next year within each bin (out-of-sample). Statistics are calculated by strategy, then averaged within bins, then averaged across sorting years. Decay is the percentage decrease in mean return out-of-sample relative to in-sample. We omit decay for bin 4 because the mean return in-sample is negligible. Out-of-sample returns are calculated using only data from 2004-2020.
  \end{singlespace}
  \begin{centering}
  \vspace{-3ex}
  \par\end{centering}
  \centering{}\setlength{\tabcolsep}{1.0ex} \small
  \begin{center}
  \begin{tabular}{crrrrrrrrrrrr} \toprule 
  % \multicolumn{13}{c}{Panel (b): Out-of-Sample Returns 2004-2020} \\ 
  In- &   & \multicolumn{5}{c}{Equal-Weighted Long-Short Deciles} &   & \multicolumn{5}{c}{Value-Weighted Long-Short Deciles} \\ \cmidrule{3-7}\cmidrule{9-13}Sample &   & \multicolumn{2}{c}{Past 30 Years (IS)} &   & \multicolumn{2}{c}{Next Year (OOS)} &   & \multicolumn{2}{c}{Past 30 Years (IS)} &   & \multicolumn{2}{c}{Next Year (OOS)} \\ \cmidrule{3-4}\cmidrule{6-7}\cmidrule{9-10}\cmidrule{12-13}Bin &   & \multicolumn{1}{c}{Return} & \multicolumn{1}{c}{\multirow{2}[2]{*}{t-stat}} &   & \multicolumn{1}{c}{Return} & \multicolumn{1}{c}{Decay} &   & \multicolumn{1}{c}{Return} & \multicolumn{1}{c}{\multirow{2}[2]{*}{t-stat}} &   & \multicolumn{1}{c}{Return} & \multicolumn{1}{c}{Decay} \\   &   & \multicolumn{1}{c}{(bps pm)} &   &   & \multicolumn{1}{c}{(bps pm)} & \multicolumn{1}{c}{(\%)} &   & \multicolumn{1}{c}{(bps pm)} &   &   & \multicolumn{1}{c}{(bps pm)} & \multicolumn{1}{c}{(\%)} \\ \cmidrule{1-1}\cmidrule{3-4}\cmidrule{6-7}\cmidrule{9-10}\cmidrule{12-13}
  
  % latex table generated in R 4.2.3 by xtable 1.8-4 package
% Mon Mar  3 17:09:07 2025
 1 &  & -59.2 & -3.99 &  & -24.9 & 57.9 &  & -37.3 & -1.88 &  & -4.2 & 88.7 \\ 
  2 &  & -28.1 & -2.29 &  & -9.6 & 65.8 &  & -14.6 & -0.91 &  & -1.1 & 92.5 \\ 
  3 &  & -11.7 & -1.01 &  & 0.1 & 100.9 &  & -4.2 & -0.28 &  & -2.6 & 38.7 \\ 
  4 &  & 1.8 & 0.14 &  & 6.7 &  &  & 5.5 & 0.36 &  & -3.7 &  \\ 
  5 &  & 23.9 & 1.48 &  & 16.3 & 31.8 &  & 25.8 & 1.31 &  & 0.6 & 97.8 \\

  \bottomrule \end{tabular}%   
  \end{center} 
\end{table}

\begin{table}[!h]
  \caption{Pairwise Correlations of Data-Mined Predictors}
  \label{tab:dm-cor}
  
  \begin{singlespace}
  \noindent Data-mined predictors are represented by strategies with t-statistics greater than 2.0 in at least 10\% of the in-sample periods from Table \ref{tab:dm-oos}.  The table reports percentiles of Pearson correlation coefficients computed over pairwise-complete return observations. 
  \end{singlespace}
  \begin{centering}
  \vspace{-2ex}
  \par\end{centering}
  \centering{}\setlength{\tabcolsep}{1.7ex} \small
  \begin{center}
  % ==== begin paste
  \begin{tabular}{clccccccccccc}
  \toprule
   \multicolumn{11}{c}{Panel (a): Pairwise correlations} \\
  \midrule
  Percentile & & 1 & 5 & 10 & 25 & 50 & 75 & 90 & 95 & 99 \\
  \midrule
  % latex table generated in R 4.2.3 by xtable 1.8-4 package
% Mon Mar  3 17:13:36 2025
 Equal-Weighted &  & -0.40 & -0.23 & -0.15 & -0.04 & 0.06 & 0.18 & 0.31 & 0.41 & 0.61 \\ 
  Value-Weighted &  & -0.33 & -0.20 & -0.14 & -0.06 & 0.02 & 0.11 & 0.21 & 0.30 & 0.57 \\ 
  
  \bottomrule
  \end{tabular}
  \end{center}
\end{table}

% =====================================================
\clearpage
\newpage
\section{Data-Mined Themes in Other Samples}\label{sec:intapp-themes}

\begin{table}[!h]
\caption{Themes from Mining Accounting Ratios in 1990}\label{tab:dm-theme1990}
\small

Table reports the 20 accounting ratio numerator and stock weight (equal- or value-) combinations with the largest mean t-stats using returns in the years 1963-1990 (IS). `ew' is equal-weight, `vw' is value-weight. Strategies are signed to have positive mean returns IS. `Pct Short' is the share of strategies that short the ratio.  `t-stat' and  `Mean Return' are averages across the 65 possible denominators. 
\vspace{2ex}

\centering

\setlength{\tabcolsep}{1.0ex}

\begin{tabular}{lrrrlrr}
\toprule
 & \multicolumn{3}{c}{1963-1990 (IS)} &  & \multicolumn{1}{c}{1991-2004} & \multicolumn{1}{c}{1991-2022} \\ \cmidrule{2-4} \cmidrule{6-7}
Numerator (Stock Weight) & \multicolumn{1}{c}{Pct} & \multirow{2}{*}{t-stat} & \multicolumn{1}{c}{Mean} &  & \multicolumn{2}{c}{Mean Return} \\  
 & \multicolumn{1}{c}{Short} &  & \multicolumn{1}{c}{Return} &  & \multicolumn{2}{c}{OOS / IS} \\ 
\midrule
$\Delta$Capital surplus (ew) & 100 & 5.8 & 0.67 &  & 1.04 & 0.94\\
$\Delta$Common stock (ew) & 100 & 5.8 & 0.69 &  & 0.80 & 0.55\\
$\Delta$Liabilities (ew) & 100 & 5.7 & 0.74 &  & 0.87 & 0.56\\
$\Delta$Inventories (ew) & 100 & 5.4 & 0.65 &  & 1.44 & 0.79\\
$\Delta$Current liabilities (ew) & 100 & 5.4 & 0.60 &  & 1.04 & 0.56\\
$\Delta$Debt in current liab (ew) & 100 & 5.2 & 0.48 &  & 0.30 & 0.31\\
Stock issuance (ew) & 100 & 5.2 & 0.89 &  & 1.03 & 0.80\\
$\Delta$Long-term debt (ew) & 100 & 5.1 & 0.53 &  & 1.31 & 0.75\\
$\Delta$Notes payable st (ew) & 100 & 5.1 & 0.46 &  & 0.17 & 0.25\\
$\Delta$Interest expense (ew) & 100 & 5.1 & 0.58 &  & 1.01 & 0.80\\
$\Delta$PPE net (ew) & 100 & 4.8 & 0.73 &  & 1.41 & 0.75\\
$\Delta$PPE gross (ew) & 100 & 4.7 & 0.73 &  & 1.15 & 0.61\\
Retained earnings restatement (ew) & 100 & 4.6 & 0.54 &  & 1.38 & 0.70\\
$\Delta$Assets (ew) & 100 & 4.5 & 0.73 &  & 1.63 & 0.94\\
Stock repurchases (ew) & 0 & 4.4 & 0.38 &  & 0.27 & 0.63\\
$\Delta$Convertible debt and stock (ew) & 100 & 4.1 & 0.42 &  & 1.47 & 1.18\\
$\Delta$Capital surplus (vw) & 100 & 4.0 & 0.57 &  & 0.72 & 0.64\\
$\Delta$Cost of goods sold (ew) & 100 & 3.9 & 0.49 &  & 1.41 & 0.84\\
Long-term debt issuance (ew) & 88 & 3.9 & 0.48 &  & 1.30 & 0.71\\
$\Delta$Invested capital (ew) & 100 & 3.9 & 0.63 &  & 2.16 & 1.20\\
\bottomrule
\end{tabular}

\end{table}

\begin{table}
\caption{Themes from Mining Accounting Ratios in 2000}\label{tab:dm-theme2000}
\small

Table reports the 20 accounting ratio numerator and stock weight (equal- or value-) combinations with the largest mean t-stats using returns in the years 1963-2000 (IS).  `ew' is equal-weight, `vw' is value-weight. Strategies are signed to have positive mean returns IS. `Pct Short' is the share of strategies that short the ratio.  `t-stat' and  `Mean Return' are averages across the 65 possible denominators. 
\vspace{2ex}

\centering

\setlength{\tabcolsep}{1.0ex}

\begin{tabular}{lrrrlrr}
\toprule
 & \multicolumn{3}{c}{1963-2000 (IS)} &  & \multicolumn{1}{c}{2001-2004} & \multicolumn{1}{c}{2001-2022} \\ \cmidrule{2-4} \cmidrule{6-7}
Numerator (Stock Weight) & \multicolumn{1}{c}{Pct} & \multirow{2}{*}{t-stat} & \multicolumn{1}{c}{Mean} &  & \multicolumn{2}{c}{Mean Return} \\  
 & \multicolumn{1}{c}{Short} &  & \multicolumn{1}{c}{Return} &  & \multicolumn{2}{c}{OOS / IS} \\ 
\midrule
$\Delta$Inventories (ew) & 100 & 6.9 & 0.77 &  & 0.72 & 0.33\\
$\Delta$Long-term debt (ew) & 100 & 6.4 & 0.60 &  & 0.81 & 0.37\\
$\Delta$Common stock (ew) & 100 & 6.3 & 0.66 &  & 0.81 & 0.46\\
$\Delta$PPE net (ew) & 100 & 6.3 & 0.82 &  & 1.10 & 0.37\\
$\Delta$Current liabilities (ew) & 100 & 6.1 & 0.61 &  & 0.94 & 0.33\\
$\Delta$Interest expense (ew) & 100 & 6.1 & 0.61 &  & 0.45 & 0.58\\
$\Delta$Liabilities (ew) & 100 & 6.0 & 0.71 &  & 0.87 & 0.44\\
$\Delta$PPE gross (ew) & 100 & 5.9 & 0.78 &  & 0.87 & 0.30\\
$\Delta$Debt subordinated convertible (ew) & 100 & 5.4 & 0.71 &  & 1.15 & 0.62\\
$\Delta$Debt convertible (ew) & 100 & 5.4 & 0.61 &  & 1.72 & 0.74\\
Retained earnings restatement (ew) & 100 & 5.4 & 0.61 &  & 1.07 & 0.29\\
$\Delta$Invested capital (ew) & 100 & 5.3 & 0.83 &  & 1.55 & 0.56\\
Merger sales contrib (ew) & 100 & 5.2 & 0.53 &  & 0.93 & 0.51\\
$\Delta$Assets (ew) & 100 & 5.2 & 0.86 &  & 1.33 & 0.53\\
$\Delta$Capital surplus (ew) & 100 & 5.2 & 0.69 &  & 0.86 & 0.85\\
$\Delta$Capital expenditure (ew) & 100 & 5.2 & 0.53 &  & 1.67 & 0.64\\
$\Delta$Cost of goods sold (ew) & 100 & 5.1 & 0.58 &  & 0.66 & 0.40\\
$\Delta$Num employees (ew) & 100 & 5.0 & 0.59 &  & 1.42 & 0.52\\
$\Delta$Intangible assets (ew) & 100 & 5.0 & 0.49 &  & 1.89 & 0.61\\
$\Delta$Debt in current liab (ew) & 100 & 4.9 & 0.40 &  & 0.03 & 0.32\\
\bottomrule
\end{tabular}

\end{table}

\begin{table}
\caption{Themes from Mining Accounting Ratios in 2010}\label{tab:dm-theme2010}
\small

Table reports the 20 accounting ratio numerator and stock weight (equal- or value-) combinations with the largest mean t-stats using returns in the years 1963-2010 (IS).  `ew' is equal-weight, `vw' is value-weight. Strategies are signed to have positive mean returns IS. `Pct Short' is the share of strategies that short the ratio.  `t-stat' and  `Mean Return' are averages across the 65 possible denominators. 
\vspace{2ex}

\centering

\setlength{\tabcolsep}{1.0ex}

\begin{tabular}{lrrrlrr}
\toprule
 & \multicolumn{3}{c}{1963-2010 (IS)} &  & \multicolumn{1}{c}{2011-2014} & \multicolumn{1}{c}{2011-2022} \\ \cmidrule{2-4} \cmidrule{6-7}
Numerator (Stock Weight) & \multicolumn{1}{c}{Pct} & \multirow{2}{*}{t-stat} & \multicolumn{1}{c}{Mean} &  & \multicolumn{2}{c}{Mean Return} \\  
 & \multicolumn{1}{c}{Short} &  & \multicolumn{1}{c}{Return} &  & \multicolumn{2}{c}{OOS / IS} \\ 
\midrule
$\Delta$Long-term debt (ew) & 100 & 6.5 & 0.54 &  & 0.64 & 0.24\\
$\Delta$Inventories (ew) & 100 & 6.5 & 0.65 &  & 0.47 & 0.46\\
$\Delta$Liabilities (ew) & 100 & 6.4 & 0.68 &  & 0.52 & 0.14\\
$\Delta$Common stock (ew) & 100 & 6.3 & 0.60 &  & 0.24 & 0.40\\
$\Delta$Interest expense (ew) & 100 & 6.3 & 0.57 &  & 0.61 & 0.51\\
$\Delta$PPE net (ew) & 100 & 6.1 & 0.72 &  & 0.58 & 0.37\\
$\Delta$Current liabilities (ew) & 100 & 5.8 & 0.54 &  & 0.17 & 0.24\\
$\Delta$Debt convertible (ew) & 100 & 5.7 & 0.60 &  & 0.80 & 0.60\\
Merger sales contrib (ew) & 100 & 5.5 & 0.47 &  & 0.16 & 0.52\\
$\Delta$Assets (ew) & 100 & 5.5 & 0.81 &  & 0.40 & 0.35\\
$\Delta$Invested capital (ew) & 100 & 5.5 & 0.78 &  & 0.43 & 0.47\\
$\Delta$Intangible assets (ew) & 100 & 5.4 & 0.50 &  & 0.30 & 0.23\\
$\Delta$PPE gross (ew) & 100 & 5.3 & 0.65 &  & 0.52 & 0.45\\
$\Delta$Convertible debt and stock (ew) & 100 & 5.0 & 0.47 &  & 1.05 & 0.89\\
Retained earnings restatement (ew) & 100 & 5.0 & 0.51 &  & 0.53 & 0.19\\
$\Delta$Num employees (ew) & 100 & 4.9 & 0.54 &  & 0.25 & 0.53\\
$\Delta$Capital surplus (ew) & 100 & 4.8 & 0.65 &  & 0.54 & 0.97\\
$\Delta$Debt subordinated convertible (ew) & 100 & 4.8 & 0.63 &  & 0.34 & 0.86\\
$\Delta$Debt in current liab (ew) & 100 & 4.8 & 0.35 &  & 0.32 & 0.25\\
$\Delta$Capital expenditure (ew) & 100 & 4.7 & 0.47 &  & 0.36 & 0.89\\
\bottomrule
\end{tabular}

\end{table}

% =====================================================
\clearpage
\newpage
\section{Full Sample Risk Adjustments}\label{sec:intapp-fullsamplerisk}

As a robustness check, we present results using full sample risk adjustments instead of sample-specific alphas. In the full sample approach, betas are estimated over the entire available period from the sample start date onwards, rather than separately for in-sample and out-of-sample periods.

\begin{figure}[!h]
   \caption{Research vs Data-Mining: Full Sample Factor-Adjusted Returns}
   \label{fig:pub-vs-dm-factor-fs}
   We repeat Figure \ref{fig:pub-vs-dm-2} using full sample risk adjustments, where $\hat{\beta}_i$ is estimated using all available data from the sample start date onwards. Shaded area shows one standard error for the published predictors, clustered by calendar month and predictor. 
   \vspace{0.15in}
   
   \centering
   \subfloat[CAPM-Adjusted]{
   \includegraphics[width=0.48\textwidth]{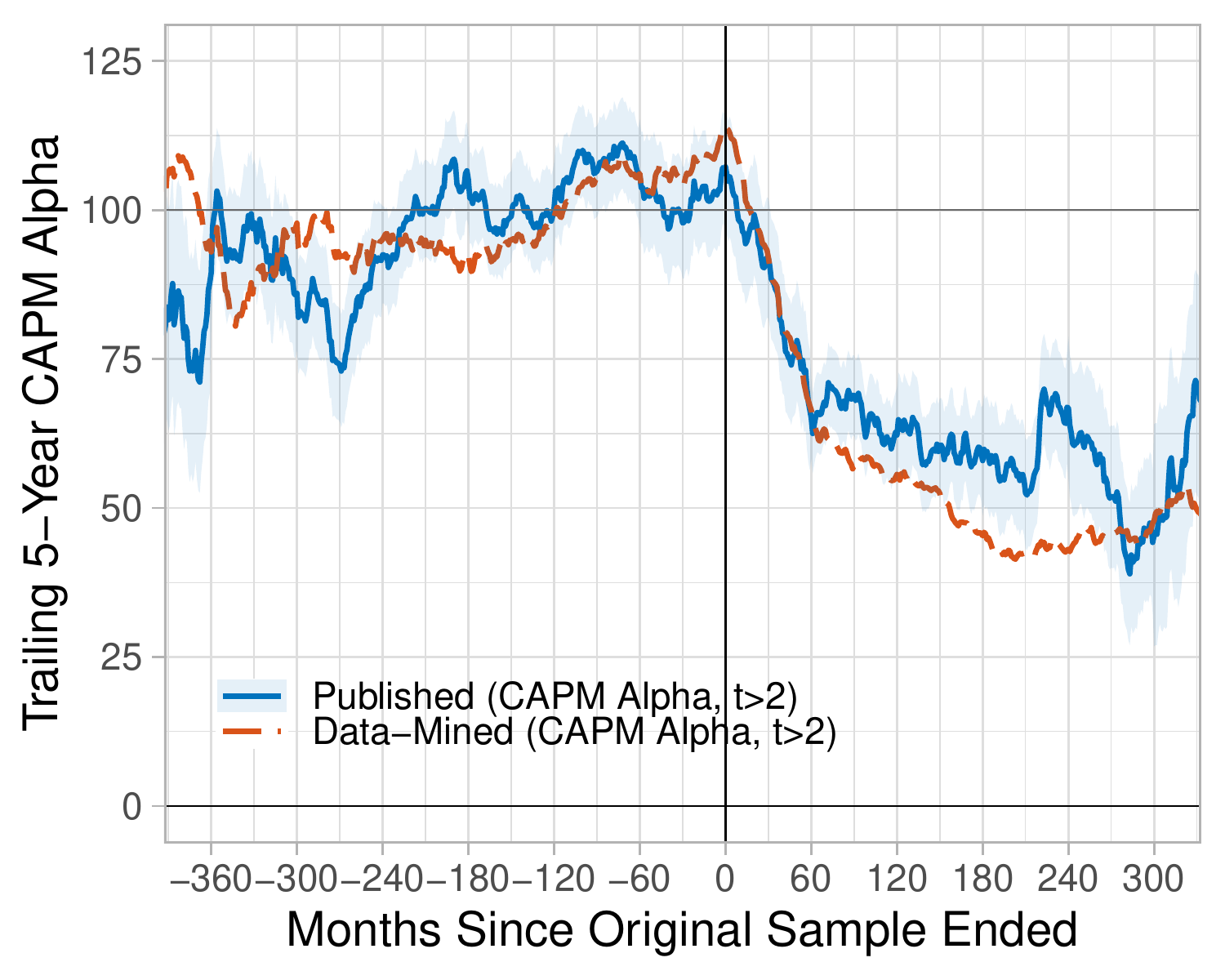} 
   }
   \subfloat[FF3-Adjusted]{
   \includegraphics[width=0.48\textwidth]{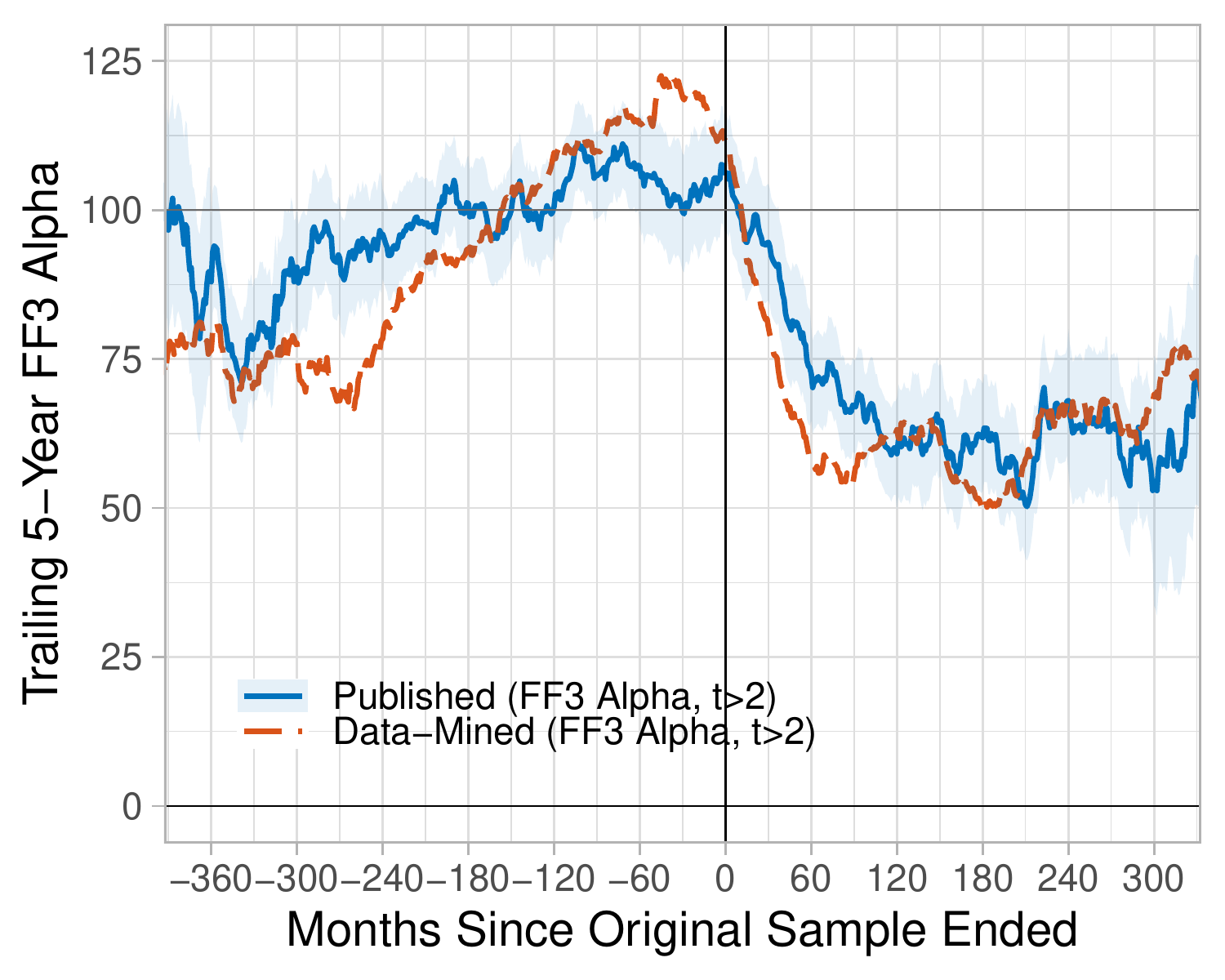}
   }
\end{figure}

\begin{table}[htbp]
\centering
\small
\caption{Full Sample Risk-Adjusted Returns: Theoretical Explanation and Modeling Formalism}
\label{tab:risk_adjusted_fs}
\setlength{\tabcolsep}{0.4ex}

\begin{tabular}{lcccccc}
\toprule
  & \multicolumn{2}{c}{Raw} & \multicolumn{2}{c}{CAPM} & \multicolumn{2}{c}{FF3}
\\
\cmidrule(lr){2-3} \cmidrule(lr){4-5} \cmidrule(lr){6-7}
Group & Return & Outperf. & Return & Outperf. & Return & Outperf.
\\
\midrule
\multicolumn{7}{l}{\textbf{Theoretical Explanation}}\\
\addlinespace[2pt]
Risk & 43 & 5 & 42 & 0 & 53 & -1
\\
 & (8) & (8) & (8) & (8) & (7) & (8)
\\[2pt]
Mispricing & 55 & 4 & 57 & 4 & 57 & -3
\\
 & (4) & (4) & (3) & (4) & (3) & (4)
\\[2pt]
Agnostic & 65 & 9 & 82 & 25 & 89 & 18
\\
 & (8) & (8) & (9) & (9) & (7) & (8)
\\[2pt]
\multicolumn{7}{l}{\textbf{Modeling Formalism}}\\
\addlinespace[2pt]
No Model & 56 & 5 & 60 & 8 & 63 & 3
\\
 & (3) & (3) & (3) & (3) & (3) & (3)
\\[2pt]
Stylized & 63 & 15 & 57 & 1 & 68 & -9
\\
 & (12) & (13) & (13) & (13) & (12) & (13)
\\[2pt]
Dynamic or Quantitative & 34 & -2 & 55 & 17 & 44 & -12
\\
 & (14) & (14) & (13) & (15) & (13) & (16)
\\[2pt]
\multicolumn{7}{l}{\textbf{Overall}}\\
\addlinespace[2pt]
All & 56 & 5 & 60 & 8 & 63 & 1
\\
 & (3) & (3) & (3) & (3) & (3) & (3)
\\[2pt]
\bottomrule
\end{tabular}
\end{table}

\begin{table}[htbp]
\centering
\small
\caption{Full Sample Risk-Adjusted Returns: Discipline and Journal Rank}
\label{tab:risk_adjusted_fs_dj}
\setlength{\tabcolsep}{0.2ex}
\begin{tabular}{lcccccc}
\toprule
  & \multicolumn{2}{c}{Raw} & \multicolumn{2}{c}{CAPM} & \multicolumn{2}{c}{FF3}
\\
\cmidrule(lr){2-3} \cmidrule(lr){4-5} \cmidrule(lr){6-7}
Group & Return & Outperformance & Return & Outperformance & Return & Outperformance
\\
\midrule
\multicolumn{7}{l}{\textbf{Discipline}}\\
\addlinespace[2pt]
Finance & 59 & 8 & 63 & 12 & 67 & 6
\\
 & (4) & (4) & (4) & (4) & (3) & (4)
\\[2pt]
Accounting & 43 & -6 & 46 & -6 & 45 & -19
\\
 & (6) & (7) & (5) & (6) & (5) & (6)
\\[2pt]
\multicolumn{7}{l}{\textbf{Journal Rank}}\\
\addlinespace[2pt]
JF, JFE, RFS & 60 & 8 & 66 & 14 & 70 & 8
\\
 & (4) & (4) & (4) & (5) & (4) & (4)
\\[2pt]
AR, JAR, JAE & 43 & -6 & 46 & -6 & 45 & -19
\\
 & (6) & (7) & (5) & (6) & (5) & (6)
\\[2pt]
Other & 53 & 8 & 57 & 5 & 62 & 1
\\
 & (6) & (6) & (6) & (6) & (6) & (7)
\\[2pt]
\bottomrule
\end{tabular}
\end{table}

\clearpage

% ==============================
\clearpage
\newpage
\section{Alternative Measures of Risk}\label{sec:intapp-riskalt}

\subsection{Risk vs Mispricing Words}\label{sec:intapp-risk-vs-words}

In Section \ref{sec:hetero-method}, the risk vs mispricing categorization is binary. To make a more continuous measure, we count risk and mispricing words in the published papers. We remove stopwords, lowercase and lemmatize all words using standard methods. 

We consider as risk words the following terms and
their grammatical variations: \textquotedbl utility,\textquotedbl{}
\textquotedbl maximize,\textquotedbl{} \textquotedbl minimize,\textquotedbl{}
\textquotedbl optimize,\textquotedbl{} \textquotedbl premium,\textquotedbl{}
\textquotedbl premia,\textquotedbl{} \textquotedbl premiums,\textquotedbl{}
\textquotedbl consume,\textquotedbl{} \textquotedbl marginal,\textquotedbl{}
\textquotedbl equilibrium,\textquotedbl{} \textquotedbl sdf,\textquotedbl{}
\textquotedbl investment-based,\textquotedbl{} and \textquotedbl theoretical.\textquotedbl{}
We also count as risk words appearances of \textquotedblleft risk\textquotedblright{}
that are not preceded by \textquotedblleft lower,\textquotedblright{}
and appearances of \textquotedblleft aversion,\textquotedblright{}
\textquotedblleft rational,\textquotedblright{} and \textquotedblleft risky\textquotedblright{}
that are not preceded by \textquotedblleft not.\textquotedblright{}

The mispricing words consist of \textquotedbl{}
\textquotedbl anomaly,\textquotedbl{} \textquotedbl behavioral,\textquotedbl{}
\textquotedbl optimistic,\textquotedbl{} \textquotedbl pessimistic,\textquotedbl{}
\textquotedbl sentiment,\textquotedbl{} \textquotedbl underreact,\textquotedbl{}
\textquotedbl overreact,\textquotedbl{}
\textquotedbl failure,\textquotedbl{} \textquotedbl bias,\textquotedbl{}
\textquotedbl overvalue,\textquotedbl{} \textquotedbl misvalue,\textquotedbl{}
\textquotedbl undervalue,\textquotedbl{} \textquotedbl attention,\textquotedbl{}
\textquotedbl underperformance,\textquotedbl{} \textquotedbl extrapolate,\textquotedbl{}
\textquotedbl underestimate,\textquotedbl{} \textquotedbl misreaction,\textquotedbl{}
\textquotedbl inefficiency,\textquotedbl{} \textquotedbl delay,\textquotedbl{}
\textquotedbl suboptimal,\textquotedbl{} \textquotedbl mislead,\textquotedbl{}
\textquotedbl overoptimism,\textquotedbl{} \textquotedbl arbitrage,\textquotedbl{}
\textquotedbl factor unlikely,\textquotedbl{} and their grammatical
variations. We further count as mispricing the terms \textquotedbl not
rewarded,\textquotedbl{} \textquotedbl little risk,\textquotedbl{}
\textquotedbl risk cannot {[}explain{]},\textquotedbl{} \textquotedbl{}
low {[}type of{]} risk,\textquotedbl{} \textquotedbl unrelated {[}to
the type of{]} risk,\textquotedbl{} \textquotedbl fail {[}to{]} reflect,\textquotedbl{}
and \textquotedbl market failure,\textquotedbl{} where the terms
in brackets are captured using regular expressions or correspond to
stopwords.

\begin{figure}[!h]
\caption{Post-Sample Returns vs Risk to Mispricing Words}
\label{fig:decay_vs_words}

Each marker represents one published predictor's mean return.  The regression line is fitted with OLS.  The full reference for each acronym can be found at \url{https://github.com/OpenSourceAP/CrossSection/blob/master/SignalDoc.csv}.  The relationship between risk words and post-sample returns is negative.
\vspace{0.15in}

\centering
\includegraphics[width=0.75\textwidth]{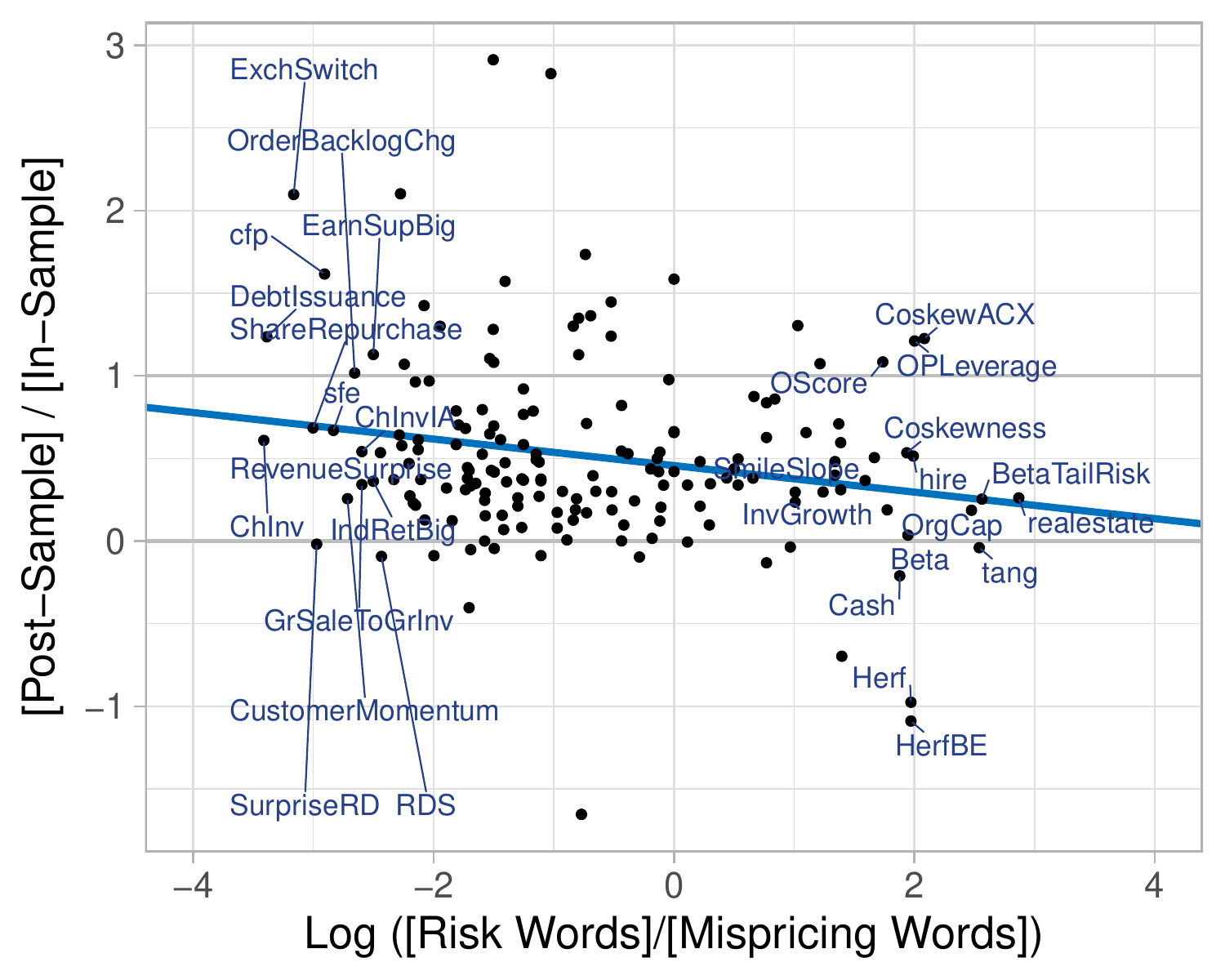}      
\end{figure}

Figure \ref{fig:decay_vs_words} plots post-sample returns against the ratio of risk words (e.g., ``utility,'' ``equilibrium'') to mispricing words (e.g., ``sentiment,'' ``underreact'') in the published papers.  The relationship is negative, consistent with Table \ref{tab:hetero-bytheory}.

\clearpage

\subsection{Factor Model Measures of Risk}\label{sec:intapp-risk-factor}

One can alternatively measure risk using factor models, as follows. For each published long-short portfolio $i$, we estimate exposure to factor $k$ using time-series regressions on the original papers' sample periods.  According to the factor models, the estimated expected return is $\sum_k \hat{\beta}_{k, i} \bar{f}_{k}$, where $\bar{f}_k$ is the original-sample mean return of factor $k$.  \citet{fama1993common} state that $\hat{\beta}_{i,k}$  with respect to their SMB and HML factors have ``a clear interpretation as risk-factor sensitivities.''  If this interpretation is both correct and stable, then the estimated expected return should remain post-sample. 

\begin{figure}[!h]
\caption{Mean Returns Post-Sample vs Factor Model Predictions}
\label{fig:risk-alt}
Each marker is one published long-short strategy.  [Post-Sample]/[In-Sample] is the mean return post-sample divided by the mean return in-sample.  [Predicted by Risk Model] is $\sum_k \hat{\beta}_{k, i} \bar{f}_{k}$, where $\bar{f}_k$ is the in-sample mean return of factor $k$ and $\hat{\beta}_{k,i}$ comes from an in-sample time series regression of long-short returns on factor realizations. FF3 and FF5 are the Fama-French 3- and 5-factor models. The blue line is the OLS fit. The axes zoom in on the interpretable region of the chart and omits outliers.  Factor models attribute a minority of in-sample predictability to risk, at best.  Post-sample decay is the distance between the horizontal line at 1.0 and the regression line, and this decay is near 50\% even for predictors that are entirely due to risk according to the CAPM and FF3.  For FF5, decay is smaller for predictors that are more than 75\% due to risk, but these predictors are rare.
\vspace{0.15in}

\centering
\subfloat[Post-Sample Return vs CAPM Prediction]{
\includegraphics[width=0.45\textwidth]{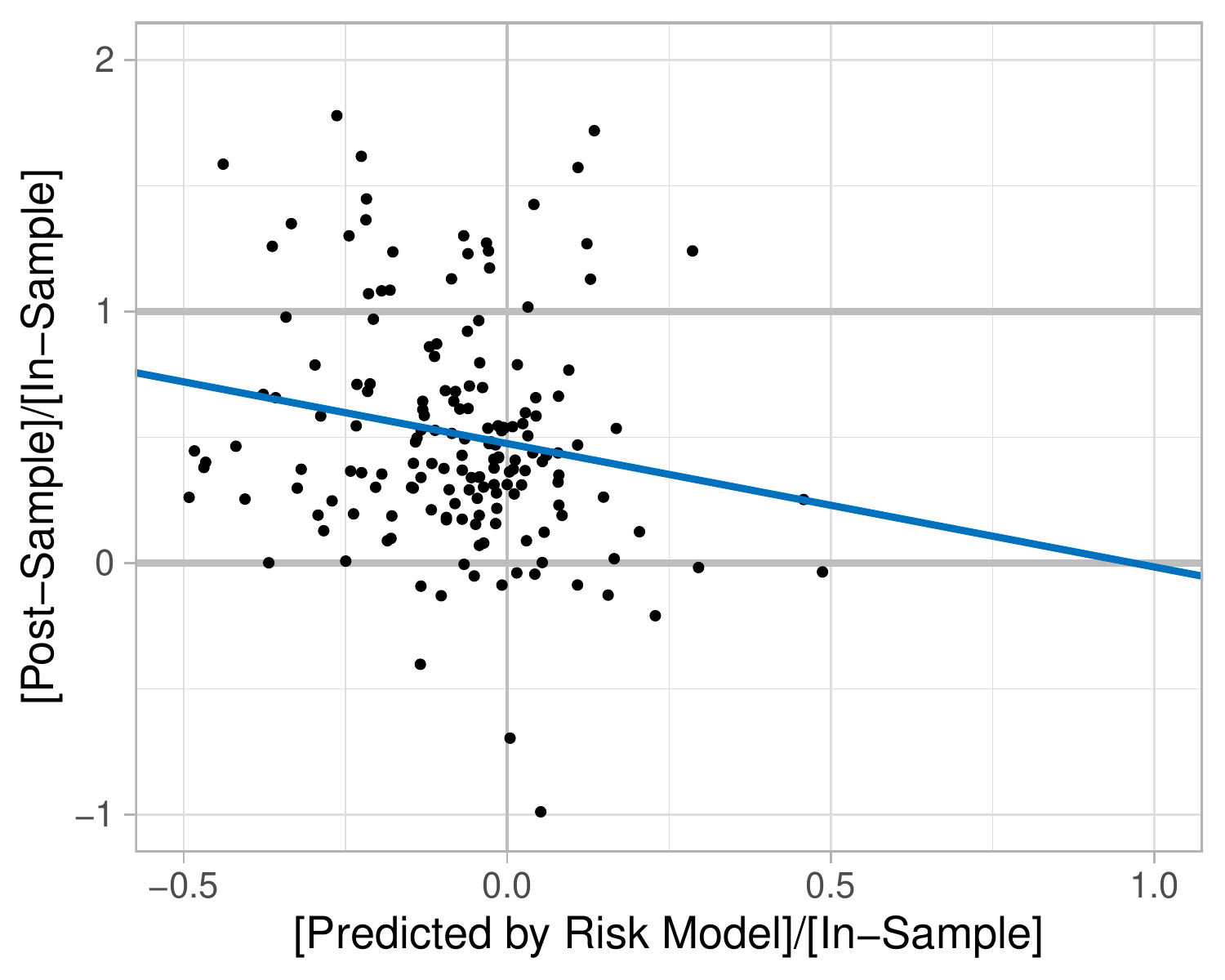} 
}
\subfloat[Post-Sample Return vs FF3 Prediction]{
\includegraphics[width=0.45\textwidth]{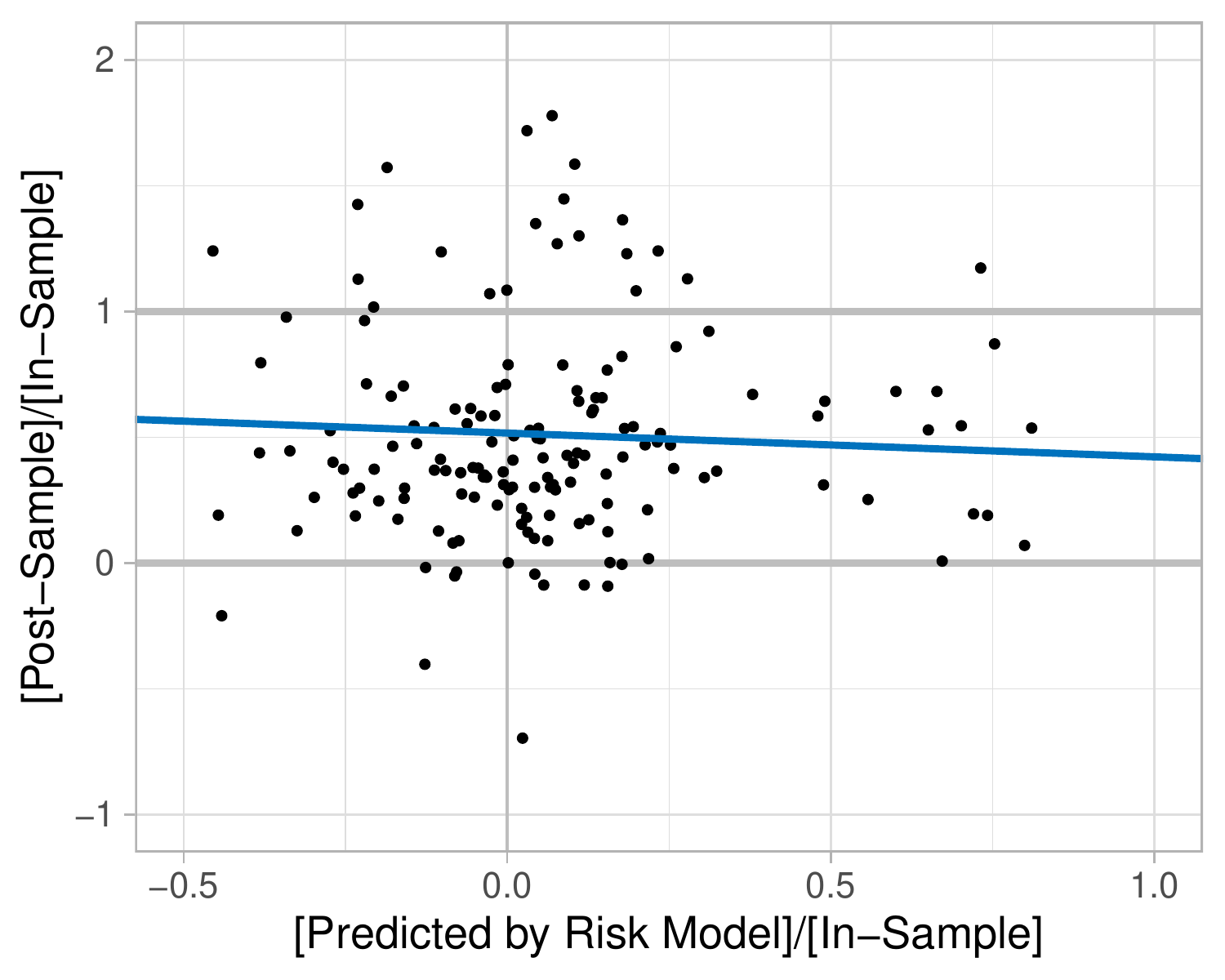}
}\\
\subfloat[Post-Sample Return vs FF5 Prediction]{
\includegraphics[width=0.45\textwidth]{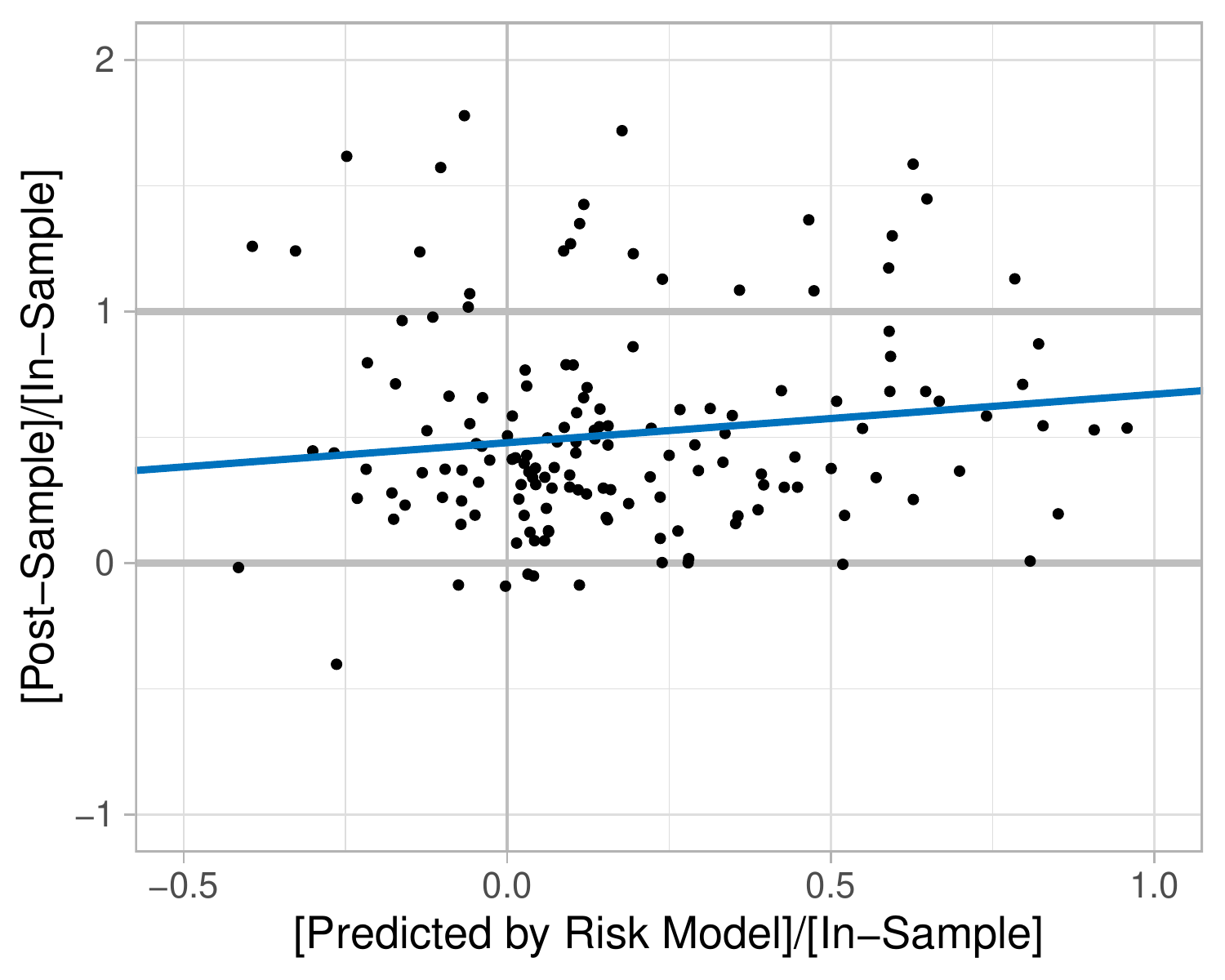} 
}
\end{figure}

Figure \ref{fig:risk-alt} plots the post-sample mean return against the factor model expected returns, using the CAPM, Fama-French 3 (FF3), or Fama-French 5 (FF5) models.  We normalize by the original-sample mean return for ease of interpretation.  With this normalization, the position on the x-axis ([Predicted by Risk Model]/[In-Sample]) represents the share of predictability due to risk.

 % --------------------------------------------

The figure shows that a minority of in-sample predictability is attributed to risk, at best.  Using the CAPM (Panel (a)), nearly all predictability is less than 25\% due to risk (to the left of the vertical line at 0.25), and many predictors have a \textit{negative} risk share.  FF3 (Panel (b)) implies more predictability is due to risk, but still the vast majority of predictors lie to the left of 0.50.

\citet{fama2015five} are more cautious than \citet{fama1993common}, and describe the risk-based ICAPM as ``the more ambitious interpretation'' of the five factor model. Under the more ambitious interpretation, FF5 implies that most predictors are less than 50\% due to risk. These results are consistent with our manual reading of the papers, which typically attribute predictability to mispricing (Table \ref{tab:hetero-count}). 

The regression lines in Figure \ref{fig:risk-alt} show negative or mildly positive relationships between factor model risk and post-sample returns. The regression fits for the CAPM and FF3 models never stray far from 50\%, implying that even predictors that are entirely due to risk are little different than the typical predictor in terms of post-sample robustness.  FF5 risk shows a stronger relationship with post-sample returns, but even the rare predictors that are 75\% due to risk decay by roughly 40\% post-sample. Moreover, the \citet{fama2015five} model may have the benefit of hindsight, as the median publication year for the \citet{ChenZimmermann2021} predictors is 2006.
    
% =====================================================
\clearpage
\newpage
\section{Additional Results on Published Predictors}\label{sec:intapp-pub-additional}

\begin{table}[ht]
  \caption{Signals by Theory and Published Journal}
  \label{tab:journal_theory}
  
  \begin{singlespace}
  \noindent This table lists the number of signals by theory and published journal.  Finance journals find risk explanations more frequently than accounting journals, but risk explanations still account for a small minority of predictors in finance journals.
  \end{singlespace}
  \begin{centering}
  \vspace{2ex}
  \par\end{centering}
  \centering
  \begin{tabular}{l|r|r|r}
    \toprule
   & Agnostic & Mispricing & Risk \\ 
    \midrule
    % latex table generated in R 4.4.1 by xtable 1.8-4 package
% Mon Mar  3 19:58:32 2025
 AR &   1 &  14 &   0 \\ 
  BAR &   0 &   1 &   0 \\ 
  Book &   2 &   0 &   0 \\ 
  CAR &   0 &   1 &   0 \\ 
  FAJ &   1 &   1 &   0 \\ 
  JAE &   2 &   8 &   0 \\ 
  JAR &   2 &   2 &   0 \\ 
  JBFA &   0 &   1 &   0 \\ 
  JEmpFin &   0 &   1 &   0 \\ 
  JF &  12 &  34 &  10 \\ 
  JFE &  11 &  19 &   6 \\ 
  JFM &   0 &   2 &   0 \\ 
  JFQA &   0 &   3 &   2 \\ 
  JFR &   0 &   0 &   1 \\ 
  JOIM &   0 &   1 &   0 \\ 
  JPE &   0 &   0 &   3 \\ 
  JPM &   1 &   0 &   0 \\ 
  MS &   0 &   2 &   2 \\ 
  Other &   1 &   1 &   0 \\ 
  RAS &   0 &   5 &   1 \\ 
  RED &   0 &   0 &   1 \\ 
  RFQA &   0 &   1 &   0 \\ 
  RFS &   0 &   6 &   7 \\ 
  ROF &   0 &   1 &   1 \\ 
  WP &   1 &   1 &   0 \\ 
  
  \bottomrule
  \end{tabular}
  \end{table}

\begin{figure}[!h]
\caption{Decay vs Journal}
Plot shows the ratio of post-sample to in-sample returns for each predictor, grouped by journal type. Journal types are Top 5 Economics (QJE, JPE), Top 3 Finance (JF, JFE, RFS), Top 3 Accounting (JAR, JAE, AR), and Other journals. Each point represents one predictor. The blue diamonds show the mean ratio within each journal group. The horizontal gray lines show ratios of 0 and 1. A ratio of 1 means the predictor maintains its full predictive power out-of-sample, while a ratio of 0 means the predictor completely fails out-of-sample. Text labels identify notable predictors and the top performers within each journal group. The blue line connects group means to highlight the pattern across journal types.
\label{fig:decay-vs-journal}

\vspace{0.15in}

\centering
\includegraphics[width=0.75\textwidth]{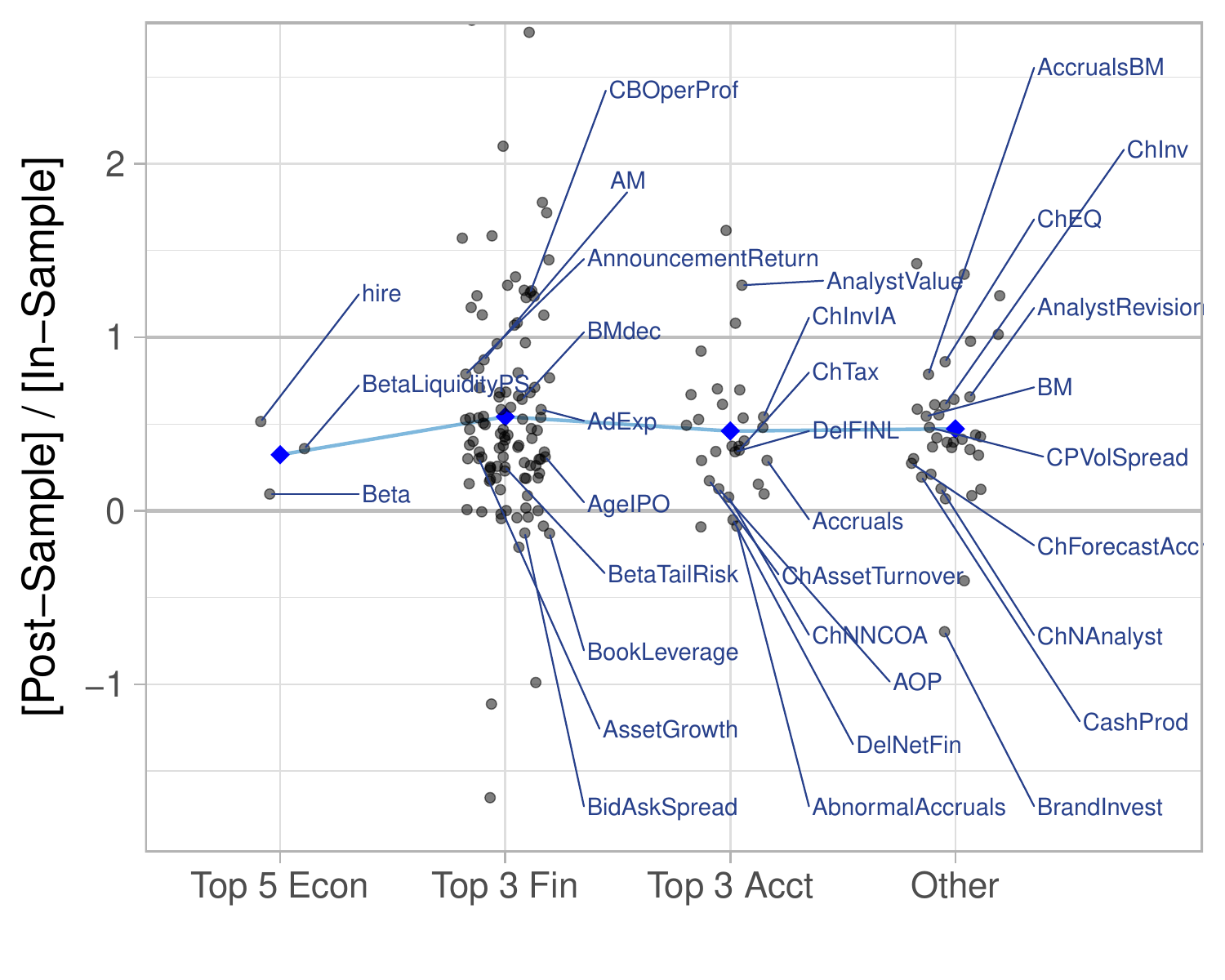}      
\end{figure}

\begin{table}[!h]\caption{Regression Estimates of Risk vs Mispricing Effects on Predictability Decay}
  \label{tab:mp-theory-no-reg}
  
  \begin{singlespace}
  \noindent We regress monthly long-short returns on indicator variables to quantify the effects of peer-reviewed risk vs mispricing explanations on predictability decay. ``Post-Sample'' is 1 if the month occurs after the predictor's sample ends and is zero otherwise. ``Post-Pub'' is defined similarly. ``Risk'' is 1 if peer review argues for a risk-based explanation (Table \ref{tab:hetero-count}) and 0 otherwise. ``Mispricing'' and ``Post-2004'' are defined similarly. Parentheses show standard errors clustered by month.  ``Null: Risk No Decay'' shows the $p$-value that tests whether risk-based returns do not decrease post-sample ((1) and (3)) or post-publication ((2) and (4)).  Risk-based predictors decay more than other predictors, but the difference is only marginally significant. The decay in risk-based predictors overall is highly significant.
  \end{singlespace}
  \begin{centering}
  \vspace{0ex}
  \par\end{centering}
  \centering{}\setlength{\tabcolsep}{2.0ex} \small
  \begin{center}
  % ==== begin paste
  % Table generated by Excel2LaTeX from sheet 'MP theory' 
  \begin{tabular}{lcccccc} \toprule 
   & \multicolumn{5}{c}{LHS:  Long-Short Strategy Return (bps pm, scaled) } \\
     \hline
  RHS Variables & (1) & (2) & (3) & (4) & (5) \\ 
     \hline
  Intercept & 71.4 & 71.4 & 71.4 & 71.4 & 73.0 \\ 
     & (3.7) & (3.7) & (3.7) & (3.7) & (3.9) \\ 
    Post-Sample & -28.9 & -25.4 & -25.5 & -22.9 & -5.6 \\ 
     & (5.6) & (7.1) & (6.6) & (10.7) & (8.2) \\ 
    Post-Pub &  & -4.2 &  & -3.0 &  \\ 
     &  & (7.8) &  & (12.5) &  \\ 
    Post-Sample x Risk & -19.1 & -7.6 & -22.5 & -10.1 & -15.6 \\ 
     & (7.7) & (10.2) & (8.5) & (12.9) & (7.6) \\ 
    Post-Pub x Risk &  & -15.2 &  & -16.4 &  \\ 
     &  & (13.3) &  & (15.9) &  \\ 
    Post-Sample x Mispricing &  &  & -4.5 & -3.2 &  \\ 
     &  &  & (5.8) & (11.0) &  \\ 
    Post-Pub x Mispricing &  &  &  & -1.8 &  \\ 
     &  &  &  & (12.0) &  \\ 
    Post-2004 &  &  &  &  & -33.7 \\ 
     &  &  &  &  & (9.9) \\ 
     \hline
  Null: Risk No Decay & $<$ 0.1\% & $<$ 0.1\% & $<$ 0.1\% & $<$ 0.1\% & $<$ 0.1\% \\ 
  \bottomrule
  \end{tabular}% 
  % ==== end paste
  \end{center} 
  \end{table}
  \clearpage
  
\begin{table}[!h]
   \caption{Model vs No Model}
   \label{tab:hetero-model}
   
   \begin{singlespace}
   \noindent This table compares predictors with any mathematical model (stylized, dynamic, or quantitative) versus those without formal models. `Raw' shows unadjusted returns. `CAPM' and `FF4' adjust for the CAPM and Fama-French three-factor model plus momentum, respectively, using sample-specific alphas. All returns are normalized to have a mean of 100 bps per month in the original papers' samples. Numbers in parentheses are standard errors clustered by calendar month and predictor.
   \end{singlespace}
   \begin{centering}
   \vspace{0ex}
   \par\end{centering}
   \centering{}\setlength{\tabcolsep}{2.0ex} \small
   \begin{center}
   \begin{tabular}{lcccccc}
      \toprule
      & \multicolumn{2}{c}{Raw} & \multicolumn{2}{c}{CAPM} & \multicolumn{2}{c}{FF4} \\
      \cmidrule(lr){2-3}\cmidrule(lr){4-5}\cmidrule(lr){6-7}
      & Return & Outperf. & Return & Outperf. & Return & Outperf. \\
      % FF4 figures copied from exhibits/ff4/Table_RiskAdjusted_AnyModelVsNoModel_ff4_t2.tex
      \midrule
      No Model  & 56 & 5 & 62 & 9 & 71 & -4 \\
                & (3) & (3) & (3) & (3) & (3) & (4) \\
      Any Model & 49 & 7 & 50 & 0 & 45 & -48 \\
                & (13) & (13) & (14) & (17) & (11) & (22) \\
      \bottomrule
   \end{tabular}
   \end{center} 
\end{table}

\begin{table}[htbp]
\centering
\small
\caption{Full Sample Risk-Adjusted Returns: Any Model vs No Model}
\label{tab:risk_adjusted_fs_anymodel}
\setlength{\tabcolsep}{2ex}
\begin{tabular}{lcccccc}
\toprule
  & \multicolumn{2}{c}{Raw} & \multicolumn{2}{c}{CAPM} & \multicolumn{2}{c}{FF3}
\\
\cmidrule(lr){2-3} \cmidrule(lr){4-5} \cmidrule(lr){6-7}
Group & Return & Outperf. & Return & Outperf. & Return & Outperf.
\\
\midrule
\multicolumn{7}{l}{\textbf{}}\\
\addlinespace[2pt]
No Model & 56 & 5 & 60 & 8 & 63 & 3
\\
 & (3) & (3) & (3) & (3) & (3) & (3)
\\[2pt]
Any Model & 49 & 7 & 56 & 9 & 56 & -11
\\
 & (13) & (13) & (13) & (14) & (13) & (15)
\\[2pt]
\bottomrule
\end{tabular}
\end{table}

% =====================================================
\clearpage
\newpage
\section{Why Do Published Predictors Decay?}\label{sec:intapp-struct-break}

Table \ref{tab:samp-split} illustrates two methods for documenting peer-reviewed predictability decay:
\begin{enumerate}
  \item Split at the end of the publication's sample period, following \citet{mclean2016does}
  \item Split in 2004 when high-speed internet became widely available, consistent with \citet{Chordia2014Have} and \citet{chen2022zeroing}
\end{enumerate}
Both approaches yield similar empirical results: a mean split date around 2000, a decay of about 50\%, with 85\% of predictors showing reduced effectiveness after the split.

\begin{table}[!h]
  \caption{Why Do Peer-Reviewed Returns Decay?}\label{tab:samp-split}
  \small
  
  Table compares splitting samples using various methods: (1) the end of the original sample period, (2) when high speed internet became widely available, and (3) by minimizing the mean squared residual a la \citet{bai1998estimating}. Each method leads to a similar average break date, magnitude of decay, and frequency of decay.  It is unclear which sample split best explains why peer-reviewed predictability decays.
  \vspace{2ex}
  
  \centering
  \setlength{\tabcolsep}{1.0ex}
  
\begin{tabular}{lcccc}
\toprule
\multicolumn{1}{c}{Event} & \multicolumn{1}{c}{Mean} & \multicolumn{2}{c}{Return (bps p.m.)} & \multicolumn{1}{c}{\% of Signals} \\  & \multicolumn{1}{c}{Date} & \multicolumn{1}{c}{Before} & \multicolumn{1}{c}{After} & \multicolumn{1}{c}{w/ Decay} \\ 
\midrule
1. Paper's Sample Ends & Feb 2000 & 72 & 37 & 85\\
2. High Speed Internet & Dec 2004 & 71 & 31 & 88\\
3. Data-Driven Break & Mar 2001 & 80 & 25 & 82\\
\bottomrule
\end{tabular}

\end{table}

Which split best explains why peer-reviewed predictability decays? To examine this, we compute data driven breaks for each predictor by minimizing the mean squared residual (as in \citet{bai1998estimating}). We then compare the data-driven breaks with the breaks specified by the two methods above.

\begin{figure}[!h]
  \caption{Data-Driven Breaks vs Paper Sample Ends}
  \label{fig:break-vs-sampend}
  Each marker is one published predictor. Data-driven breaks split the predictor's sample into two periods to minimize the mean squared residual (as in \citet{bai1998estimating}).
  \vspace{0.15in}
  
  \centering
  \includegraphics[width=0.55\textwidth]{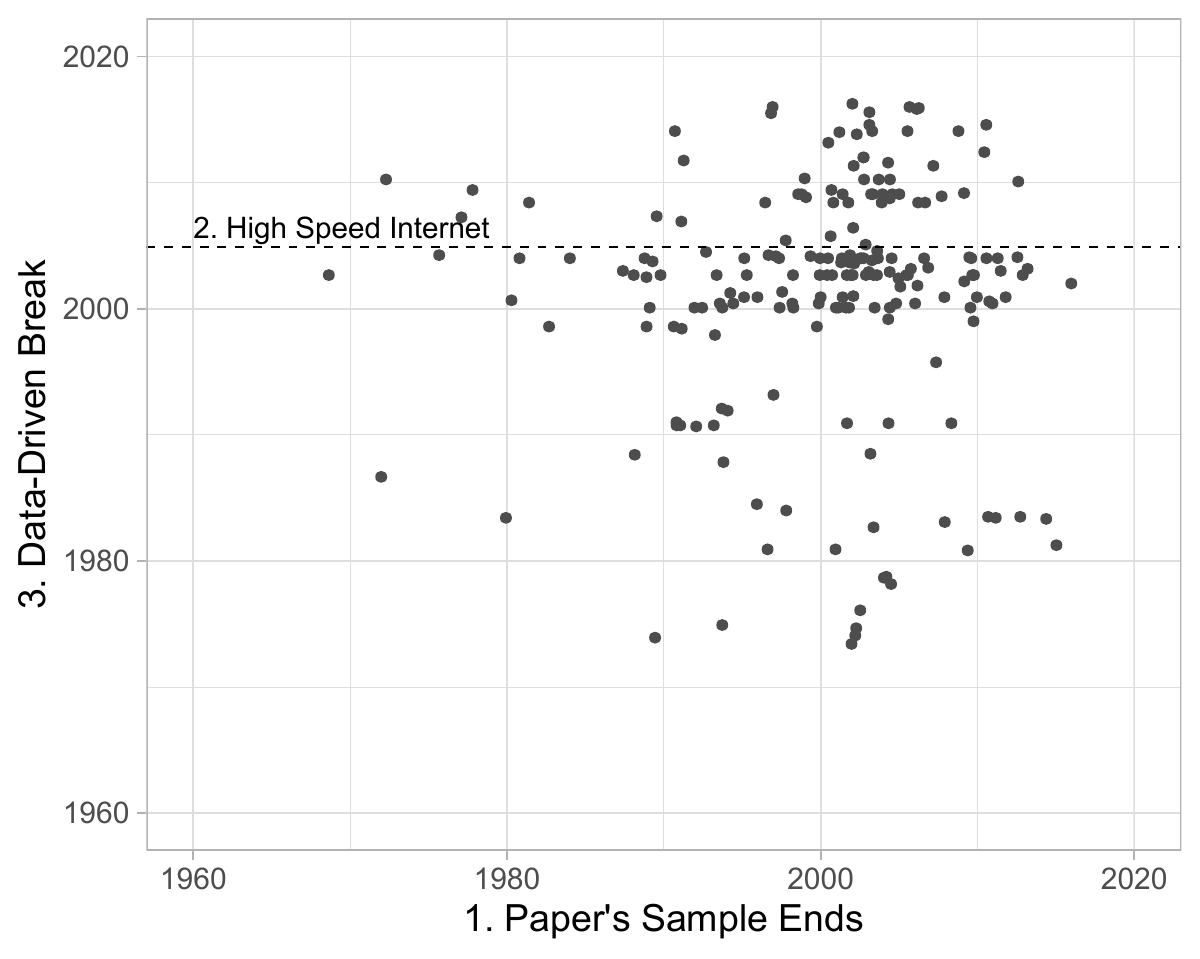}
\end{figure}

Figure \ref{fig:break-vs-sampend} shows the result. The scatter shows at best a mild relationship between the data-driven breaks and the papers' sample ends. Similarly, there is some clustering around the 2004 break, but the evidence is far from definitive. 

This result is natural given the noise in long-short returns. The typical monthly volatility is 350 bps, implying the standard error of a 60-month mean is 45 bps, making it impossible to tell if a predictor decays  in a particular 5-year period.

Thus, it is difficult to determine the fundamental cause of the relative decay of peer-reviewed and data-mined predictors. This observation leads to our focus on making inferences about post-sample performance. 

% =============================================
\end{appendices}

\clearpage{}
\newpage{}

\end{document}